\documentclass{jfm}

\graphicspath{{figures/}}

\usepackage{graphicx}
\usepackage{newtxtext}
\usepackage{newtxmath}
\usepackage{natbib}
\usepackage{hyperref}
\usepackage{enumitem}

\usepackage{amsmath}
\usepackage{amssymb}
\usepackage{mathrsfs}
\usepackage{mathtools}
\usepackage{graphicx}
\usepackage{caption}
\usepackage{subcaption}
\usepackage{color}
\usepackage{multirow}
\usepackage{caption}
\usepackage{algorithmic}
\usepackage{algorithm}

\usepackage{hyperref}
\hypersetup{
	colorlinks=true,
	linkcolor=blue,
	filecolor=magenta,      
	urlcolor=cyan,
}

\newtheorem{rem}{Remark}
\algsetup{indent=2em} 

\hypersetup{
    colorlinks = true,
    urlcolor   = blue,
    citecolor  = black,
}

\newcommand{\RomanNumeralCaps}[1]
\linenumbers

\shorttitle{Tempered Fractional LES Modeling}
\shortauthor{M. Samiee, A. Akhavan-Safaei, and M. Zayernouri}

\title{Tempered Fractional LES Modeling}

\author{Mehdi Samiee\aff{1,2}, Ali Akhavan-Safaei\aff{1,2},
 \and Mohsen Zayernouri\aff{1,3}\corresp{\email{zayern@msu.edu}}}

\affiliation{\aff{1}Department of Mechanical Engineering, Michigan State University, East Lansing, MI 48824, USA
\aff{2}Department of Computational Mathematics, Science and Engineering, Michigan State University, East Lansing, MI 48824, USA
\aff{3}Department of Statistics and Probability, Michigan State University, East Lansing, MI 48824, USA}

\begin{document}

\maketitle

\begin{abstract}
The presence of nonlocal interactions and intermittent signals in the homogeneous isotropic turbulence grant multi-point statistical functions a key role in formulating a new generation of large-eddy simulation (LES) models of higher fidelity. We establish a tempered fractional-order modeling framework for developing nonlocal LES subgrid-scale models, starting from the kinetic transport. We employ a tempered \textit{L\'evy}-stable distribution to represent  the source of turbulent effects at the kinetic level, and we rigorously show that the corresponding turbulence closure term emerges as the tempered fractional Laplacian, $(\Delta+\lambda)^{\alpha} (\cdot)$, for $\alpha \in (0,1)$, $\alpha \neq \frac{1}{2}$, and $\lambda>0$ in the filtered Navier-Stokes equations. Moreover, we prove the frame invariant properties of the proposed model, complying with the subgrid-scale stresses. To characterize the optimum values of model parameters and infer the enhanced efficiency of the tempered fractional subgrid-scale model, we develop a robust algorithm, involving two-point structure functions and conventional correlation coefficients. In an \textit{a priori} statistical study, we evaluate the capabilities of the developed model in fulfilling the closed essential requirements, obtained for a weaker sense of the ideal LES model \citep{Meneveau1994}. Finally, the model undergoes the \textit{a posteriori} analysis to ensure the numerical stability and pragmatic efficiency of the model.

\end{abstract}


\section{Introduction}
\label{Sec 0}
\noindent With the recent notable developments in computer technologies and, by extension, in the computational mechanics, there is a rapidly growing interest toward using large eddy simulations (LES) in a wide range of applications. Over the past decade, LES modeling has received an increasing attraction from scientific communities as a powerful and promising tool in connection with turbulence phenomena \citep[][]{Piomelli2014, Bouffanais2010}. In LES, one resolves the large energy-containing eddies by modeling the interplay between large and subgrid scale motions. Due to the tendency of small scales to homogeneous and universal dynamics, LES offers more accurate predictions comparing with the results of resolving the Reynolds-averaged Navier-Stokes (RANS) equations \citep[][]{Holgate2019, Zhiyin2015}. Furthermore, it lightens the burden of computational costs imposed by accurately capturing the dissipative scales, which renders LES more affordable than direct numerical simulations (DNS). 

Concurrent with the recent computational advancements, a marked shift occurred toward using artificial intelligence (AI) as an effective and tractable tool in turbulence modeling due to their significant capabilities in discovering anomalous structures and reproducing nonlocal statistical properties \citep[][]{Beck2020}. 
In this paradigm, turbulence modeling comes into two divisions:

\vspace{0.02 in}
\noindent \textbf{Machine learning based approaches:} 
They introduce advancements in prediction capabilities and reconstructing turbulence structures. Several assorted machine learning (ML) algorithms were proposed for turbulence closure problems including kernel regression and a deep neural network \citep[][]{Pawar2020, Sirignano2020, Portwood2020}. Essentially, pure machine learning based approaches are limited by the representativeness training dataset though they appear to be simpler for implementation. Moreover, to pinpoint complex patterns, large volumes of data are required for the algorithms to learn physical constraints (e.g., frame invariance) and statistical properties, which secondarily makes further complications like optimizing of data compression \citep[][]{Chao2020}. This reveals the significance of physics based models in mentoring the AI approaches and pushing hybrid models as a new direction \citep[][]{Taghizadeh2020, Willard2020, Jouybari2020, Patra2018}, which exploit ML algorithms with a significant reduction in the required input data.

\vspace{0.02 in}
\noindent \textbf{Physics based approaches:} 
They introduce a mathematical representation of physical structures through a number of parameters with a sufficient amount of information. Contrary to ML based approaches, physics based models do not involve large volumes of data although they are inherently limited by the model incompleteness or the complexity of parameterizing physical structures \citep[][]{Chao2020}. Accordingly, it is markedly essential to entail the underlying statistical properties in formulating and inferring an optimum model in a numerically rigorous framework, which links a variety of research disciplines like turbulence, numerical and statistical analysis, and data science. This approach employs principles of physics and borrows insights from the statistical analysis to form a model for real phenomena, which can also be used to guide the ML algorithms \citep[see e.g.,][]{You2021, Kurz2020, AkhavanScalar2020}. 

Establishment of such a physically-consistent LES model ties strongly with characterization of nonlocal turbulence mechanisms and a better understanding of \textit{anomalous} structures. As a puzzling feature, the non-Gaussian behavior of turbulent dynamics is linked to the spatial intermittency of small-scale motions, which is embodied in the form of very thin and elongated vortices \citep[][]{Vincent1991, Laval2001}. Technically, the nonlocal closure of Navier-Stokes (NS) equations, originated from the Green's function of the Laplacian operator for solving Poisson pressure equation, induces long-range interactions (nonlocal triadic structures) in spectral space of homogeneous turbulence \citep[][]{Sagaut2008}. In a preliminary investigation of isotropic turbulence \citep[][]{She1990}, the significant role of highly vortical structures, typically tube-like, was disclosed on generating nonlocal dynamics and coherence of turbulence. Supported by \citep[][]{Laval2001}, nonlocality as a crucial element in generating intermittent structures has tendency to prevail the local interactions by orders of magnitude. She and Leveque attempted to express self-similar structures in terms of a squence of moment ratios for the energy dissipation field \citep[][]{She1994}. 
Recently, Mishra and Girimaji \citep[][]{Mishra2019} studied the role of pressure on nonlocal mechanisms in incompressible turbulent flows and identified the intercomponent of energy transfer by the rapid pressure strain correlation. For more information, the reader is referred to \citep[][]{Buaria2020, Pang2020, Akhavan-Safaei2020, Hamlington2008}.

From this perspective, an ideal subgrid-scale (SGS) model represent correctly statistics of the filtered real turbulence at the resolved levels. Given the dependence of an ideal model on an infinite-dimensional set of multi-point statistics, it would be more practical to define a weaker set of conditions in study of SGS parameterization \citep[][]{Sagaut2008}. As one of the earliest studies on statistical analysis of LES, \cite{Meneveau1994} derived a closed set of necessary, yet mild, conditions to fulfill \textit{a priori} consistency in SGS quantities. More generally, the Karman-Howarth (KH) theorem for anisotropic turbulent flows were studied in \citep[][]{Hill2002} by eliminating pressure velocity correlations to determine the two-point structure function equations. By proposing a hyper-eddy viscosity term in \citep[][]{Cerutti2000}, SGS dissipation spectrum were measured in locally isotropic turbulence to assess ability of classical two-point closures in prediction of the mean energy transfer. Recently, some of the prevailed challenges in developing an optimal LES model was reviewed succinctly by \citep[][]{Moser2020}. This review presents a clear set of statistical characteristics in performing an \textit{a priori} analysis and providing adequate information for optimizing SGS models. 
   
Statistical descriptions of an ideal closure model derive a desire for developing nonlocal approaches in terms of two-point high-order structure functions in a rigorous mathematical framework. The eddy damped quasi-normal Markovian (EDQNM) approach, described in \citep[][]{Briard2016}, undertakes closing of SGS motions in spectral space by involving high-order statistical moments. As a functional approach, direct interaction approximation pushes the non-Markovanized stochastic models to the direction of turbulence closure problem, whose solutions are constructed in a fraction form \citep[][]{Shivamoggi2019}. Furthermore, multifractal models \citep[][]{Yang2017, Burton2005} suggest a potential realizable strategy to accurately capture anomalous scaling exponents, observed in turbulent velocity increments. In addressing statistical local and nonlocal interactions, this progress proceeds with modeling turbulent effects at the kinetic level. \cite{Premnath2009} developed a framework for applying dynamic procedure in the lattice-Boltzmann method for LES of inhomogeneous and anisotropic turbulent flows.  A new collision approach were proposed by \citep[][]{Jacob2018} for LES of weakly compressible flows using two forms of the modified Bhatanagar-Gross-Krook (BGK) collision operators. For more comprehensive review of the literature, we refer the reader to \citep[][]{Jin2018, Sagaut2010}.  

Focusing on the key ideas of \textit{(i)} describing of anomalous structures in turbulence and \textit{(ii)} nonlocal closure modeling, fractional calculus appears to be a tractable mathematical tool due to their power-law or logarithmic types of kernel. As an alternative approach to standard methods, they leverage their inherent potentials in representing long-range interactions, self-similar structures, sharp peaks, and memory effects in a variety of application \citep[see][]{Burkovska2020, Kharazmi2019, Zayernouri2013}. \cite{Egolf2017} generalized Reynolds shear stresses in local zero-equation to the fractional counterparts. Furthermore, Epps and Cushmann-Roisin derived fractional NS equations from the Boltzmann transport equation in \citep[][]{Epps2018}, which supply profound understanding of turbulent nonlocal effects at the kinetic level. For more information, \cite{Egolf2020} provided a comprehensive overview of fractional and nonlocal turbulence, spanning from coherent structures to state-of-the-art ideas on closure modeling in canonical flows. Recently, \cite{DiLeoni2020} contributed in fractional LES modeling by developing a two-point correlation based model in a robust physically-meaning framework.

In the class of nonlocal models, \cite{SamieePoF2020} laid out a mathematical framework for developing fractional models, which starts treating turbulence effects at the kinetic level. In a precise derivation, the proposed distribution function in the closed form of filtered collision operator, turns into a fractional model in the LES equations. Throughout a data-driven approach, \cite{AkhavanScalar2020} extended the fractional modeling to the LES of scalar turbulence using two-point correlation functions between the SGS scalar flux and filtered scalar gradient.

In the specific case of isotropic turbulent flows, cascading of energy from large to small scales expresses a self-similar behavior in the inertial range, and then it falls exponentially into the dissipation range. Inspired by such real physics phenomena, we focus on developing a nonlocal model by employing a tempered heavy-tailed distribution within the proposed fractional framework, which contributes in tempered fractional SGS (TFSGS) modeling. Such a tractable fractional operator offers a great flexibility in characterizing nonlocal structures in the turbulent inertial and dissipation ranges through fractional and tempering parameters. To achieve the enhanced performance of the proposed model, we also present an optimization algorithm, involving two-point structure functions. Regarding the best approximation of an ideal physics based model, the optimized TFSGS model restores many essential statistical properties of SGS stresses and presents an \textit{a priori} consistency in the dissipation spectrum.

The paper is organized as follows. In section \ref{Sec 1}, we introduce some preliminaries of tempered fractional calculus. We outline a mathematical framework in section \ref{Sec 2} to develop the tempered fractional model from the Boltzmann transport equation and derive the corresponding forms for SGS quantities. Within a statistical framework, we present a two-point structure based algorithm to infer the optimal behavior of the tempered fractional model in section \ref{Sec 3}. Using the DNS database of an stationary isotropic turbulent flow, we evaluate the statistical \textit{a priori} analysis and perform a comparative study on the two-point structure functions in section \ref{Sec 4}. Moreover, we study numerical stability of the LES solutions through an \textit{a posteriori} investigation in section \ref{Sec 4}. Lastly, section \ref{Sec 6} summarize the findings with a conclusion.

%
%
%

\section{Preliminaries on Tempered Fractional Calculus}
\label{Sec 1}

\noindent Fractional calculus introduces well-established mathematical tools for an accurate description of anomalous phenomena, ubiquitous in a wide range of applications from bio-tissues \citep[][]{Naghib2018, Ionescu2017} and material science \citep[][]{Suzuki2020, Suzuki2019, Meral2010} to vibration \citep[][]{SuzukiV2020}, porous media \citep[][]{Samiee2020, Zaky2020, Xie2019} and turbulence \citep[][]{AkhavanScalar2020, DiLeoni2020, Epps2018}. As alternative approaches to the standard nonlinear models, fractional models offer a great potential in capturing heavy-tailed distributions, self-similar structures, nonlocal interactions, and memory effects. This potential is substantially indicated by power-law or logarithmic kernels of convolution type in the corresponding fractional operators. From the stochastic point of view, fractional transport models arise from the heavy-tailed distribution functions in modeling the underlying super- or sub-diffusive motions of particles in complex heterogeneous systems at the microscopic level \citep[][]{SamieeJCPII2019}. Nevertheless, common patterns in nature follow finite variance dynamics, which urges the role of tempered fractional calculus as a more sophisticated approach in representing natural cut-offs in real applications and retaining their finite statistical properties. 

Recalling from \citep[][]{Sabzikar2015, Zayernouri2015}, we begin by the definitions of the left- and right-sided tempered fractional derivatives respectively as 
\begin{equation}
\label{eq1-1}
\prescript{}{}{\mathcal{D}}_{\pm x}^{\alpha, \lambda} u(x) = \frac{\alpha}{\Gamma(1-\alpha)}  \int_{0}^{\infty} \frac{u(x)-u(x\mp s)}{s^{\alpha+1} e^{\lambda s} }\,\, ds, 
\end{equation}
where the fractional derivative order, $\alpha \in (0,1)$, and the tempering parameter, $\lambda>0$. Also, $\Gamma(\cdot)$ represents a Gamma function. For $\alpha \in (1,2)$, the corresponding fractional derivatives are given by:
\begin{equation}
\label{eq1-3}
\prescript{}{}{\mathcal{D}}_{\pm x}^{\alpha, \lambda} u(x) = \frac{\alpha(\alpha-1)}{\Gamma(2-\alpha)}  \int_{0}^{\infty} \frac{u(x \mp s)-u(x) \pm s \frac{du(x)}{dx} }{s^{\alpha+1} e^{\lambda s} }\,\,ds. 
\end{equation}
The link between the derivatives in \eqref{eq1-3} and \eqref{eq1-7} and their counterparts in the Riemann-Liouville sense are described by
\begin{eqnarray}
	\label{eq1-7}
	\prescript{RL}{}{\mathcal{D}}_{\pm x}^{\alpha, \lambda} u(x) &=&
	\prescript{}{}{\mathcal{D}}_{\pm x}^{\alpha, \lambda} u(x) + \lambda^{\alpha}u(x) 
	\\
	\label{eq1-8}
	\prescript{RL}{}{\mathcal{D}}_{\pm x}^{\alpha, \lambda} u(x) &=&
	\prescript{}{}{\mathcal{D}}_{\pm x}^{\alpha, \lambda} u(x) + \lambda^{\alpha}u(x) \pm \alpha \lambda^{\alpha-1} \frac{du(x)}{dx}. 
\end{eqnarray}
In particular, for $n \ge 0$ the tempered integer-order derivatives are reduced as 
\begin{equation}
	\label{eq1-4}
	\prescript{RL}{}{\mathcal{D}}_{+x}^{n, \lambda} u(x) = e^{-\lambda x} \frac{d^n \left( e^{\lambda x} u(x)  \right)}{dx^n},
\end{equation}
which recover the classic integer-order derivatives as $\lambda \rightarrow$ 0. 

\vspace{0.1 in}
Let $\mathcal{F}  \left[ u \right] (\xi)$ denote the Fourier transform of $u$, where $\xi$ is the Fourier numbers. Then, we obtain $$\mathcal{F}  \left[ \prescript{RL}{}{\mathcal{D}}_{\pm x}^{\alpha, \lambda} u(x) \right] = (\lambda\pm \mathfrak{i} \, \xi)^{\alpha} \, \mathcal{F}  \left[ u \right] (\xi).$$ In this context, the corresponding Fourier transform of the left- and right-sided tempered fractional integrals are given by $$\mathcal{F}  \left[ \prescript{RL}{}{\mathcal{I}}_{\pm x}^{\alpha, \lambda} u(x) \right] (\xi) = (\lambda\pm \mathfrak{i} \, \xi)^{-\alpha} \, \mathcal{F}  \left[ u\right] (\xi).$$ Evidently, tempered integrals and derivatives functions as inverse operators when $u$ possesses sufficient regularity \citep[see][]{Zhang2018, Sabzikar2015}. Moreover, tempered fractional operators preserve semi-group property, which prepares a useful and rigorous framework for further numerical considerations. 

\subsection{Tempered fractional Laplacian}
\noindent Denoted by $(\Delta+\lambda)^{\alpha} (\cdot)$, we define the tempered fractional Laplacian of the integral form as 
\begin{eqnarray}
\label{eq1-9}
(\Delta+\lambda)^{\alpha} u(\boldsymbol{x}) &=& C_{d,\alpha} \,  \mathrm{P.V.} \int_{\mathbb{R}^d }\frac{u(\boldsymbol{x})-u(\boldsymbol{s})}{e^{\lambda\vert\boldsymbol{x}-\boldsymbol{s}\vert}\vert \boldsymbol{x}-\boldsymbol{s}\vert^{2\alpha+d}} d\boldsymbol{s},
\end{eqnarray}
where $C_{d,\alpha} = \frac{-\Gamma(\frac{d}{2})}{2 \pi^{\frac{d}{2}}\Gamma(-2\alpha)} \frac{1}{\prescript{}{2}{F}^{}_{1}(-\alpha, \frac{d+2\alpha-1}{2};\frac{d}{2};1)}$ for $\alpha \in (0,1)$ and $\alpha \neq \frac{1}{2}$. In particular, for $d=1$  $(\Delta+\lambda)^{\alpha}$ is reduced to the so-called \textit{Riesz} fractional form, described by 
\begin{eqnarray}
\label{eq1-10}
(\Delta+\lambda)^{\alpha} u(x) &=& (-1)^{\lfloor 2\alpha \rfloor +1}\frac{\prescript{RL}{}{\mathcal{D}}_{+x}^{\alpha, \lambda} u(x) + \prescript{RL}{}{\mathcal{D}}_{- x}^{\alpha, \lambda} u(x)}{2}
\nonumber
\\
\nonumber
&=& C_{\alpha} \,\, \mathrm{P.V.} \int_{\mathbb{R} }\frac{u(x)-u(s)}{e^{\lambda\vert x-s\vert}\vert x-s\vert^{2\alpha+1}} ds,
\end{eqnarray}
where $C_{\alpha} = \frac{-\Gamma(\frac{1}{2})}{2 \pi^{\frac{1}{2}}\Gamma(-2\alpha)} \frac{1}{\cos(\pi\alpha)}$ \citep[see][]{Zhang2018}. 
In Appendix \ref{appA}, we detail the derivation of the Fourier transform of $(\Delta+\lambda)^{\alpha} u(\boldsymbol{x}) $, formulated as
\begin{eqnarray}
\label{eq1-11}
\mathcal{F} \big {[}(\Delta+\lambda)^{\alpha} u(\boldsymbol{x}) \big {]} (\boldsymbol{\xi}) &=& \mathfrak{C}_{d,\alpha} \times 
\\
\nonumber 
&&
\left( \lambda^{2\alpha} - (\lambda^2+\xi^2)^{\alpha} \prescript{}{2}{F}^{}_{1}(-\alpha, \frac{d+2\alpha-1}{2};\frac{d}{2};\frac{\xi^2}{\xi^2+\lambda^2}) \right)\mathcal{F}  \left[ u \right] (\boldsymbol{\xi}),
\end{eqnarray}
in which $\mathfrak{C}_{d,\alpha} = \frac{1}{\prescript{}{2}{F}^{}_{1}(-\alpha, \frac{d+2\alpha-1}{2};\frac{d}{2};1)}$ and $\xi = \vert \boldsymbol{\xi} \vert$. For $d=3$, we define $\mathfrak{C}_{\alpha} = \frac{1}{\prescript{}{2}{F}^{}_{1}(-\alpha, \frac{2+2\alpha}{2};\frac{3}{2};1)}$. It is worth noting that when $\lambda$ approaches $0$, we recover the usual fractional Laplacian in both integral or Fourier forms.

\section{Boltzmannian Framework}
\label{Sec 2}
\noindent The kinetic Boltzmann transport (BT) is a formal framework for describing fluid particle motions over a wide range of flow physics (e.g., rarefied gas flows and turbulence). This framework offers a great potential for statistical description of turbulent small-scales towards a better understanding of coherent structures in turbulence yet at the kinetic level. As an alternative approach in turbulent closure modeling, reconciling SGS terms in the BT and the Navier-Stokes (NS) equations can conceivably give rise to a rigorous physics based model at the continuum level. 

Within the BT framework proposed in \citep[][]{SamieePoF2020}, we develop a SGS model, respecting the statistical and physical properties of turbulent unresolved-scale motions.

\subsection{Subgrid-scale modeling}
\label{Sec 2-1}
\noindent In the description of incompressible turbulent flows, we consider large eddy simulation (LES) equations \citep[][]{Pope2001}, governing the dynamics of the resolved-scale flow variables,
\begin{eqnarray}
\label{Sec2-1}
\frac{\partial \boldsymbol{\bar{V}}}{\partial t}+ \boldsymbol{\bar{V}}\cdot \nabla \boldsymbol{\bar{V}}&=&-\frac{1}{\rho} \, \nabla \bar{p}+ \nu \, \nabla \cdot \mathsfbi{\bar{S}} - \nabla \cdot \mathsfbi{T}^{\mathcal{R}},
\end{eqnarray}
where in the index form $\boldsymbol{\bar{V}}(\boldsymbol{x},t)=\bar{V}_i$ and $\bar{p}(\boldsymbol{x},t)$ represent the velocity and the pressure fields for $i=1,2,3$ and $\boldsymbol{x}=x_i$. Moreover, $\nu$ and $\rho$ denote the kinematic viscosity and the density, respectively. Considering $\mathcal{L}$ as the filter width, the filtered field is obtained in the form of $\boldsymbol{\bar{V}}= G \ast \boldsymbol{V}$, where $G = G(\boldsymbol{x})$ denotes the kernel of a spatial isotropic filtering type and $*$ is the convolution operator. By implementing the filtering operation, we decomposes the velocity field, $\boldsymbol{V}$, into the filtered (resolved), $\boldsymbol{\bar{V}}$, and the residual, $\boldsymbol{v}$, components. In \eqref{Sec2-1}, the filtered strain rate, $\mathsfbi{\bar{S}}$, and the SGS stress tensor, $\mathsfbi{T}^{\mathcal{R}}$, are defined by $\mathsfi{\bar{S}}_{ij}=\frac{1}{2}(\frac{\partial V_i}{\partial x_j}+\frac{\partial V_j}{\partial x_i})$ and $\mathsfi{T}^{\mathcal{R}}_{ij}=\overline{V_i  V_j}-\bar{V}_i\bar{V}_j$. 

Since the filtering operator cannot commute with the nonlinear terms in the NS equations, SGS stresses must be modeled in terms of the resolved velocity field. As a common yet reliable approach, Smagorinsky \citep[][]{smagorinsky1963general} offered modeling the SGS stresses borrowing the Boussinessq approximation from the kinetic theory such that $\mathsfbi{T}^{\mathcal{R}} = -2 \nu_{R} \mathsfbi{\bar{S}}$ and $\nu_{R}$ is indicated by $\nu_{R}= (C_s \mathcal{L})^2\,\vert \mathsfbi{\bar{S}} \vert$, where $\vert \mathsfbi{\bar{S}} \vert = \sqrt{2\mathsfi{\bar{S}}_{ij}\mathsfi{\bar{S}}_{ij}}$ and $C_s$ is the Smagorinsky (SMG) constant. 

\subsection{The BGK equation and the closure problem}
\label{Sec 2-2}
\noindent Starting from the Boltzmann kinetic theory \citep[][]{Soto2016}, the evolution of mass distribution function $f$ is governed by the Boltzmann transport (BT) equation as 
\begin{equation}
\label{Sec2-2}
\frac{\partial f}{\partial t} + \boldsymbol{u}\cdot \nabla f = \Omega(f),
\end{equation}
in which $f(t,\boldsymbol{x},\boldsymbol{u})\,d\boldsymbol{x}d\boldsymbol{u}$ represent the probability of finding mass of particles, located within volume $d\boldsymbol{x}d\boldsymbol{u}$ centered on a specific location, $\boldsymbol{x}$, and speed, $\boldsymbol{u}$, at time $t$. It is worth noting that in the particle phase space $\boldsymbol{x}$, $\boldsymbol{u}$, and $t$ are independent variables. Technically, the left-hand side of \eqref{Sec2-2} concerns the streaming of non-reacting particles in absence of any body force and the right-hand side represent the collision operator. The most common form of $\Omega(f)$ with a single collision is the so-called BGK approximation \citep[][]{Soto2016}, given by
\begin{equation}
\label{Sec2-2-1}
\Omega(f)=-\frac{f-f^{eq}}{\tau},
\end{equation}
where $\tau$ represent the single relaxation time. In the case of incompressible flows with a roughly constant temperature, $\tau$ is assumed to be independent of macroscopic flow field velocity and pressure. Moreover, under the circumstances of thermodynamic equilibrium of particles, $f^{eq}(\Delta)$ serves as 
\begin{equation}
\label{Sec2-3}
f^{eq}(\Delta) = \frac{\rho}{U^3}F(\Delta),
\end{equation}
where $F(\Delta)=e^{-\Delta/2}$, $\Delta=\frac{\vert \boldsymbol{u}-\boldsymbol{V}\vert^2}{U^2}$ as an isotropic Maxwellian distribution and $U$ denotes the agitation speed. More specifically, $U=\sqrt{3\textit{k}_B T/m}$, in which $\textit{k}_B$, $T$, and $m$ represent the Boltzmann constant, room temperature, and the molecular weight of air. 

By recalling the basics of BT equation from \citep[][]{Epps2018, SamieePoF2020}, we introduce the following quantities: $L$ as the macroscopic length scale, $l_s$ as the microscopic characteristic length associated with the Kolmogorov length scale, $l_m$ as the average distance, traveled by a particle between successive collisions. Furthermore, we define $\boldsymbol{x}^{\prime}$ as the location of particles before scattering, where $\boldsymbol{x}$ is the current location. Thus, $\boldsymbol{x}^{\prime}=\boldsymbol{x}-(t-t^{\prime})\boldsymbol{u}$, where $\boldsymbol{u}$ is assumed to be constant during $t-t^{\prime}$. The analytical solution of \eqref{Sec2-2} and \eqref{Sec2-2-1} is given by
\begin{eqnarray}
\label{Sec2-4}
f(t,\boldsymbol{x},\boldsymbol{u}) &=& \int_{0}^{\infty} e^{-s} \,f^{eq}(t-s\tau,\boldsymbol{x}-s\tau \boldsymbol{u},\boldsymbol{u}) \,  ds
\nonumber
\\
&=&\int_{0}^{\infty} e^{-s} f^{eq}_{s,s}(\Delta) ds,
\end{eqnarray}
where $s\equiv\frac{t-t^{\prime}}{\tau}$ and $f^{eq}_{s,s}(\Delta)=f^{eq}(t-s\tau,\boldsymbol{x}-s\tau \boldsymbol{u},\boldsymbol{u})$. 

\vspace{0.05 in}
\begin{rem}
\label{remark 1}
In order to develop an LES model within the kinetic transport framework, we constrain our attention to the BT equation with the BGK collision approximation, involving a single relaxation time. Moreover, we follow Assumption 1 in \citep[][pp. 4]{SamieePoF2020} in the further derivations to establish a physical connection between the collision operator and the convective terms at the continuum level.
\end{rem}
\vspace{0.05 in}

In description of turbulence effects at the kinetic level, we decompose $f$ into the filtered, $\bar{f}$, and residual values, $f'$, where $\bar{f}=G*f$. As defined previously, $G$ represents the kernel of any generic spatial isotropic filtering type. Then, the filtered kinetic transport for $\bar{f}$ suffices:
\begin{equation}
\label{Sec2-5}
\frac{\partial \bar{f}}{\partial t} + \boldsymbol{u}\cdot \nabla \bar{f} = - \frac{\bar{f}-\overline{f^{eq}(\Delta)}}{\tau},
\end{equation}
in which $u$ is independent of $t$ and $\boldsymbol{x}$. Ensuing \eqref{Sec2-4}, the analytical solution of \eqref{Sec2-5} is described by 
\begin{equation}
\label{Sec2-6}
\bar{f} (t,\boldsymbol{x}, \boldsymbol{u}) =\int_{0}^{\infty} e^{-s} \,  \overline{f^{eq}_{s,s}(\Delta)}\, ds,
\end{equation}
where $\overline{f^{eq}_{s,s}(\Delta)}=\overline{f^{eq}_{}\big (\Delta(t-s\tau,\boldsymbol{x}-s\tau \boldsymbol{u},\boldsymbol{u}) \big )}$. Let define $\bar{\Delta}:=\frac{\vert \boldsymbol{u}-\bar{\boldsymbol{V}} \vert^2}{U^2}$. Due to the nonlinear character of the collision operator \citep[][]{Gir2007}, the filtering operation does not commute with $\Omega$, which yields the following inequality as
\begin{equation}
\label{Sec2-7}
\overline{f^{eq}(\Delta)} = \frac{\rho}{U^3} \, \overline{e^{-\Delta/2}} \neq \frac{\rho}{U^3} \, e^{-\bar{\Delta}/2}=f^{eq}(\bar{\Delta}).
\end{equation}
This inequality gives rise to the so-called turbulence \textit{closure problem} at the kinetic level. From the mathematical standpoint, the SGS motions stem from the convective nonlinear terms in the NS equations, which resembles with the corresponding advective term of the BT equation. Therefore, it seems natural to recognize $\boldsymbol{u}\cdot \nabla f$ responsible for the unresolved turbulence effects in the BT equation, they manifest implicitly via the filtered collision operator though. That is, the filtered collision term in the right-side of \eqref{Sec2-5} undertakes not only molecular collisions, but also the embedded SGS motions. By emphasizing on the importance of modeling $\overline{f^{eq}(\Delta)}$ in the filtered collision operator, we review some different approaches in treating nonlinear effects. 

\vspace{0.1 in}
\noindent \textbf{Classical approaches:} As a common practice in modeling the SGS closures, the attentions were directed toward with eddy-viscosity approximations by employing a modified relaxation time, $\tau^{\star}$, in the BT equation \citep[e.g.,][]{Sagaut2010}. Therefore, the proposed filtered BT equation reads as
\begin{equation}
\label{Sec2-7-1}
\frac{\partial \bar{f}}{\partial t} + \boldsymbol{u}\cdot \nabla \bar{f} = - \frac{\bar{f}-f^{eq}(\bar{\Delta})}{\tau^{\star}}.
\end{equation}
In this approach, the inequality in \eqref{Sec2-7} is disregarded through using $\tau^{\star}$, which renders the SGS model inappropriate for reproducing many features of the SGS motions. Nevertheless, there some non-eddy viscosity models within the lattice Boltzmann framework, which make use of \eqref{Sec2-7} to propose more consistent SGS model. For more details, the reader is referred to \citep[][]{Chen2004, Premnath2009, malaspinas2012consistent}.

\vspace{0.05 in}
\noindent \textbf{Fractional approach:} In the proposed framework in \citep{SamieePoF2020}, the modeling of turbulence nonlinear effects begins with closing the filtered collision operator, where the multi-exponential behavior of $\overline{f^{eq}(\Delta)}$ is approximated properly by a heavy-tailed distribution function. Therefore, the $\overline{f^{eq}(\Delta)}$ in \eqref{Sec2-5} is described by
\begin{equation}
\label{Sec2-8}
\overline{f^{eq}(\Delta)} - f^{eq}(\bar{\Delta}) \simeq  f^{\beta}(\bar{\Delta}),
\end{equation}
where $f^{\beta}(\bar{\Delta}) = \frac{\rho}{U^3}F^{\beta}(\Delta)$ and $F^{\beta}(\Delta)$ denotes an isotropic \textit{L\'evy} $\beta$-stable distribution. By taking the first moment of \eqref{Sec2-5}, one derives the corresponding fractional Laplacian operator, termed as fractional SGS (FSGS) model, at the continuum level, where 
\begin{equation}
\label{Sec2-8-1}
(\nabla \cdot \mathsfbi{T}^{R})=\mu_{\alpha} (-\Delta)^{\alpha} \bar{\boldsymbol{V}},
\end{equation}
where $\mu_{\alpha}=\frac{\rho (U\tau)^{2\alpha}\Gamma(2\alpha+1)}{\tau}\, \frac{2^{2\alpha} \Gamma(\alpha+d/2)}{\pi^{d/2} \Gamma(-\alpha)} \, \mathrm{c}_{\alpha}$ for $\alpha \in (0,1)$ and $\mathrm{c}_{\alpha}$ is a real-valued constant. In principle, the choice of distribution function in \eqref{Sec2-8} gives rise to a nonlocal operator of the resolved flow field in \eqref{Sec2-1} as an SGS model. 

Despite the notable potentials of the FSGS model in maintaining some important physical and mathematical properties of the SGS stresses, it 
lacks a finite second-order statistical moment. To control this statistical barrier in the FSGS model and to achieve more congruence between both sides of \eqref{Sec2-8}, we seek a finite-variance alternative for the \textit{L\'evy} $\beta$-stable distribution by employing the tempered counterpart and thereby a more flexible and predictive fractional operator in the LES equations in the following subsection.

\subsection{Tempered fractional SGS modeling}
\label{Sec 2-3}
\noindent Multi-exponential functions express a power-law behavior in the moderate range of distribution and eventually relaxes into an exponential decay \citep[see][]{Evin2016}. By engaging more exponential terms to a multi-exponential function, the corresponding power-law behavior extends toward long ranges; however, it is bound to vanish exponentially at the tail of the distribution, enforced by nature of the physical phenomenon. As a rich class of stochastic functions for fitting into realistic phenomena, tempered stable distributions \citep[][]{Sabzikar2015} resemble a sheer power-law at the moderate range and then converge to an exponential decay. 

Inspired by this argument, we propose to model $\overline{f^{eq}(\Delta)}$ with a coefficient of tempered \textit{L\'evy} $\beta$-stable distribution, denoted by $f^{\beta, \lambda}(\bar{\Delta})$, within the proposed fractional framework as
\begin{equation}
\label{Sec2-9}
\overline{f^{eq}(\Delta)} - f^{eq}(\bar{\Delta}) \simeq f^{Model}(\bar{\Delta}) = \mathrm{c}_{\beta,\lambda}\,  f^{\beta, \lambda}(\bar{\Delta}),
\end{equation}
where $ \mathrm{c}_{\beta,\lambda}$ is a real-valued constant number. Moreover, we consider $\beta\in (-1-\frac{d}{2}, \, -\frac{d}{2})$, $\lambda >0$ and $d=3$ represents dimension of the physical domain. Therefore the filtered BT equation reads as
\begin{eqnarray}
\label{Sec2-10}
\frac{\partial \bar{f}}{\partial t} + \boldsymbol{u}\cdot \nabla \bar{f} &=& - \frac{\bar{f}-f^{eq}(\bar{\Delta})+f^{eq}(\bar{\Delta})-\overline{f^{eq}(\Delta)}}{\tau} 
\nonumber
\\
&\simeq& - \frac{\bar{f}-f^{eq}(\bar{\Delta})-f^{Model}(\bar{\Delta})}{\tau}.
\end{eqnarray}
For the sake of simplicity, we take $f^{*}(\bar{\Delta}) = f^{eq}(\bar{\Delta})+f^{Model}(\bar{\Delta})$. The approximation in \eqref{Sec2-10} conceivably provides a good fit into the filtered collision operator by maintaining the significant statistical features of $\overline{f^{eq}(\Delta)}$ and sets a physically richer starting point for developing a more expressive nonlocal LES model at the continuum level. 

In this regard, the macroscopic variables associated with the flow field can be reconstructed according to 
\begin{eqnarray}
\label{Sec2-11}
\bar{\rho} &=& \int_{\mathbb{R}^d} \bar{f}(t,\boldsymbol{x},\boldsymbol{u}) d\boldsymbol{u},
\\
\bar{V}_i&=&\frac{1}{\rho} \int_{\mathbb{R}^d} u_i \, \bar{f}(t,\boldsymbol{x},\boldsymbol{u}) d\boldsymbol{u}, \quad i=1,2,3,
\end{eqnarray}
where $\bar{\rho}=\rho$ for an incompressible flow. To establish the connection between the kinetic description and the filtered NS equation, we proceed with deriving the macroscopic form of \eqref{Sec2-10} by multiplying it with $\boldsymbol{u}$, and integrating over the kinetic momentum, which yields
\begin{eqnarray}
\label{Sec2-12}
\int_{\mathbb{R}^d} \Big ( \boldsymbol{u}  \frac{\partial \bar{f}}{\partial t} + \nabla \cdot (\boldsymbol{u}^2 \bar{f}) \Big ) d\boldsymbol{u} = 0 \quad 
&\Longrightarrow &  \quad  \rho \frac{\partial \bar{\boldsymbol{V}}}{\partial t} + \nabla \cdot \int_{\mathbb{R}^d} \boldsymbol{u}^2 \bar{f} d\boldsymbol{u} = 0.
\end{eqnarray}
Recalling the assumptions in Remark \ref{remark 1} that $\int_{\mathbb{R}^d} \boldsymbol{u} \Big ( \frac{\bar{f}-f^{*}(\bar{\Delta})}{\tau}  \Big ) d\boldsymbol{u} = 0$ due to microscopic reversibility of particle collisions. Following the derivations in \citep[][pp. 5-6]{SamieePoF2020}, we add and subtract $\bar{\boldsymbol{V}}\bar{\boldsymbol{V}}$ to the advection term and accordingly, \eqref{Sec2-12} is found to be
\begin{equation}
\label{Sec2-13}
\rho \Big (\frac{\partial \bar{\boldsymbol{V}}}{\partial t} + \nabla \cdot \bar{\boldsymbol{V}} \bar{\boldsymbol{V}} \Big ) = -\nabla \cdot \mathsfbi{\varsigma},
\end{equation}
where $\mathsfbi{\varsigma}$ in the index form is expressed as
\begin{equation}
\label{Sec2-14}
\mathsfi{\varsigma}_{ij}=\int_{\mathbb{R}^d} (u_i-\bar{V}_i)(u_j-\bar{V}_j) \, \bar{f} \, d\boldsymbol{u}.
\end{equation} 
Comparing \eqref{Sec2-1} and \eqref{Sec2-13}, it turns out that pressure term, viscous and SGS stresses all trace back to $\nabla \cdot \mathsfbi{\varsigma}$, where $\mathsfi{\varsigma}_{ij}=-\bar{p} \, \delta_{ij}+\mathsfi{T}^{shear}_{ij}+\mathsfi{T}^{\mathcal{R}}_{ij}$. By plugging \eqref{Sec2-6} into the kinetic definitions of each term in $\mathsfi{\varsigma}_{ij}$, we obtain
\begin{eqnarray}
\label{Sec2-15}
\bar{p} \, \delta_{ij} &=& -\int_{\mathbb{R}^d} (u_i-\bar{V}_i)(u_j-\bar{V}_j) \, f^{*}(\bar{\Delta}) \, d\boldsymbol{u} \int_{0}^{\infty} e^{-s} ds,
\\
\label{Sec2-16}
\mathsfi{T}_{ij}^{shear} &=& \int_{0}^{\infty}  \int_{\mathbb{R}^d} (u_i-\bar{V}_i)(u_j-\bar{V}_j)
\times \left(f^{eq}_{s,s}(\bar{\Delta})-f^{eq}(\bar{\Delta})\right) \, e^{-s} d\boldsymbol{u}\,  ds = 2 \mu \mathsfi{\bar{S}}_{ij},
\end{eqnarray}
where $\mu=\rho \, U^2\tau$. Similarly, by employing $f^{Model}$ in \eqref{Sec2-9}, we attain
\begin{eqnarray}
\label{Sec2-17}
\mathsfi{T}_{ij}^{R} &=&\mathrm{c}_{\beta,\lambda}\, \int_{0}^{\infty}  \int_{\mathbb{R}^d} (u_i-\bar{V}_i)(u_j-\bar{V}_j) \left(f^{\beta,\lambda}_{s,s}(\bar{\Delta})-f^{\beta,\lambda}(\bar{\Delta})\right) \, e^{-s} d\boldsymbol{u}\,  ds
\nonumber
\\
&=&\frac{\rho \, \mathrm{c}_{\beta,\lambda}}{U^3} \int_{0}^{\infty}  \int_{\mathbb{R}^d} (u_i-\bar{V}_i)(u_j-\bar{V}_j) \left(F^{\beta,\lambda}(\bar{\Delta}_{s,s})-F^{\beta,\lambda}(\bar{\Delta})\right) e^{-s} d\boldsymbol{u}\,  ds,
\end{eqnarray}
in which $\bar{\Delta}_{s,s} =  \bar{\Delta}(t-s\tau,\boldsymbol{x}-s\tau \boldsymbol{u}, \boldsymbol{u})$. As discussed in \citep[][Appendix]{SamieePoF2020}, the temporal shift can be detached from $f^{\beta,\lambda}_{s,s}(\bar{\Delta})$ and then $\bar{\Delta}_{s,s}$ is simplified to $\bar{\Delta}_s = \bar{\Delta}(t,\boldsymbol{x}-s\tau \boldsymbol{u}, \boldsymbol{u})$. Therefore,
\begin{eqnarray}
\label{Sec2-18}
\mathsfi{T}_{ij}^{R} &=& \frac{\rho \, \mathrm{c}_{\beta,\lambda}}{U^3} \int_{0}^{\infty}  \int_{\mathbb{R}^d} (u_i-\bar{V}_i)(u_j-\bar{V}_j) \left(F^{\beta,\lambda}(\bar{\Delta}_{s})-F^{\beta,\lambda}(\bar{\Delta})\right) e^{-s} d\boldsymbol{u}\,  ds.
\end{eqnarray}

The strategy to evaluate $\mathsfi{T}_{ij}^{R}$ is to decouple the particle speed into time and displacement by employing $\boldsymbol{u}=\frac{\boldsymbol{x}^{\prime}-\boldsymbol{x}}{s\tau}$ and approximate the asymptotic behavior of $F^{\beta,\lambda}(\bar{\Delta})$ with a tempered power-law distribution. In a detailed discussion in Appendix \ref{appB}, we show that
\begin{eqnarray}
\label{Sec2-19}
\mathsfi{T}_{ij}^{R} &=& \mathrm{c}_{\alpha,\lambda} \, \bar{\nu}_{\alpha} \sum_{k=0}^{\mathcal{K}} \bar{\phi}_{k}^{\mathcal{K}}(\alpha) \int_{\mathbb{R}^d} (x_i -x_i^{\prime}) \, (x_j -x_j^{\prime}) \frac{(\boldsymbol{x} -\boldsymbol{x}^{\prime})\cdot (\bar{\boldsymbol{V}}- \bar{\boldsymbol{V}}^{\prime} ) }{\vert \boldsymbol{x} -\boldsymbol{x}^{\prime}\vert^{2\alpha+5} \, e^{\bar{\lambda}_k \vert \boldsymbol{x} -\boldsymbol{x}^{\prime} \vert}}     d\boldsymbol{x}^{\prime}, 
\end{eqnarray}
in which $\bar{\nu}_{\alpha} = (2\alpha+3) (\rho \, C_{\alpha} \tau^{2\alpha-1} U^{2\alpha})$ for $\alpha \in (0,\frac{1}{2})\cup(\frac{1}{2},1)$ and recalling $\bar{\lambda}_k =  \frac{k}{\tau \, U}\lambda$.  Moreover, $\bar{\phi}_{k}^{\mathcal{K}}(\alpha)$ is indicated in \eqref{app2-3-1}. Eventually, we disclose the integral form of $\nabla \cdot \mathsfbi{T}^{\mathcal{R}}$ as
\begin{equation}
\label{Sec2-19-1}
(\nabla \cdot \mathsfbi{T}^{\mathcal{R}})_j = \mathrm{c}_{\alpha,\lambda} \, \bar{\nu}_{\alpha} \sum_{k=0}^{\mathcal{K}} \frac{(2\alpha+\bar{\lambda}_k)}{(2\alpha+3)} \bar{\phi}_{k}^{\mathcal{K}}(\alpha) \int_{\mathbb{R}^d} \frac{(\bar{V}_j -\bar{V}_j^{\prime}) }{\vert \boldsymbol{x} -\boldsymbol{x}^{\prime}\vert^{2\alpha+3} \, e^{\bar{\lambda}_k \vert \boldsymbol{x} -\boldsymbol{x}^{\prime} \vert}}  d\boldsymbol{x}^{\prime},
\end{equation} 
where $ \nu_{\alpha} = \mathrm{c}_{\alpha,\lambda} \bar{\nu}_{\alpha} $. Reminding the integral representation of a tempered fractional Laplacian in \eqref{eq1-9}, we formulate the divergence of the SGS stresses as follows
\begin{equation}
\label{Sec2-21}
(\nabla \cdot \mathsfbi{T}^{\mathcal{R}})_j = \nu_{\alpha} \sum_{k=0}^{\mathcal{K}} \phi_{k}^{\mathcal{K}}(\alpha,\lambda) (\Delta+\bar{\lambda}_k)^{\alpha} \bar{V}_j,
\end{equation}
where $\phi_{k}^{\mathcal{K}}(\alpha,\lambda) =  \frac{(2\alpha+\bar{\lambda}_k)}{(2\alpha+3)} \bar{\phi}_{k}^{\mathcal{K}}(\alpha)$. Evidently, by setting $\mathcal{K} = 0$, we find $\bar{\phi}_{k}^{\mathcal{K}}(\alpha)=\Gamma(2\alpha)$ and the new operator in \eqref{Sec2-21} reduces to a fractional Laplacian, which recovers the FSGS model. 

\begin{rem}
In terms of the explicit Fourier form of the tempered operator, $(\Delta+\bar{\lambda})^{\alpha}(\cdot)$, the TFSGS model maintains the high-order accuracy of scheme in LES solutions similar to the eddy-viscosity models without including any computational cost.
\end{rem}

Inferring from \eqref{Sec2-21}, our choice in the kinetic description of turbulent effects reflects in the form of a tempered fractional operator through a rigorous connection between the filtered BT and NS equations. More specifically, we adopt $\mathcal{K} = 1$ and hence the tempered fractional SGS (TFSGS) model can be formulated as
\begin{eqnarray}
\label{Sec2-21-2}
\nabla \cdot \mathsfbi{T}^{\mathcal{R}} = \nu_{\alpha}\, \left[ \phi_{0}^{1}(\alpha)\,\left(- (-\Delta)^{\alpha} \right) \bar{\boldsymbol{V}} + \phi_{1}^{1}(\alpha)\,  \left(\Delta+\bar{\lambda}_1 \right)^{\alpha} \bar{\boldsymbol{V}} \right],
\end{eqnarray}
where $\phi_{0}^{1}(\alpha) =  \frac{1}{(2\alpha+3)} \left( \Gamma(2\alpha+1)-\Gamma(2\alpha) \right) $ and $\phi_{1}^{1}(\alpha,\lambda) =  \frac{(2\alpha+\lambda)}{(2\alpha+3)} \, \Gamma(2\alpha-1)$. Accordingly, the governing LES equations read as 
\begin{eqnarray}
\label{Sec2-22}
\frac{\partial \bar{V}_i}{\partial t}+\frac{\partial \bar{V}_i\,\bar{V}_j}{\partial x_j}&=&-\frac{1}{\rho}\frac{\partial \bar{p}}{\partial x_i}+\nu \, \Delta \bar{V}_i -\nu_{\alpha} \sum_{k=0}^{1} \phi_{k}^{1}(\alpha,\lambda) (\Delta+\bar{\lambda}_k)^{\alpha} \, \bar{V}_i,
\end{eqnarray}
where $\alpha \in (0,\frac{1}{2})\cup(\frac{1}{2},1)$, $\lambda >0$, and $\nu_{\alpha} = \frac{\mu_{\alpha}}{\rho}$.  


\begin{rem}
\label{remark 2}
As a generator of tempered \textit{L\'evy} stable processes, the tempered fractional Laplacian is proven to be rotationally and Galilean invariant \citep[see][]{Cairoli2016, Huang2015, Kaleta2015}. Therefore, by having $\nu_{\alpha}$ and $\phi_{k}^{1}$ as real-valued functions of $\alpha$ and $\lambda$, the TFSGS model also adopts the frame invariance property in a consistent fashion with the SGS stresses.
\end{rem}

\subsection{TFSGS formulations for the SGS stresses}
\noindent To study the key role of tempering fractional operators in recovering turbulent statistical structures, it is essential to establish a straightforward form of the modeled SGS stresses. Due to some numerical complications in evaluating the integral in \eqref{Sec2-19-1}, we settle to proceed with the Fourier representation of the TFSGS model. Employing the definition of $\mathcal{I}^{\alpha}$ ($\alpha$-Riesz potential) from \citep[][]{Stein1970}, it is possible to verify that
\begin{equation}
\label{Sec2-23}
\nabla\cdot \mathsfbi{T}^R = (\Delta+\lambda)^{\alpha} \boldsymbol{\bar{V}} = \nabla\cdot\nabla \mathcal{I}^{\alpha = 1}\left[ \nu_{\alpha} \sum_{k=0}^{1} \phi_{k}^{1}(\alpha,\lambda) (\Delta+\bar{\lambda}_k)^{\alpha} \boldsymbol{\bar{V}}  \right].
\end{equation}
Inspired by \eqref{Sec2-23}, we introduce $\mathcal{R}_j^{\alpha,\lambda} (\cdot) = \nabla_j \mathcal{I}^{\alpha=1} (\Delta+\lambda)^{\alpha} (\cdot)$ as a tempered fractional operator such that 
\begin{equation}
\label{Sec2-24}
\mathsfi{T}^R_{ij} = \frac{\nu_{\alpha}}{2}\sum_{k=0}^{1} \phi_{k}^{1}(\alpha,\lambda) \left[ \mathcal{R}_j^{\alpha,\bar{\lambda}_k} V_i + \mathcal{R}_i^{\alpha,\bar{\lambda}_k} V_j  \right],
\end{equation}
where $\mathcal{F} \left[ I^{\alpha=1} \right] = \frac{1}{\xi^2}$ and $\mathcal{F} \left[ \nabla_j \right] (\boldsymbol{\xi}) = - \mathfrak{i} \, \xi_j$ and $\mathfrak{i}$ denotes an imaginary unite. Following \eqref{eq1-11} into \eqref{Sec2-24}, we find the Fourier form of $\mathcal{R}_j^{\alpha,\lambda}$ as
\begin{equation}
\label{key}
\mathcal{F} \left[ \mathcal{R}_j^{\alpha,\lambda} \right] (\boldsymbol{\xi}) = \mathfrak{C}_{\alpha} \frac{ - \mathfrak{i} \, \xi_j}{\xi^2} \left( \lambda^{2\alpha} - (\lambda^2+\xi^2)^{\alpha} \prescript{}{2}{F}^{}_{1}(-\alpha, 1+\alpha;\frac{3}{2};\frac{\xi^2}{\xi^2+\lambda^2}) \right).
\end{equation}

\section{Statistical Analysis}
\label{Sec 3}
\noindent In pursuit of an ideal SGS model, nonlinearity induced by the convective terms and nonlocality imparted by the pressure term in the NS equations contribute to a synthetic hierarchy of transport equations and multi-point descriptions of SGS terms, as shown in \citep{Sagaut2008}. The infinitely-extended hierarchical triangle of nonlinearity and nonlocality brings up the idea of indicating a set of weaker, and yet significant, statistical conditions and make the ideal LES model more attainable, as endorsed by \citep{Moser2020}. To identify such statistical features, \cite{Meneveau1994} developed a rigorous framework via a statistical \textit{a priori} analysis and formulated some sufficient conditions for the assessment of LES models. As one of the candidates for evaluating SGS models, we give a brief review of the argued formulations in \citep{Meneveau1994} and introduce an optimization strategy, which enables the TFSGS model to correctly generate the requisite statistical conditions.

Hereafter, we consider the following notations in study of the SGS fields. Let $\mathsfbi{T}^{R,D}$ and $\mathsfbi{T}^{R,*}$ denote the SGS stresses, implied by the true DNS data and the SGS model, respectively. We also take $\boldsymbol{r}$ as the displacement vector between two points in the correlation functions and $\boldsymbol{e}$ denotes the unit vectors of the axes in the Cartesian coordinates. Then, $r = \vert \boldsymbol{r} \vert$. As discussed in \citep{Meneveau1994}, performing an ensemble-average of the filtered NS equations offers a set of necessary conditions for an LES simulation to ensure the equality of mean velocity profiles and the second-order moments, listed as:

\vspace{0.05 in}
\begin{enumerate}[label=(\alph*)]
\item \hspace{0.02 in} $\langle\mathsfi{T}^{R,D}_{ij}\rangle = \langle \mathsfi{T}^{R,*}_{ij}\rangle$,

\item \hspace{0.02 in} $\langle\bar{V}_i \mathsfi{T}^{R,D}_{ij}\rangle =\langle\bar{V}_i \mathsfi{T}^{R,*}_{ij}\rangle$,

\item \hspace{0.02 in} $\langle\mathsfi{\bar{S}}_{ij} \mathsfi{T}^{R,D}_{ij}\rangle = \langle\mathsfi{\bar{S}}_{ij} \mathsfi{T}^{R,*}_{ij}\rangle$,
\end{enumerate} 
\vspace{0.05 in}
in which conditions (b) and (c) are inferred from the ensemble-averaged SGS transport equation. 

Focusing on the nonlocality axis of the closure triangle for a homogeneous isotropic turbulent (HIT) flow, one obtains the so-called Karman-Howarth (KH) equation as
\begin{equation}
\label{Sec3-1}
\left[ \frac{\partial}{\partial t}-2\nu \, \left(\frac{\partial^2}{\partial r^2} + \frac{4}{r}\frac{\partial}{\partial r}\right) \right] B_{LL}(r,t) - \left(\frac{\partial}{\partial r} + \frac{4}{r}\right) \, B_{LLL}(r,t) = \left(\frac{\partial}{\partial r} + \frac{4}{r}\right) \, G_{LLL}(r,t),
\end{equation}
for sufficiently large $\mathcal{L}\gg \eta$, where $L$ represents the longitudinal direction. Additionally, we denote by $B_{LL}(r,t)=\langle \bar{V}_L(\boldsymbol{x},t) \, \bar{V}_L(\boldsymbol{x}+\boldsymbol{r}\cdot \boldsymbol{e},t) \rangle$ and $B_{LLL}(r,t)=\left\langle \left[\bar{V}_L(\boldsymbol{x},t)\right]^2 \bar{V}_L(\boldsymbol{x}+\boldsymbol{r}\cdot \boldsymbol{e},t) \right\rangle$ the second- and third-order velocity correlation functions, respectively and $G_{LLL}(r,t)=\langle \mathsfi{T}^{R}_{LL}(\boldsymbol{x},t) \bar{V}_L(\boldsymbol{x}+\boldsymbol{r}\cdot \boldsymbol{e},t) \rangle$ refers to the stress-strain correlation function. Technically, the third-order correlation function in \eqref{Sec3-1} is subdivided into $B_{LLL}$ stemming from the resolved velocity field and $G_{LLL}(r,t)$ coming from the SGS stresses. It turns out from \eqref{Sec3-1} that the SGS model should undergo a correct prediction of $G_{LLL}(r,t)$ to re-generate $B_{LL}$ and $B_{LLL}$ accurately. Referring to \citep[][pp. 819-820]{Meneveau1994}, we arrive at the following equation
\begin{equation}
\label{Sec3-2}
\left\langle [\bar{V}_L(\boldsymbol{x}+\boldsymbol{r}\cdot \boldsymbol{e},t)-\bar{V}_L(\boldsymbol{x},t)]^3 \right\rangle + 6\, G_{LLL}(r,t) = 6 \left\langle\mathsfi{\bar{S}}_{LL} \mathsfi{T}^{R}_{LL} \right\rangle \, r,
\end{equation}
which exhibits the only sufficient condition for modeling third-order structure in an HIT flow. Therefore, by satisfying the equality of SGS dissipation via conditions (c), modeling $G_{LLL}$ remains the only requisite for capturing the third-order structure functions. 

This finding reveals the significance of condition (c), which intrinsically ties with the stress-strain correlation function, represented by $D_{LL}(r,t) = \langle\mathsfi{\bar{S}}_{LL}(\boldsymbol{x}+\boldsymbol{r}\cdot \boldsymbol{e},t) \mathsfi{T}^{R}_{LL} \rangle$. Using the conversation in \citep[][pp. 317]{Cerutti2000}, $D_{LL}$ is derived in terms of $G_{LLL}$ as 
\begin{equation}
\label{Sec3-2-1}
D_{LL}(r,t) = \frac{7}{2} \frac{dG_{LLL}(r,t)}{dr}+\frac{4 G_{LLL}(r,t)}{r}+\frac{r}{2}\frac{d^2 G_{LLL}(r,t)}{dr^2}.
\end{equation}
Emphasizing the role of tempering parameter in modulating the turbulent dissipation range, we therefore adopt $D_{LL}(r,t)$ as a key quantity in optimizing the TFSGS model to address condition (c) and capture the nonlocal structures in \eqref{Sec3-2}. It must be noted that in evaluating the aforementioned conditions and high-order structures, $\mathsfbi{T}^{R}$ represents either $\mathsfbi{T}^{R,D}$, obtained by filtering the instant DNS database, or $\mathsfbi{T}^{R,*}$, implied by implementing any model to the true resolved velocity field.


\subsection{Optimization strategy}
\label{subsec 3-1}

\noindent Devising a robust optimization framework is an inevitable element in predictive fractional and tempered fractional modeling \citep[see][]{Burkovska2020, Pang2020}. Regarding the given set of conditions for the closure problem, we find conditions (a) and (c) practically crucial in developing an approach for estimating the parameters and coefficient associated with the TFSGS model while condition (b) can be substantially recovered by imposing \eqref{Sec3-2}, where $G_{LLL}(r,t)\vert_{r=0}=\langle\bar{V}_i \mathsfi{T}^{R}_{ij}\rangle$. As we learn from the one-point correlation analysis in \citep[][]{SamieePoF2020} and the further section, correlations between the SGS stresses, obtained by the DNS data and the model, highly rely on $\alpha$ and $\lambda$ in the TFSGS model rather plays a central role in capturing the SGS dissipation energy and nonlocal structure functions. This approach provides the basis for an optimal estimation of the fractional exponents ($\alpha$ and $\lambda$) by employing the normalized $D_{LL}$ and $\varrho_{i}$, defined in Algorithm \ref{alg:1}.

\floatname{algorithm}{Algorithm}
\renewcommand{\algorithmicrequire}{\textbf{Input:}}
\renewcommand{\algorithmicensure}{\textbf{Output:}}

\begin{algorithm}[ht!]
	\caption{Estimation of the optimal model parameters for a specific $\mathcal{L}$}
	\label{alg:1}
	\begin{algorithmic}[-1]
		\STATE \hspace{-0.25in} \textbf{INPUT}:  \hspace{0.34in} $\bar{V}_i$, ${T}^{R,D}_{ij}$, $\bar{V}_L(\boldsymbol{x}+\boldsymbol{r}\cdot \boldsymbol{e},t)$
		\vspace{0.05 in}
		
		\STATE \hspace{-0.25in} \textbf{OUTPUT}:  \hspace{0.2in} $\alpha^{opt}$, $\lambda^{opt}$, $ \mathrm{c}_{\alpha,\lambda}$ 
		\vspace{0.05 in}
		
		\STATE \hspace{-0.25in} \textbf{PROCESS}:
		\STATE \hspace{-0.25in} 1. \hspace{0.05in} Find $\alpha^{opt}$ where the maximum of $\varrho_{ii} = \Big{\langle} \varrho \Big{[} \mathrm{T}^{R,D}_{ii},\, \mathrm{T}^{R,TF}_{ii} \Big{]}\Big{\rangle}$ occurs. 
		\STATE \hspace{-0.25in} 2. \hspace{0.05in} Find $\lambda^{opt}$ where $\big{[} \frac{D_{LL}(r,t)}{D_{LL}(0,t)}\big{]}^{TF}$ fits into $\big{[} \frac{D_{LL}(r,t)}{D_{LL}(0,t)}\big{]}^{D}$ for the inferred fixed $\alpha^{opt}$.
		\STATE \hspace{-0.25in} 3. \hspace{0.05in} Quantify the model constant such that $ \mathrm{c}_{\alpha,\lambda}  =  \frac{\langle\mathsfi{\bar{S}}_{ij} \mathsfi{T}^{R,D}_{ij}\rangle}{\langle\mathsfi{\bar{S}}_{ij} \mathsfi{T}^{R,\,N}_{ij}\rangle} $, given $\alpha^{opt}$ and $\lambda^{opt}$. 
		
		\vspace{0.05 in}
		\STATE \hspace{-0.25in} \textbf{VALIDATION ANALYSIS}:
		\vspace{0.03 in}
		
		\STATE \hspace{-0.25in} 1. \hspace{0.05in} Evaluate $G_{LLL}(r)$ for the modeled and true SGS stresses to check if \eqref{Sec3-2} is validated.
		\vspace{0.03 in}
		
		\STATE \hspace{-0.25in} 2. \hspace{0.05in} Perform a comparative study on the PDFs of  $\langle\mathsfi{\bar{S}}_{ij} \mathsfi{T}^{R}_{ij}\rangle$ and $\langle \mathsfi{T}^{R}_{ii}\rangle$.
	\end{algorithmic}
\end{algorithm}

By fixing the values of fractional exponents, it is possible to accurately quantify the model coefficient and thereby re-producing the third-order structure in \eqref{Sec3-2} via modeling $G_{LLL}$. In Algorithm \ref{alg:1}, we schematically present the proposed method for optimizing the parameters associated with the TFSGS model at a given flow Reynolds number (Re) and a specific $\mathcal{L}$.


It must be noted that in step 3, we define 
\begin{equation}
\label{Sec3-3}
\mathsfi{T}^{R,\,N}_{ij}= \frac{\mathsfi{T}^{R,\,TF}_{ij}}{\mathrm{c}_{\alpha,\lambda} }=\frac{\bar{\nu}_{\alpha}}{2}\sum_{k=0}^{1} \phi_{k}^{1}(\alpha,\lambda) \big {[} \mathcal{R}_j^{\alpha,\bar{\lambda}_k} V_i + \mathcal{R}_i^{\alpha,\bar{\lambda}_k} V_j  \big {]}.	
\end{equation}
Moreover, superscripts $``D"$ and $``TF"$ represent the values obtained by filtering the true DNS data and the TFSGS model, respectively.


\section{\textit{A Priori / Posteriori} Analyses}
\label{Sec 4}

\noindent 
To attain the optimal behavior of the TFSGS model, we follow the steps in Algorithm \ref{alg:1} by performing an \textit{a priori} analysis and evaluate the capabilities of the TFSGS model in generating the statistical features of turbulent flows. 
%
%
%

\subsection{DNS Database and LES Platform}
\label{Subsec 4-1}
\noindent In terms of \textit{a priori} tests, we conduct the numerical simulation of a forced HIT flow employing the open-source pseudo-spectral NS solver for a triply periodic domain the code of which is presented at \citep[][]{DNS2020}. It should be noted that in the next section, the LES solver is successfully prepared using this DNS code and the statically stationary DNS dataset presented here is filtered and used as the initial conditions for the final \textit{a posteriori} assessments. 

Using the NS solver, we performed DNS of a stationary HIT flow with $320^3$ resolution for a periodic computation domain as $\Omega=[0,2\pi]^3$ and the large scale forcing occurs at $0 < \vert \boldsymbol{\xi} \vert \le 2$ to maintain turbulence statistics stationary. Here, $\boldsymbol{\xi}$ represents the vector of Fourier wave numbers and $\xi_{max}=\sqrt{2}\,N/3$ is the maximum wave number solved numerically, where $N=320$ is the number of grid points. In this case, $\xi_{max}\,\eta_{k}=1.6>1$ certifies that all the scales of motion are well-resolved, where $\eta_{k}$ refers to the Kolmogorov length scale. We detail the flow parameters and some of the statistical properties at Table \ref{Table1}, in which $\varepsilon$ and $K_{tot}$ denote the expected values of dissipation rate and turbulent kinetic energy, respectively. Moreover, $Re_{\lambda}=\frac{u^\prime_{rms} \, l_{\lambda}}{\nu}$ and $l_{\lambda}=\sqrt{15 \nu \,  {u'}_{rms}^2/\varepsilon}$ represent the Taylor Reynolds number and micro-scale length, respectively, where $u'_{rms}=\sqrt{2K_{tot}/3}$. The simulation undergoes running for $30$ eddy turn-over times, $\tau_{\mathscr{L}}$, to construct $40$ sample snapshots as our database. Due to the present homogeneity and isotropy in the HIT flow, we find the database adequate for obtaining the required statistics in the further analysis. The kurtosis and skewnewss values of the diagonal components of velocity gradient tensor are also presented in Table \ref{Table1}, supporting non-Gaussianity of turbulent structures.

\begin{table}
	\centering
	\caption{Flow parameters and statistical properties in the DNS of a forced HIT flow.}
	\vspace{-0.1 in}
	\label{Table1}
	\begin{tabular}{c c c c c c c c c}
		\hline  
		\vspace{-0.1 in}
		\\
		$Re_{\lambda}$ & $u'_{rms}$   & $K_{tot}$   & $\nu$  & $\varepsilon$  & $\tau_{\mathscr{L}}$ & skewness & kurtosis
		\vspace{0.1 in}
		\\
		& $(m/{sec})$ & $(m^2/{sec}^2)$ & $ (m^2/{sec})$ & $(m^2/{sec}^3)$  & $(sec)$ &  & 
		\\
		\cline{1-8}     
		\vspace{-0.1 in}
		\\
		$190$ &  $0.67$&  $0.68$ & $0.001$ & $0.1$  & $4.2$ & $-0.5$ & $6.5$
	\end{tabular}
	\vspace{0.1 in}
\end{table}

For the purpose of crunching heavy DNS database in the statistical analysis, we develop an LES platform in Python with a focus on efficiency in obtaining two-point correlations and the ease of dealing with the fractional operators. This Platform consists of three chief components: filtering the DNS database, implementation of LES models and optimization, and executing the final analysis. To overcome the burden of timely filtering process especially in two-point correlation analysis, we introduce the scalable multi-threaded filtering code using \texttt{Numpy}, \texttt{threading}, and \texttt{astropy.convolution} packages. Further steps in finding the the optimum model parameters and applying the Fourier form of the fractional models are developed employing highly efficient Intel\textregistered{} MKL library. 


\subsection{Optimal estimation of fractional parameters}
\label{Subsec 4-2}
\noindent In order to optimize the efficiency of the TFSGS model, we developed a flexible and rigorous strategy in Algorithm \ref{alg:1}. The proposed algorithm is equipped with verification and validation mechanisms through the conventional correlation coefficients and two-point structure functions. Recalling from section \ref{subsec 3-1} that $\mathsfi{T}^{R,D}_{ij}$ denotes the true SGS stresses obtained by filtering the well-resolved DNS data. Moreover, $\mathsfi{T}^{R,*}_{ij}$ represents the general form of modeled SGS values, where $*$ can be replaced by $TF$ or $SM$ in the TFSGS or SMG models, respectively. 

\begin{figure}
	\centering
\begin{subfigure}[b]{0.32\textwidth}
	\centering
	\includegraphics[width=1.0 \linewidth]{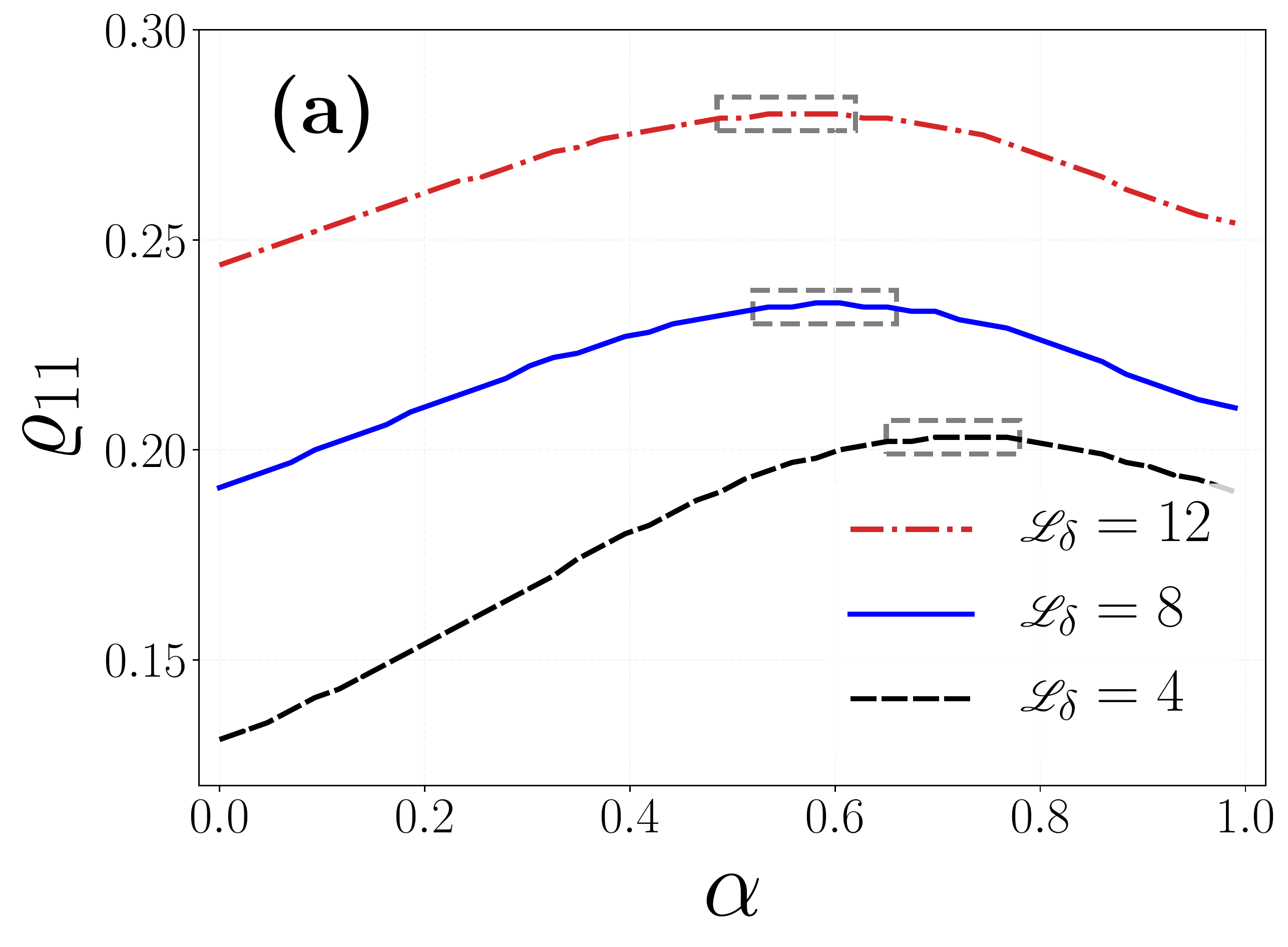}
	\subcaption*{}
\end{subfigure}
\begin{subfigure}[b]{0.32\textwidth}
	\centering
	\includegraphics[width=1.0 \linewidth]{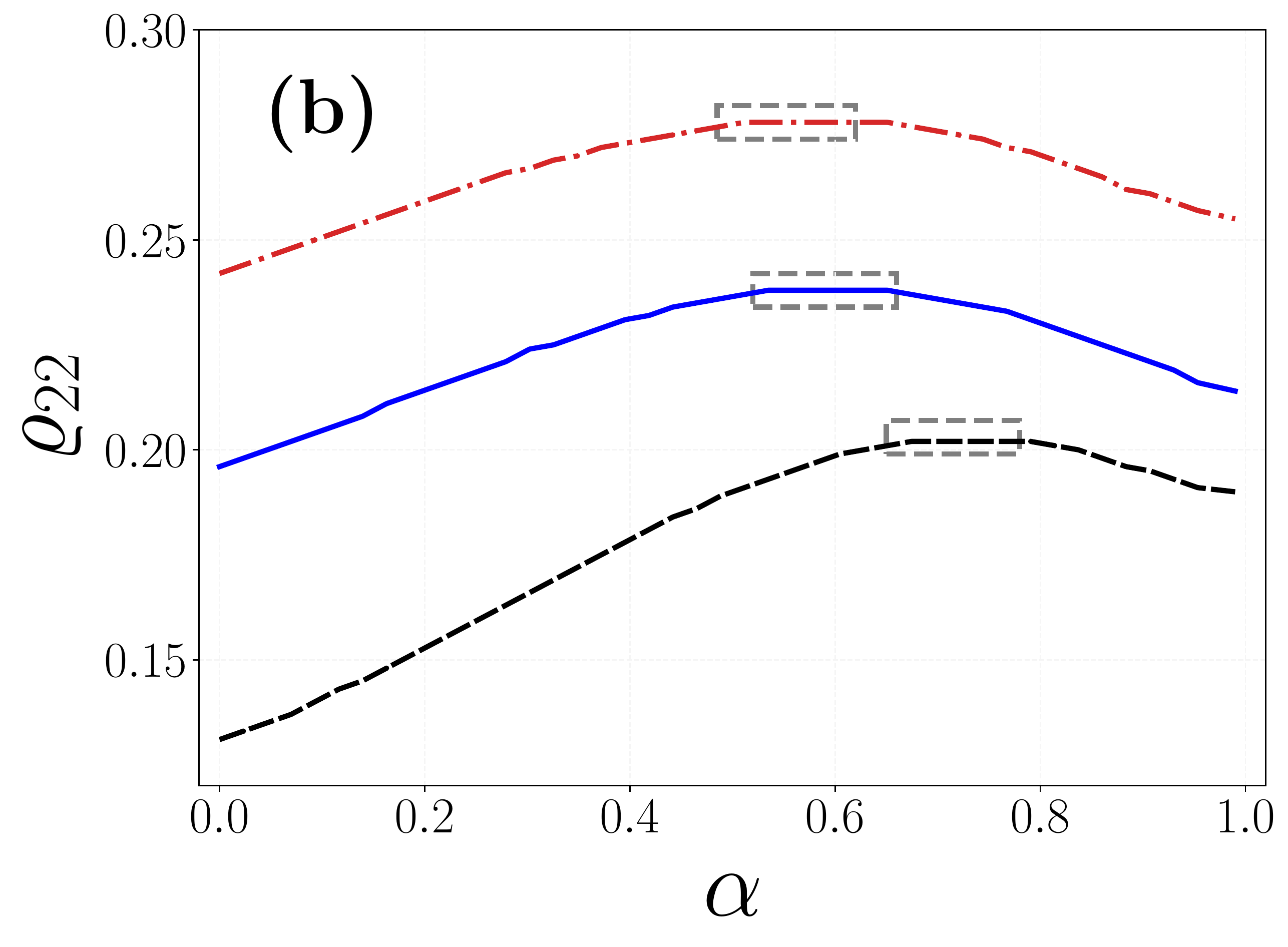}
	\subcaption*{} 
\end{subfigure}
\begin{subfigure}[b]{0.32\textwidth}
		\centering
	\includegraphics[width=1.0 \linewidth]{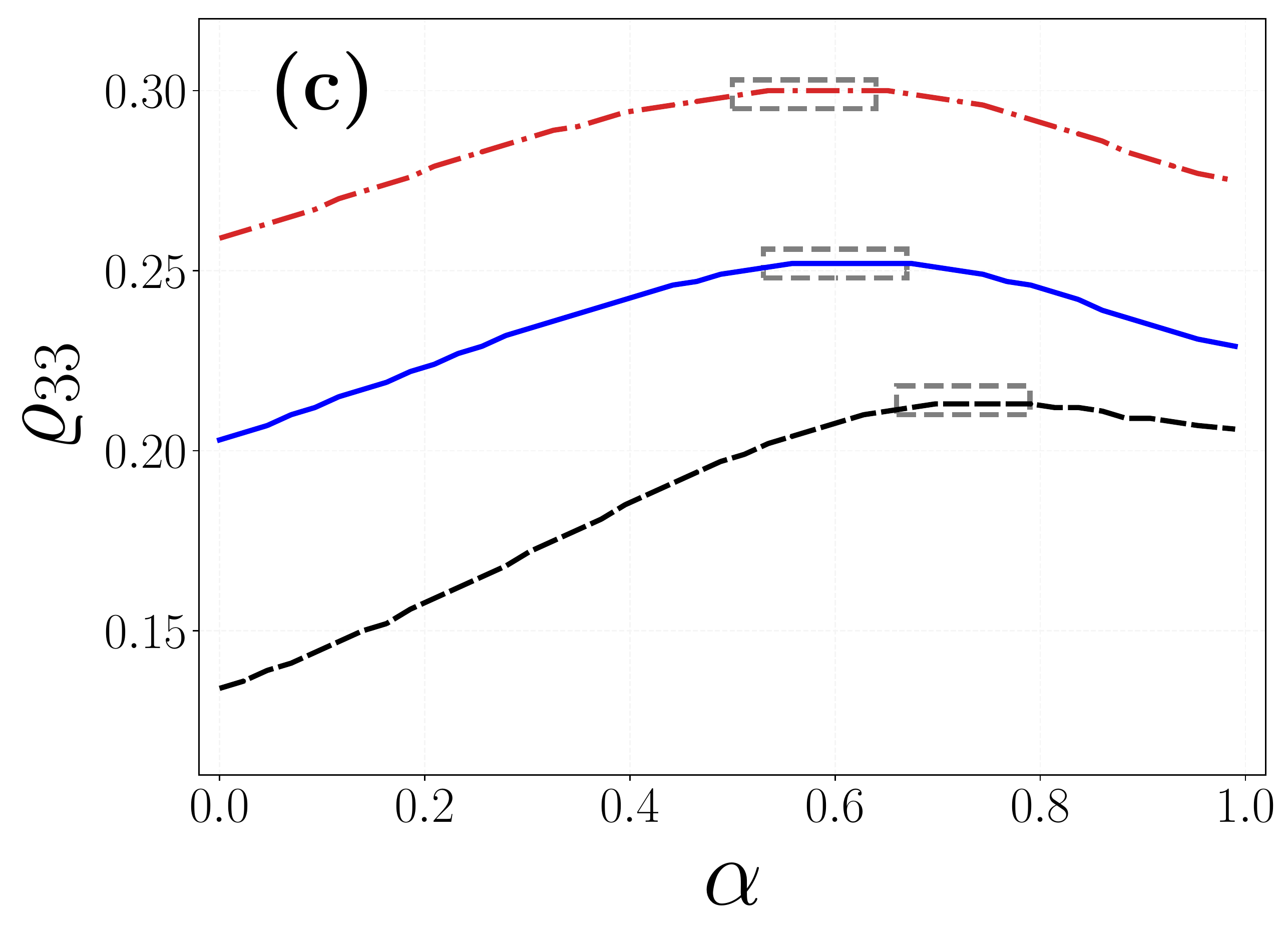}
	\subcaption*{} 
\end{subfigure}
\vspace{-0.25in}
\caption{variation of the correlation coefficients \textbf{(a)} $\varrho_{11}$, \textbf{(b)} $\varrho_{22}$, and \textbf{(c)} $\varrho_{33}$ versus $\alpha\in(0,1)$ for $\mathcal{L}_{\delta}=4,\,8,\,12$ by setting $\lambda\simeq0$ in \eqref{Sec3-3}. The maximum values lie in the dashed boxes.
}
\label{Figure1}
\end{figure}

The first step in Algorithm \ref{alg:1} concerns detecting optimum value of fractional exponent, $\alpha^{opt}$, where the maximum of ensemble-averaged correlation between $\mathsfi{T}^{R,D}_{ii}$ and $\mathsfi{T}^{R,TF}_{ii}$, denoted by $\varrho_{ii}$, occurs. Our premise is that the tempering parameter, $\lambda$, does not make any noticeable changes in $\varrho_{ii}$, namely less than 3 percents, which is endorsed by the results of Table \ref{Table2}. In the absence of $\lambda$, we plot the variations of $\varrho_{ii}$ versus $\alpha^{opt}$ in Figure \ref{Figure1} for $i=1,2,3$, in which each dashed box specifies the interval of $\alpha$ yielding the maximum of $\varrho_{ii}$. Without any loss of accuracy, we adopt $\alpha^{opt}=0.76, \, 0.58, \, 0.51$ as the corresponding minimum value in each specified interval for $\mathcal{L}_{\delta}= \frac{\mathcal{L}}{2 \delta x}=4,\,8,\,12$, respectively, where $\delta x=\frac{2\pi}{N}$ represents the computational grid size.

From the kinetic perspective, by enlarging $\mathcal{L}_{\delta}$, $\overline{f^{eq}(\Delta)}$ in \eqref{Sec2-5} demonstrates an increasingly multi-exponential pattern, which can be better described by a power-law distribution function. This argument accounts for the prediction enhancement in Figure \ref{Figure1}, achieved by the TFSGS model and the abduct reduction of $\alpha^{opt}$ versus $\mathcal{L}_{\delta}$ \citep[see][page 10]{SamieePoF2020}. Theoretically, the tempered power-law distribution can resemble a power-law or a Gaussian distribution by letting $\lambda$ go to $0$ or $\infty$, respectively. This grounds for the TFSGS model to span the gap between the FSGS model, representing self-similar behavior of the inertial range, and the SMG model, renowned for its dissipative characteristics. The results in Table \ref{Table2} support this line of reasoning by a row of correlation quantities for the given filter widths, particularly at $\mathcal{L}_{\delta}=12$.

\renewcommand{\arraystretch}{1.1}
\begin{table}
	\centering
	\caption{The ensemble-averaged correlation coefficients ($\varrho_{ii}$) between SGS stresses obtained by the filtered DNS data ($\mathsfi{T}^{R,D}_{ii}$) and the TFSGS model ($\mathsfi{T}^{R,TF}_{ii}$) for $i=1,2,3$. In the fractional models, $\alpha^{opt}$ is set as $0.76$, $0.58$, and $0.51$ for $\mathcal{L}_{\delta}=4,\,8,\,12$ respectively.} 
	\vspace{0.05in}
	\begin{tabular}{c c c c c c c c c c c  c c c c c c c c c c c c c}
		\label{Table2}
		&&\multicolumn{6}{c}{$\mathcal{L}_{\delta}=4$} &\multirow{2}{*}{}&  \multirow{2}{*}{}& \multicolumn{6}{c}{$\mathcal{L}_{\delta}=8$} & \multirow{2}{*}{}&  \multirow{2}{*}{} & \multicolumn{6}{c}{$\mathcal{L}_{\delta}=12$}
		\\ \cline{3-8}  \cline{11-16} \cline{19-24}
		& & FSGS  &    & \multicolumn{2}{c}{TFSGS} &   & SMG  &  &  & FSGS  &     &  \multicolumn{2}{c}{TFSGS}&    & SMG  &  &  & FSGS &     & 
		\multicolumn{2}{c}{TFSGS} & & SMG  
		\\
		\cline{3-3} \cline{5-6} \cline{8-8} \cline{11-11} \cline{13-14} \cline{16-16} \cline{19-19} \cline{21-22} \cline{24-24}
		$\lambda$  & &  $0$   &  & $0.1$  & $4$ & & $-$   &  &  & $0$   &   & $0.35$ & $5$ &  & $-$   &  &  & $0$   &   & $0.45$ & $5$ &  & $-$   
		\\
		\cline{3-3} \cline{5-6} \cline{8-8} \cline{11-11} \cline{13-14} \cline{16-16} \cline{19-19} \cline{21-22} \cline{24-24}
		$\varrho_{11}$  & &  $0.21$   &   &$0.20$ & $0.19$ & & $0.20$   &  &  & $0.23$  &  & $0.24$ & $0.22$ &  & $0.22$   &  &  & $0.29$   &   & $0.28$ & $0.26$ &  & $0.26$   
		\\
		$\varrho_{22}$  & &  $0.21$   &   &$0.20$ & $0.20$ & & $0.21$   &  &  & $0.24$  &   & $0.24$ & $0.23$ &  & $0.23$   &  &  & $0.29$   &    & $0.28$ & $0.26$ &  & $0.26$  
		\\
		$\varrho_{33}$  & &  $0.22$   &    &$0.21$ & $0.21$ & & $0.21$   &  &  & $0.25$  &   & $0.25$ & $0.24$ &  & $0.23$   &  &  & $0.30$   &   & $0.30$ & $0.29$ &  & $0.28$
		\\
		&&&&&&&&&&&&&&&&&&&&&&&
		\end{tabular}
\end{table}

\begin{figure}
	\centering
	\begin{subfigure}[b]{0.50\textwidth}
		\centering
		\includegraphics[width=1 \linewidth]{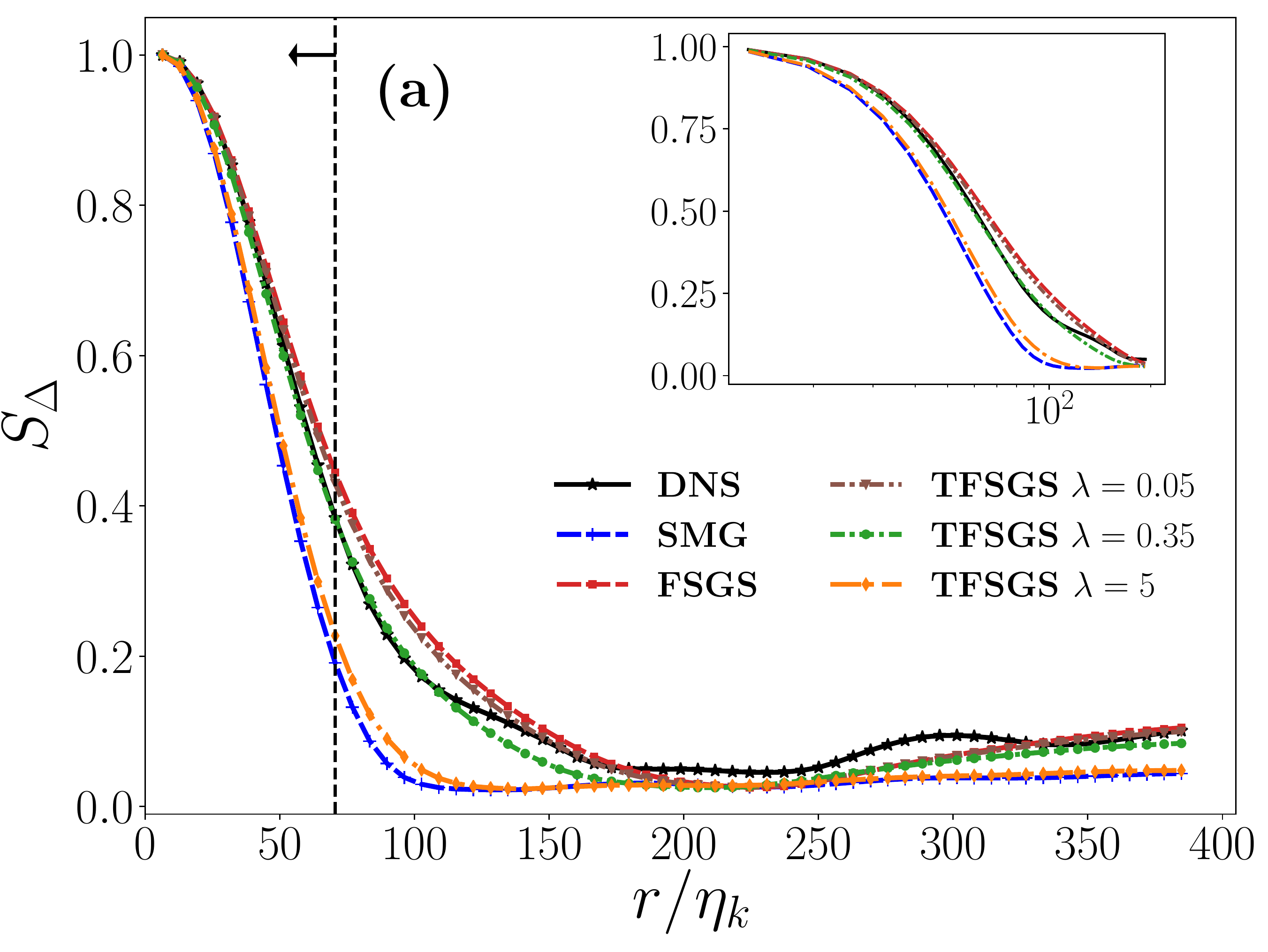}
		\subcaption*{}
		\label{Figure2-a}
	\end{subfigure}
	\hspace{0.02in}
	\begin{subfigure}[b]{0.4\textwidth}
		\centering
		\includegraphics[width=1 \linewidth]{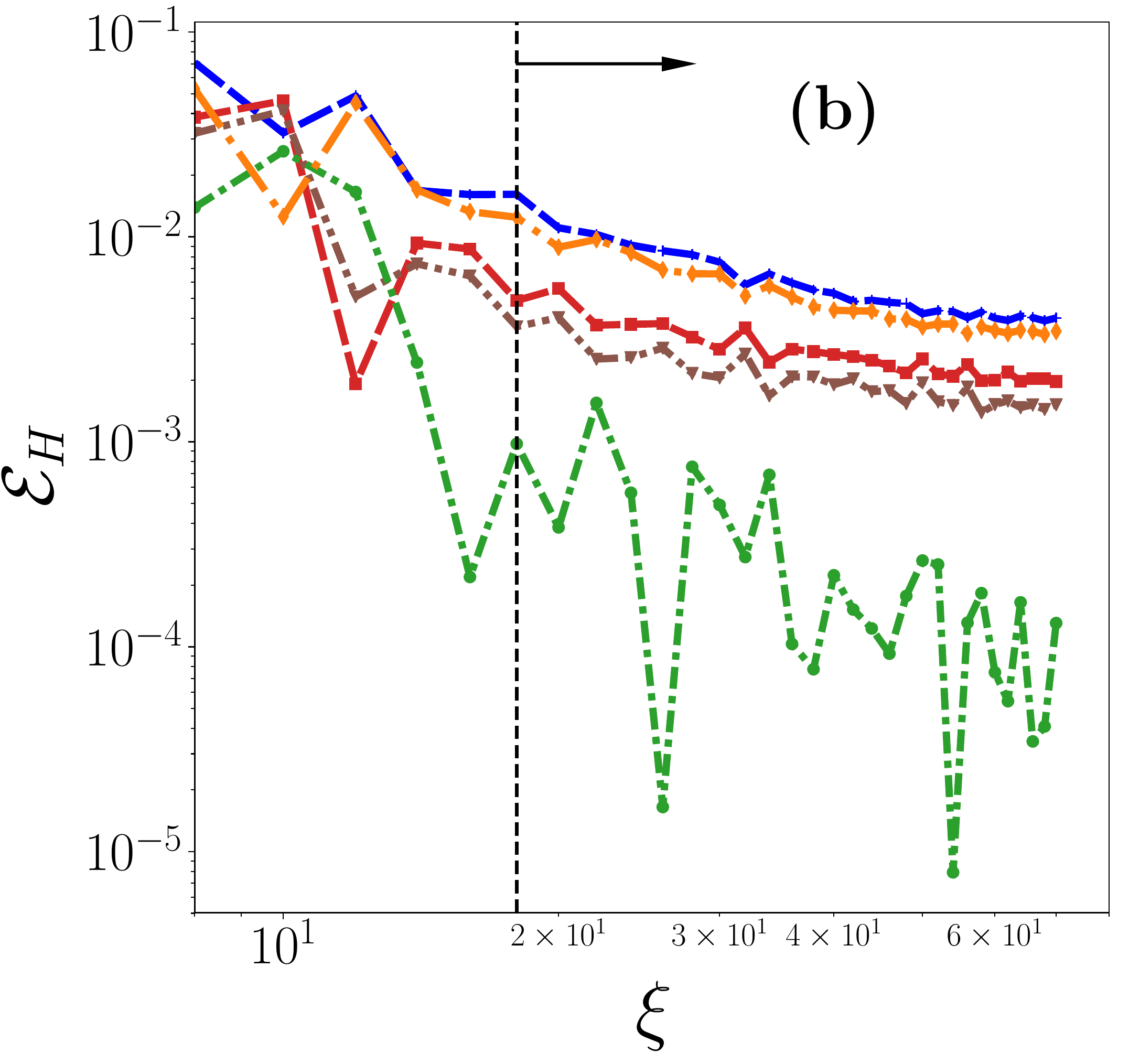}
		\subcaption*{} 
		\label{Figure2-b}
	\end{subfigure}
	\vspace{-0.25in}
	\caption{\textbf{(a)} Comparing results of the normalized two-point stress-strain rate correlation functions ($S_{\Delta}$) and \textbf{(b)} error analysis of the longitudinal dissipation spectrum ($\mathcal{E}_{H}$) for fractional SGS and Smagorinsky models, where $\mathcal{L}_{\delta}=8$ and $\xi$ represents the radius of Fourier wave numbers. In both plots, the arrows point to the dissipation range.}
	\label{Figure2}
\end{figure}

On this background, we proceed with the second step in Algorithm \ref{alg:1} to indicate $\lambda^{opt}$ through a comparative study of the normalized strain-stress correlation function, defined as $S_{\Delta}^{} =  \frac{D_{LL}(r,t)}{D_{LL}(0,t)}$. 
With the knowledge of $D_{LL}(r,t)$, we extend the two-point correlation analysis to the spectral space by evaluating the instantaneous radial dissipation spectrum, given by $\hat{\mathcal{D}}(\xi)=\mathcal{F} \big {[}D_{LL}(r,t) \big {]}(\xi)$. To evaluate the error between the dissipation spectrum, obtained by the true DNS data and the LES models at high wave numbers, we define $$\mathcal{E}_{H} = \frac{\Big {\vert} \big {[} \hat{\mathcal{D}}(\xi)\big {]}^{*} - \big {[}\hat{\mathcal{D}}(\xi)\big {]}^{D}\Big {\vert} }{ \bigg {\vert} \big {[}\hat{\mathcal{D}}(\xi)\big {]}^{D} \bigg {\vert}}, $$ where ${\vert} \cdot  {\vert} $ represents norm of the vector. Figure \ref{Figure2} \textcolor{blue}{(\textbf{a})} displays $S_{\Delta}$ versus the spatial shift, $r$, for a logarithmic sequence of $\lambda$ spanning three orders of magnitude in the TFSGS model, where $\mathcal{L}_{\delta}=8$. As stated earlier, the proposed model can take a journey from the FSGS to the SMG models by tuning $\lambda$. Evidently, the true quantities of $S_{\Delta}$, colored by black, are well-predicted by the proposed model with $\lambda^{opt}=0.45$, where $\alpha^{opt}=0.58$ is fixed. Figure \ref{Figure2} \textcolor{blue}{(\textbf{b})} confirms our findings quantitatively in a plot of $\mathcal{E}_{H}$ versus radius of wave numbers, $\xi$, with log-scale axes. In fact, this plot implies accuracy of the TFSGS model in capturing the two-point structure function at the dissipation range, pointed by an arrow.


\begin{figure}
	\centering
	\begin{subfigure}[b]{0.45\textwidth}
		\centering
		\includegraphics[width=1 \linewidth]{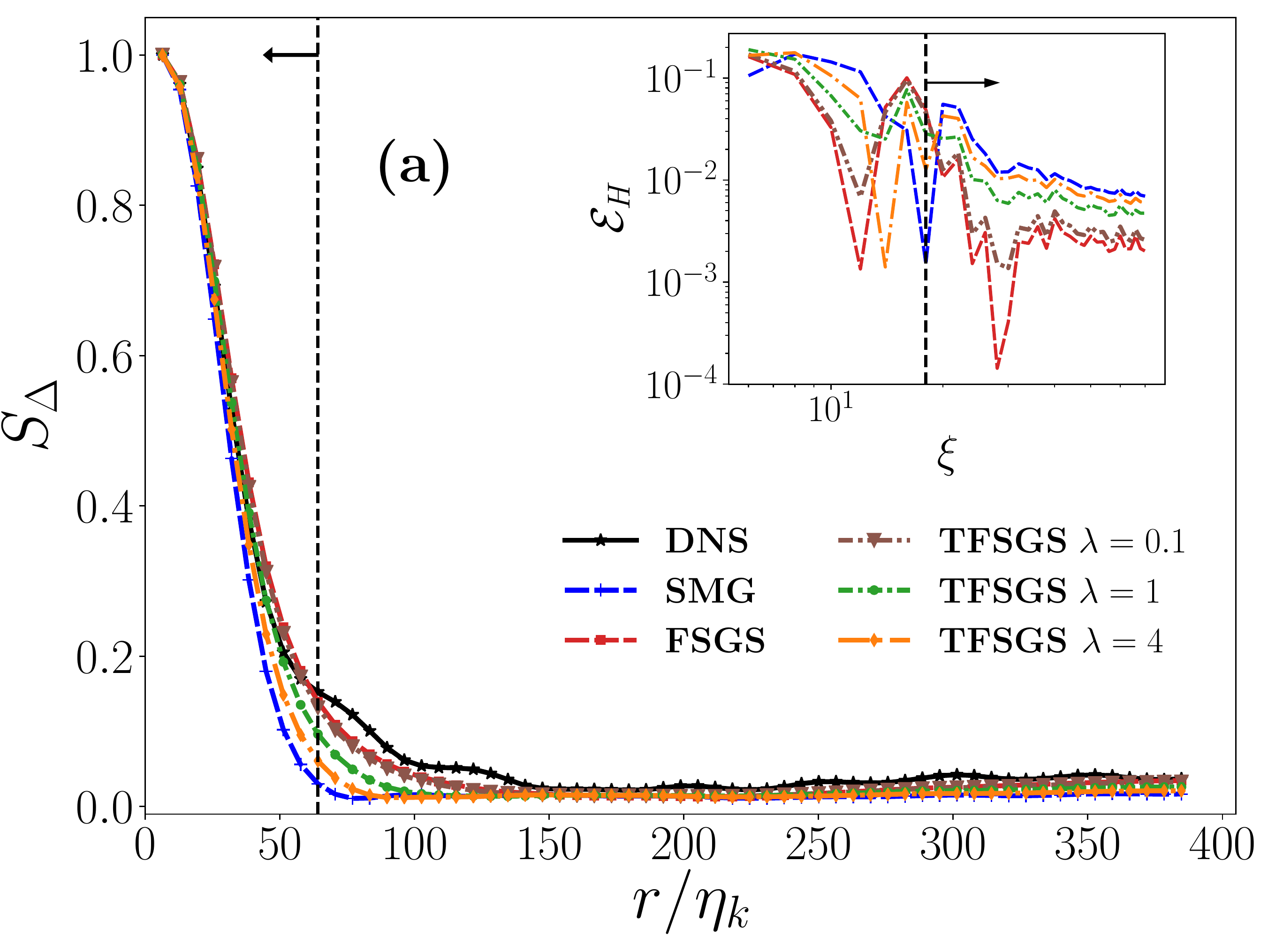}
		\subcaption*{}
	\end{subfigure}
	\hspace{0.02in}
	\begin{subfigure}[b]{0.45\textwidth}
		\centering
		\includegraphics[width=1 \linewidth]{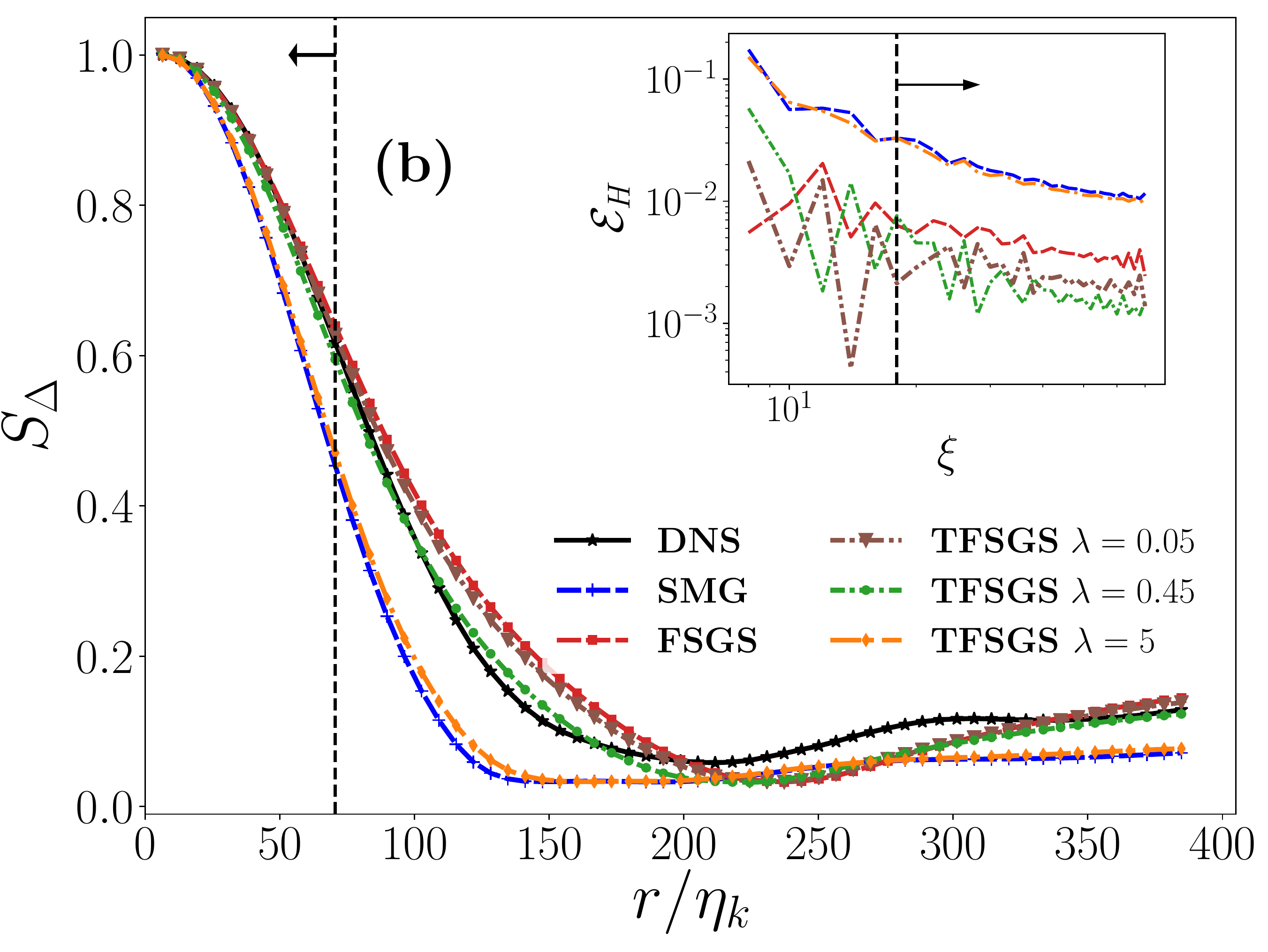}
		\subcaption*{} 
	\end{subfigure}
	\vspace{-0.25in}
	\caption{Comparing results of the normalized two-point stress-strain rate correlation functions for \textbf{(a)} $\mathcal{L}_{\delta}=4$ and \textbf{(b)} $\mathcal{L}_{\delta}=12$. The inset plots illustrate the scaled error of longitudinal dissipation spectrum ($\mathcal{E}_{H}$) versus the radius of Fourier wave numbers ($\xi$). The arrows point to the dissipation range.}
	\label{Figure3}
\end{figure}

Employing the same analysis for $\mathcal{L}_{\delta}=4, \, 12,$ we infer the optimal behavior of the TFSGS model, evaluated for a logarithmic range of $\lambda$ with a fixed $\alpha^{opt}$, in Figure \ref{Figure3}. The inset plots show $\mathcal{E}_{H}$ versus $\xi$ using log-scale on both axes to magnify the dissipation range at high wave numbers. Interestingly, at $\mathcal{L}_{\delta}=4$ the FSGS model is dissipative enough to outperform the tempered model in capturing the true $S_{\Delta}$ in Figure \ref{Figure3} \textcolor{blue}{(\textbf{a})}. With all this in mind, these results certify the importance of tempering in correct regeneration of two-point correlation functions particularly at larger filter widths ($\mathcal{L}_{\delta}=8, \, 12$). Moreover, the SMG model, resembling the TFSGS with $\lambda \sim 5$, exhibits a relatively steeper slope at the dissipation range, which is rooted in the diffusive form of its operator. In this context, tempering plays a crucial role in characterizing dissipation structures by comparing the widening gaps between the asymptotic cases ($\lambda = 0.01$ and $5$) in Figure \ref{Figure3} \textcolor{blue}{(\textbf{b})}. This brings up the TFSGS model as a superior physics based model in comparison with its counterparts, i.e., the SMG and the FSGS models.

Given the values of $\alpha^{opt}$ and $\lambda^{opt}$, we proceed lastly with quantifying $\mathrm{c}_{\alpha,\lambda}$ as prescribed in Algorithm \ref{alg:1}. Under statistically stationary circumstances of the flow field, $\mathrm{c}_{\alpha,\lambda}$ remains fairly unchanged for each $\mathcal{L}_{\delta}$ of interest, as reported in Table \ref{Table3}. It should be noted that $\mathrm{c}_{\alpha,\lambda}$ is only part of the fractional coefficient, described in \eqref{Sec3-3}, in order to scale up the model in a constant $Re_{\lambda}$ and $\mathcal{L}_{\delta}$.

\renewcommand{\arraystretch}{1.3}
\begin{table}
	\centering
	\caption{Optimized parameters associated with the TFSGS model in terms of Algorithm \ref{alg:1} for differing filter widths.} 
	\vspace{0.05in}
	\label{Table3}
	\begin{tabular}{ c c c c c c c c c  c c c c c c c c c c}
		\multicolumn{5}{c}{$\mathcal{L}_{\delta}=4$} &\multirow{2}{*}{}&  \multirow{2}{*}{}& \multicolumn{5}{c}{$\mathcal{L}_{\delta}=8$} & \multirow{2}{*}{}&  \multirow{2}{*}{} & \multicolumn{5}{c}{$\mathcal{L}_{\delta}=12$}
		\\ \cline{1-5}  \cline{8-12} \cline{14-19}
		$\alpha$  &    & $\lambda$ &   & $\mathrm{c}_{\alpha,\lambda}$  &  &  & $\alpha$  &    & $\lambda$ &   & $\mathrm{c}_{\alpha,\lambda}$  &  &  & $\alpha$  &    & $\lambda$ &   & $\mathrm{c}_{\alpha,\lambda}$  
		\\
		\cline{1-1} \cline{3-3} \cline{5-5} \cline{8-8} \cline{10-10} \cline{12-12} \cline{15-15} \cline{17-17} \cline{19-19}
		$0.76$  &    & $\simeq 0$ &   & $ 2.08$  &  &  & $0.58$  &    & $0.35$ &   & $0.88$  &  &  & $0.51$  &    & $0.45$ &   & $0.048$ 
		\\	
		&&&&&&&&&&&&&&&&&&	
	\end{tabular}
\end{table}

\subsection{Interpretation of two-point structure functions}
\label{Subsec 4-3}

\noindent The third-order structure functions, arising from the KH equations, provide insights about the statistics of unresolved scales and their strong interactions with large scale motions. As discussed previously in section \ref{Sec 3}, $G_{\Delta} = \frac{G_{LLL}}{\epsilon \mathcal{L}}$, representing the scaled two-point velocity-stress correlation function, is introduced as a sufficient condition for precise regeneration of third-order structure functions and an \textit{a priori} consistency in LES modeling. Following the derivation of the longitudinal Taylor maicroscale in \citep[][chapter 6]{Pope2001}, $D_{LL}(r)$ seems to be directly connected to the first-order derivative of $G_{LLL}(r)$ at the dissipation range. This offers the capability of the optimum edition of TFSGS model in capturing $G_{\Delta}$ and thereby fulfilling the essential conditions in \eqref{Sec3-2}.

\begin{figure}
	\centering
	\includegraphics[width=0.8 \linewidth]{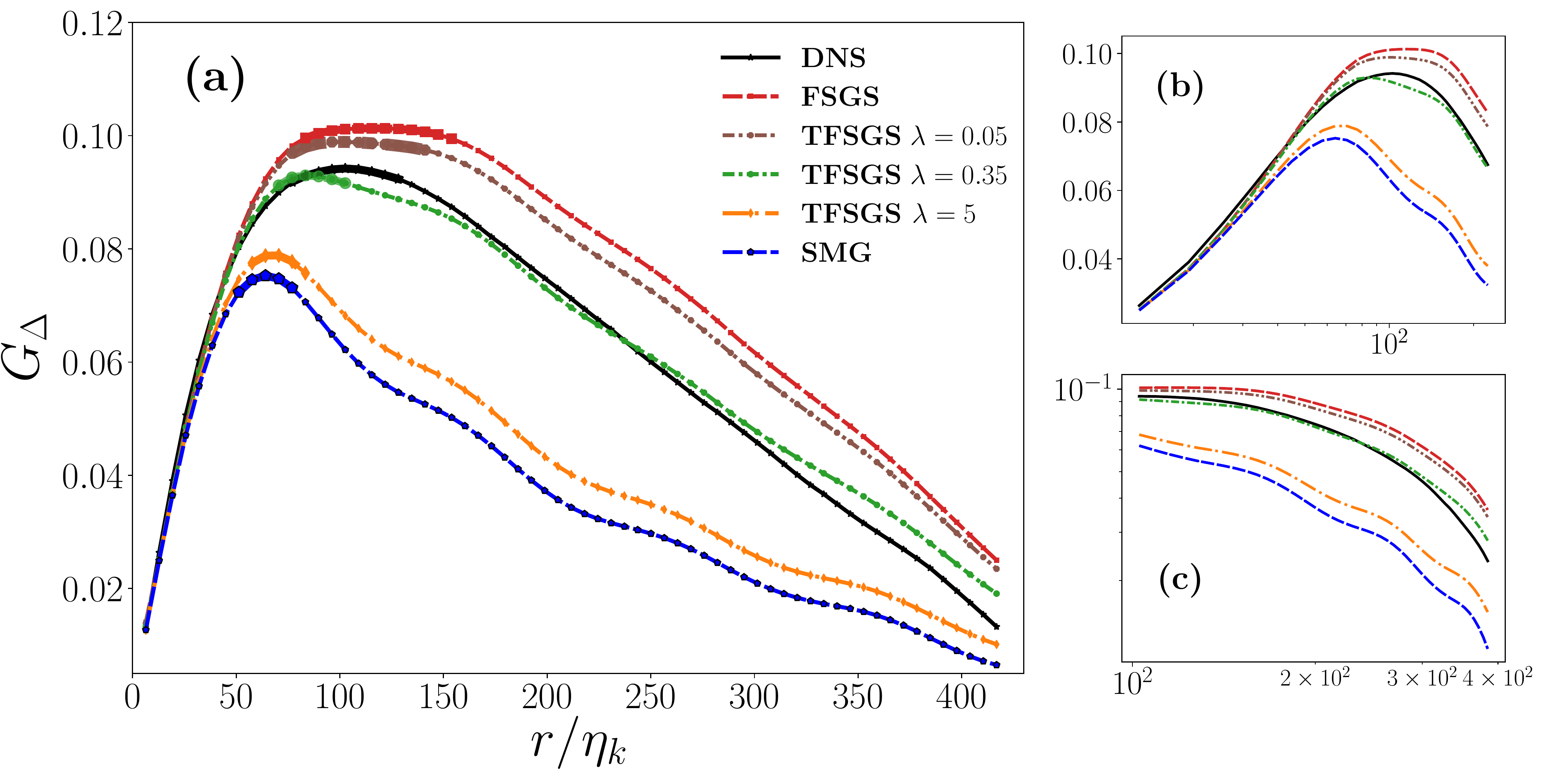}
	\caption{Two-point velocity-stress correlation function in a stationary HIT flow for $\mathcal{L}_{\delta}=8$ using box filtering. The segment of balance region (BR) has been thickened up in \textbf{(a)} for all the graphs. The dissipation and the inertial ranges have been enlarged in plots \textbf{(b)} with semi-logarithmic scale on the x-axis and \textbf{(c)} with logarithmic scale on the both axes, respectively.}
	\label{Figure 4}
\end{figure}

\begin{figure}
	\centering
	\begin{subfigure}[b]{0.45\textwidth}
		\centering
		\includegraphics[width=1 \linewidth]{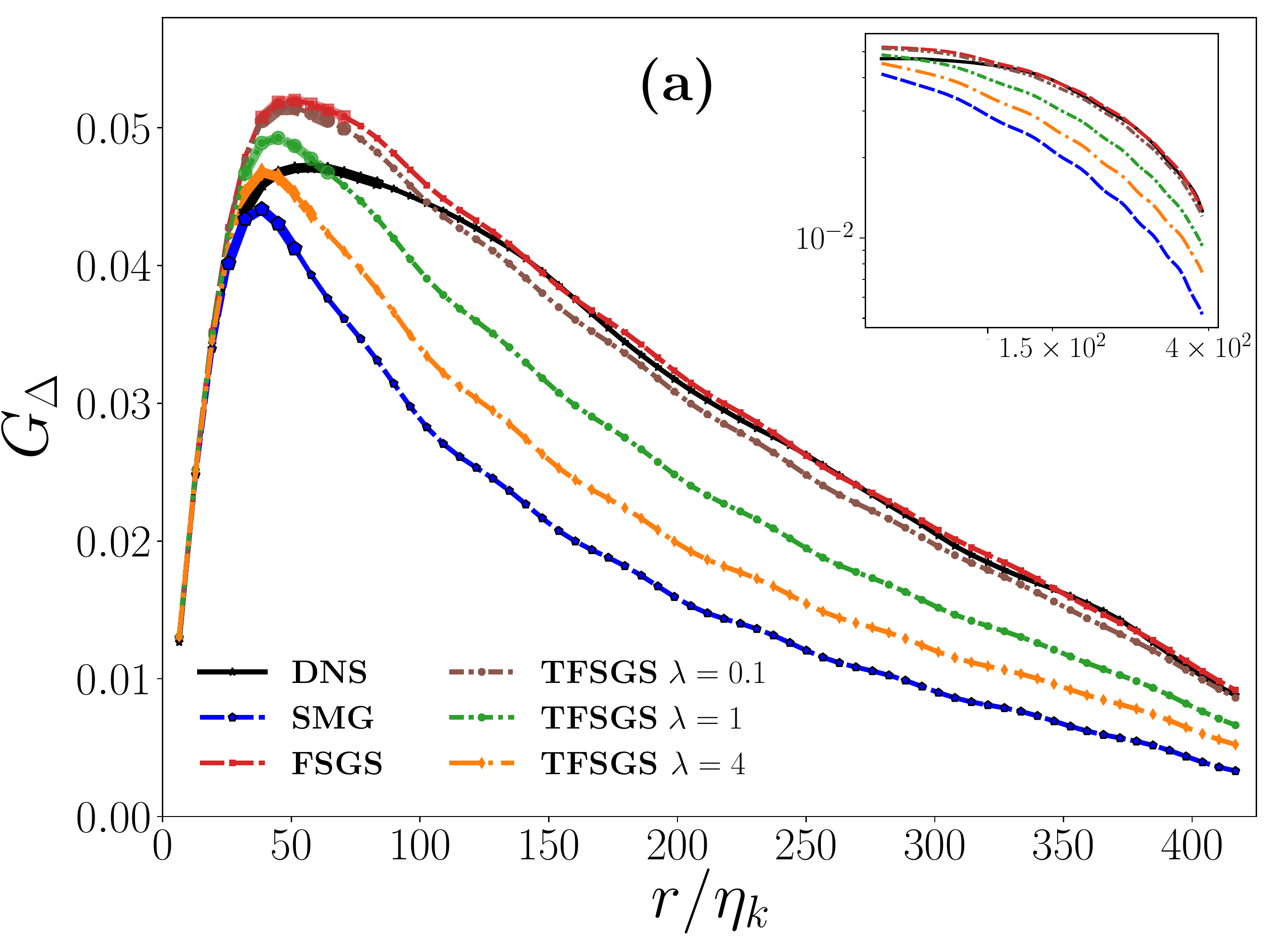}
		\subcaption*{}
	\end{subfigure}
	\hspace{0.02in}
	\begin{subfigure}[b]{0.45\textwidth}
		\centering
		\includegraphics[width=1 \linewidth]{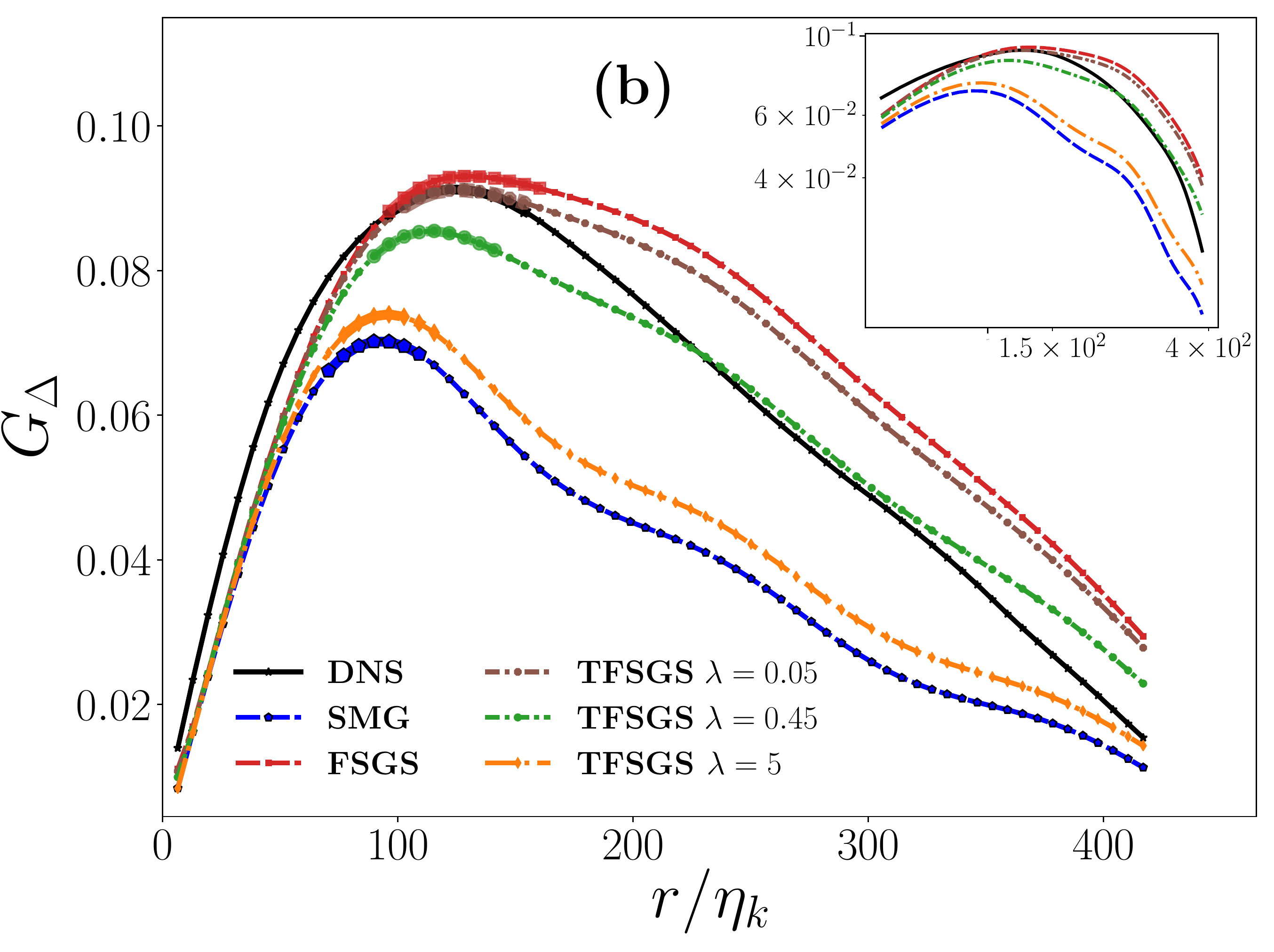}
		\subcaption*{} 
	\end{subfigure}
	\vspace{-0.25in}
	\caption{Two-point velocity-stress correlation function in a stationary HIT flow for \textbf{(a)} $\mathcal{L}_{\delta}=4$ and \textbf{(b)} $\mathcal{L}_{\delta}=12$ using box filtering. The inertial range has been enlarged in the inset plots with logarithmic scale on the both axes.}
	\label{Figure 5}
\end{figure}

In the first stage of the statistical analysis, we perform a comprehensive study on $G_{\Delta}$ in Figure \ref{Figure 4} \textcolor{blue}{(\textbf{a})} for $\mathcal{L}_{\delta} = 8$, in which the dissipation and inertial ranges are magnified in Figure \ref{Figure 4} \textcolor{blue}{(\textbf{b})} with semi-logarithmic scale on the x-axis and Figure \ref{Figure 4} \textcolor{blue}{(\textbf{c})} with logarithmic scale on both axes, respectively. The balance regions (BR), including extremum points, are thickened up in all the graphs in Figure \ref{Figure 4} \textcolor{blue}{(\textbf{a})}. BR also indicates the transitional zone between dissipation and inertial ranges. Aligned with the right side of the \eqref{Sec3-2}, the trend of $G_{\Delta}$ at small-scale interactions appear to be a linear function of spatial displacement, $r$, suggested by \citep[][Figure 2]{Meneveau1994}. The results in Figure \ref{Figure 4} \textcolor{blue}{(\textbf{a})} and more accurately in Figure \ref{Figure 4} \textcolor{blue}{(\textbf{b})} offer that the optimum TFSGS model well-predict the true DNS quantities at the left side of BR, not only the slope of $G_{\Delta}$ but also the maximum of $G_{\Delta}$ occurring at a relatively close $r$. This spotlights the importance of step three of Algorithm \ref{alg:1} in tuning slope of $G_{\Delta}$ at the dissipation range and effective role of the tempering parameter in fitting the BR, associated with the filtered DNS data. In practice, increasing $\lambda$ pushes BR toward the left side to preserve the increasing linear correlation as a notion of more dissipative behavior. These findings are endorsed qualitatively for the other filter widths in Figure \ref{Figure 5}, considering $\lambda^{opt}$ in Table \ref{Table3}.

In analysis of $G_{\Delta}$ at the inertial range, the graph, associated with $\lambda^{opt}$, shows a favorable match with true points, colored by black, in Figures \ref{Figure 4} and \ref{Figure 5}. For the purpose of clarity, the inertial ranges are magnified in log-log scale plots in Figure \ref{Figure 4} \textcolor{blue}{(\textbf{c})} for $\mathcal{L}_{\delta}=8$ and the inset plots in Figure \ref{Figure 5} for $\mathcal{L}_{\delta}=4, \, 12$, respectively. Motivated by these results, tempered fractional modeling seems to be faithful in fitting structures at the dissipation and the inertial ranges and also estimating the correct value of $r$, associated with the extremum points. Inevitably, enlarging $\mathcal{L}_{\delta}$ accounts for inaccuracies in fitting the tail of graphs as observed in \ref{Figure 5} \textcolor{blue}{(\textbf{b})}. Notwithstanding, the mid-range interactions are acceptably predicted by the optimized TFSGS model. 

With an overview of the present results, the TFSGS model stands out as a structure based approach, which reasonably covers the gap between the FSGS and the SMG models. In $\mathcal{L}_{\delta}=4$, $\lambda^{opt}$ is found to be very close to zero, which renders tempering nonessential in capturing two-point structures. As we increase $\mathcal{L}_{\delta}$, this gap starts widening up and tempering mechanism acts more dynamically in finding the true BR and fitting the dissipation structures. This argument confirms that the tempered fractional approach displays a great potential for parameterizing structure function especially at larger filter widths while retaining fairly acceptable accuracy. 

\subsection{PDF of SGS stresses}
\label{Subsec 4-4}

\begin{figure}
	\centering
	\begin{subfigure}[b]{0.32\textwidth}
		\centering
		\includegraphics[width=1.0 \linewidth]{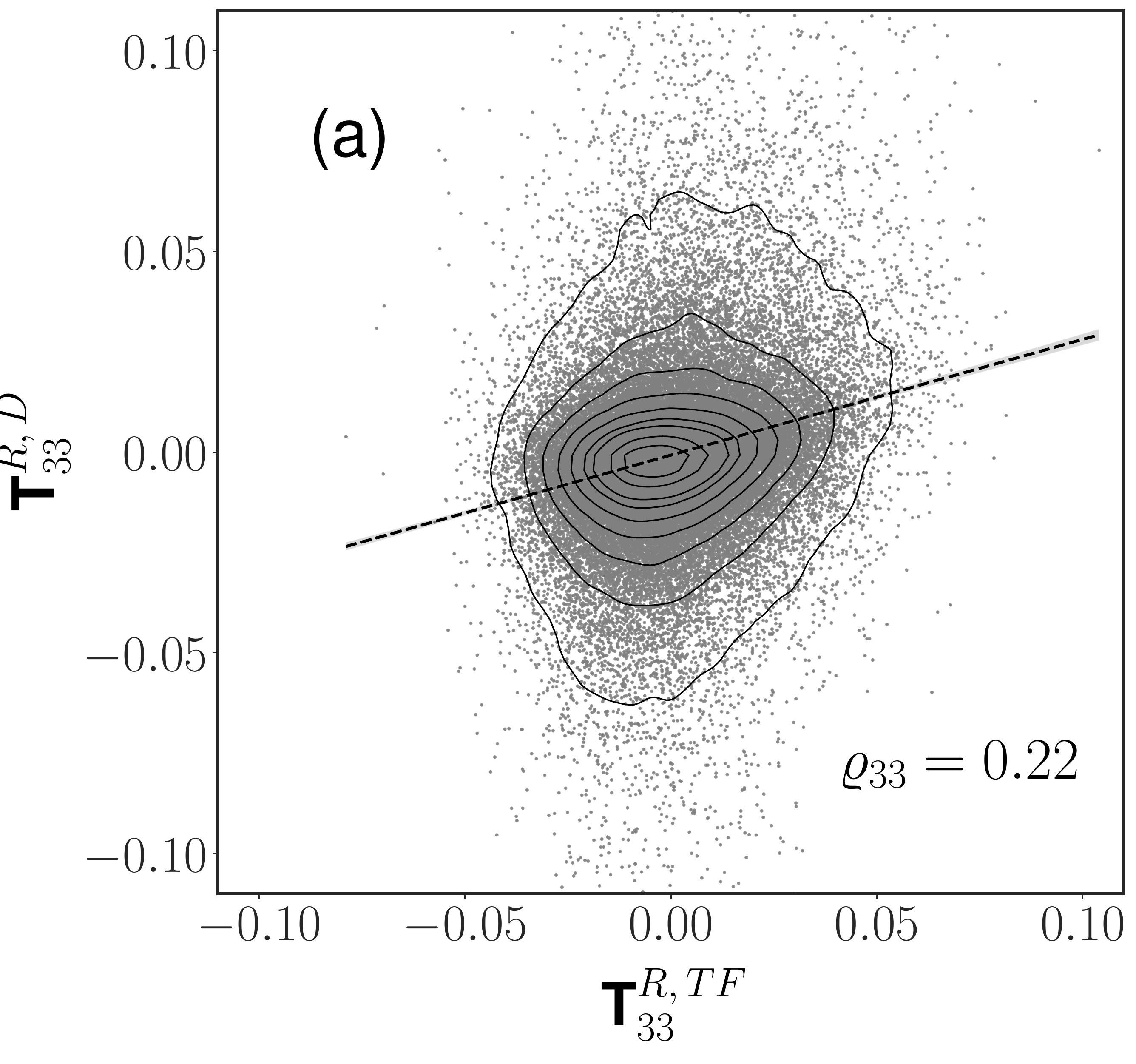}
		\subcaption*{}
	\end{subfigure}
	\begin{subfigure}[b]{0.32\textwidth}
		\centering
		\includegraphics[width=1.0 \linewidth]{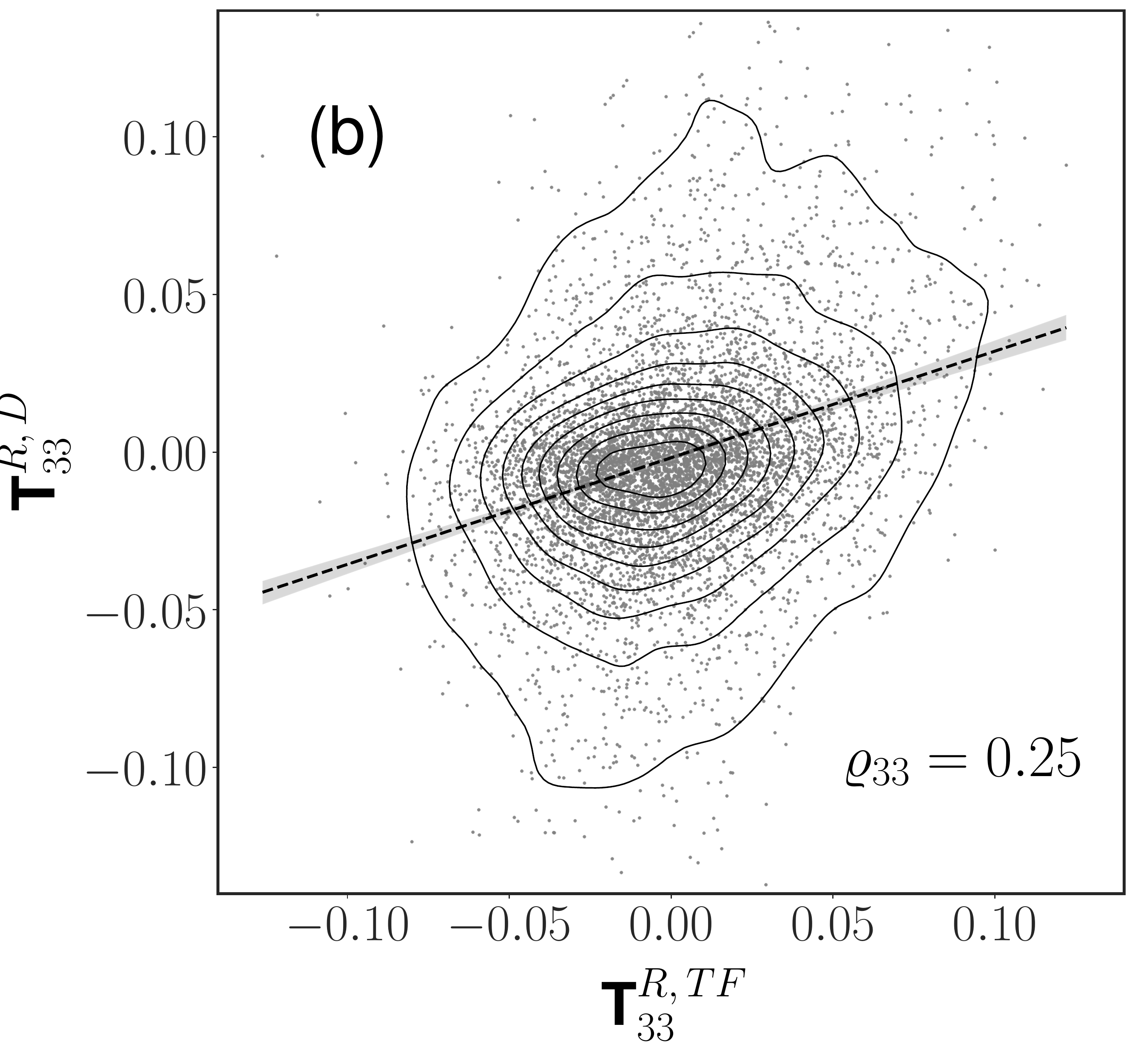}
		\subcaption*{} 
	\end{subfigure}
	\begin{subfigure}[b]{0.32\textwidth}
		\centering
		\includegraphics[width=1.0 \linewidth]{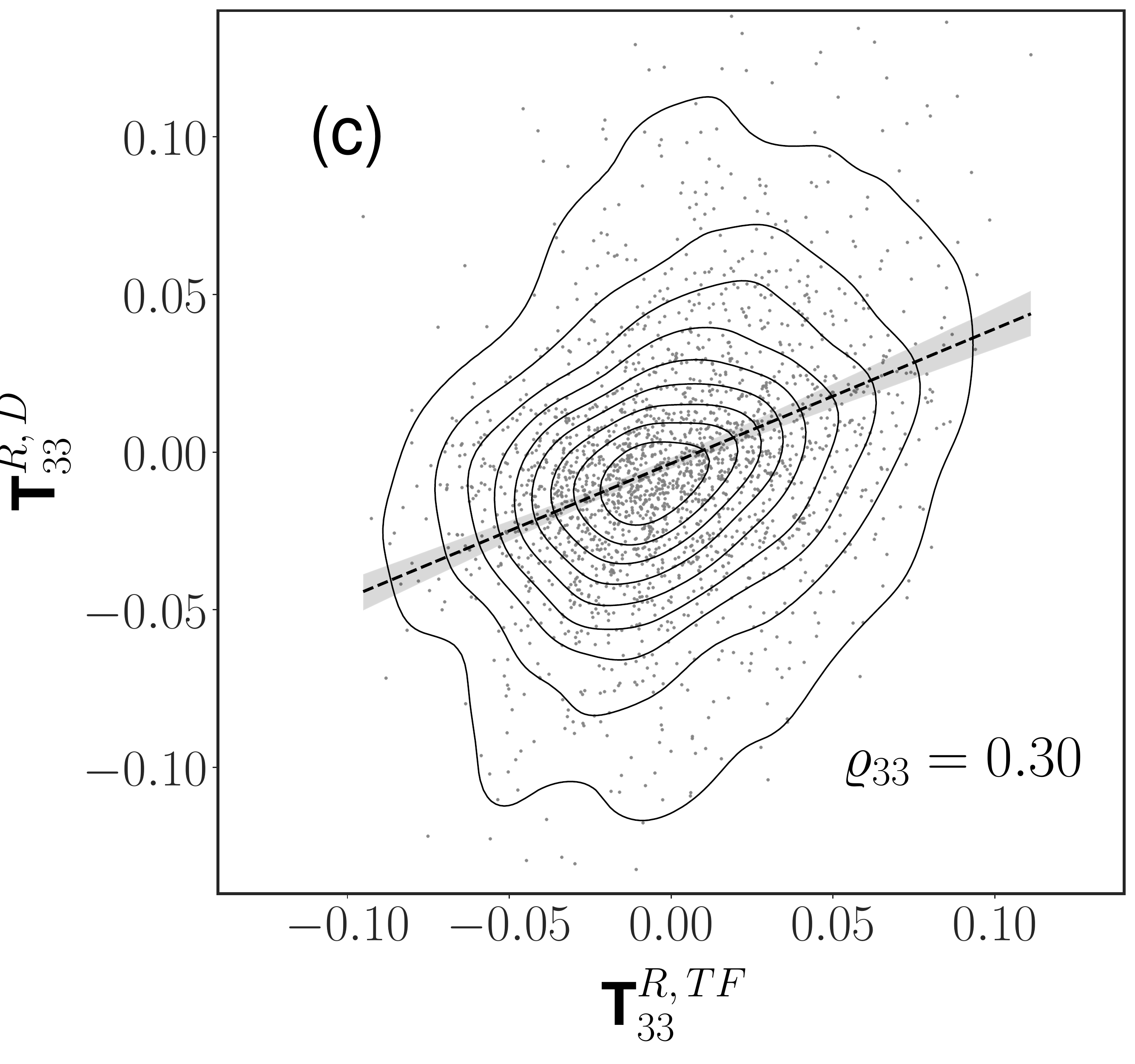}
		\subcaption*{} 
	\end{subfigure}
	\vspace{-0.25in}
	\caption{Scatter plots of the SGS stresses obtained by the filtered DNS data ($\mathsfi{T}^{R,\,D}_{33}$) versus the modeled stresses ($\mathsfi{T}^{R,\,TF}_{33}$) using optimized parameters in Table \ref{Table3} for \textbf{(a)} $\mathcal{L}_{\delta}=4$, \textbf{(b)} $\mathcal{L}_{\delta}=8$, and \textbf{(c)} $\mathcal{L}_{\delta}=12$.}
	\label{Figure 6}
\end{figure}

\begin{figure}
	\centering
	\begin{subfigure}[b]{0.32\textwidth}
		\centering
		\includegraphics[width=1.0 \linewidth]{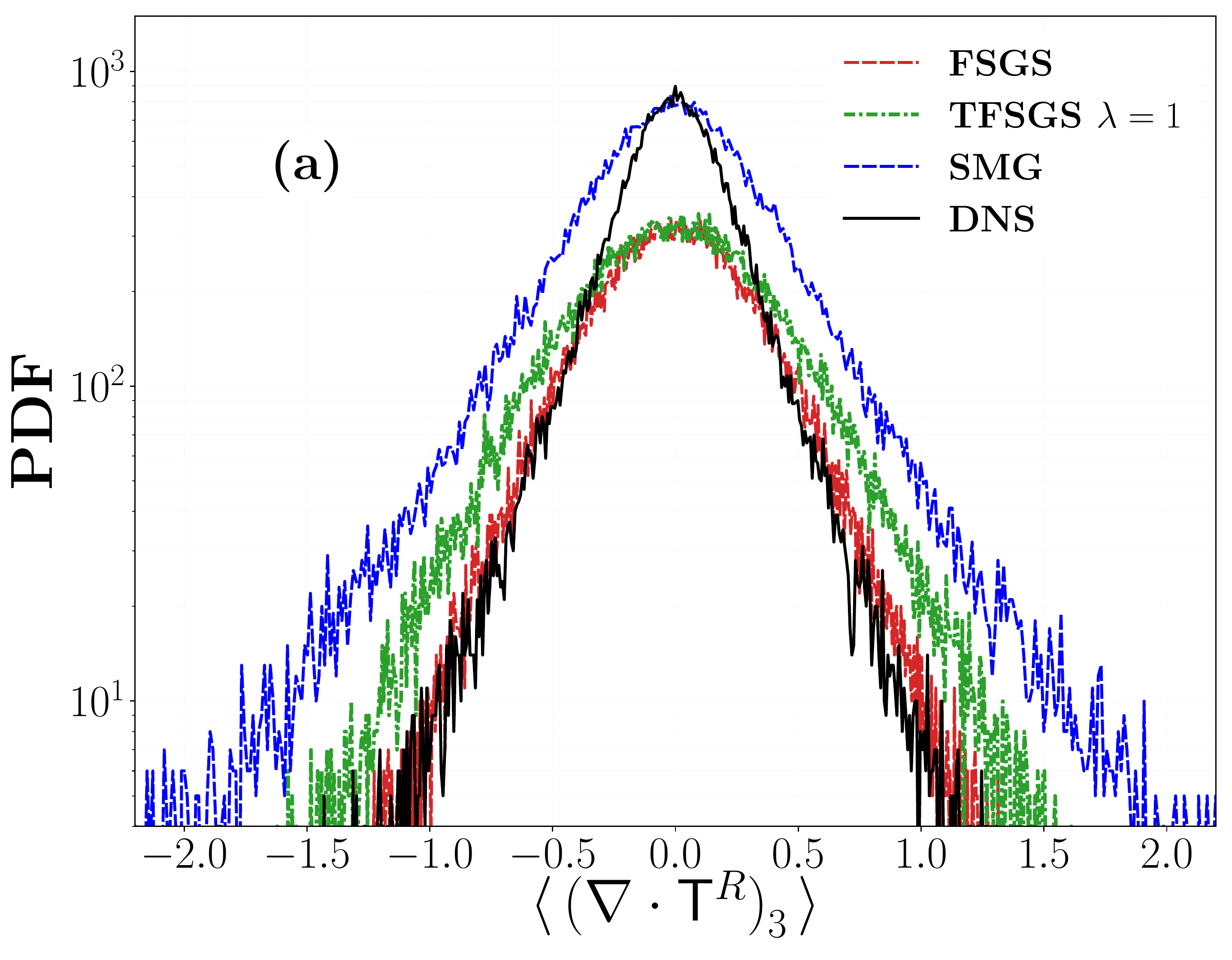}
		\subcaption*{}
	\end{subfigure}
	\begin{subfigure}[b]{0.32\textwidth}
		\centering
		\includegraphics[width=1.01 \linewidth]{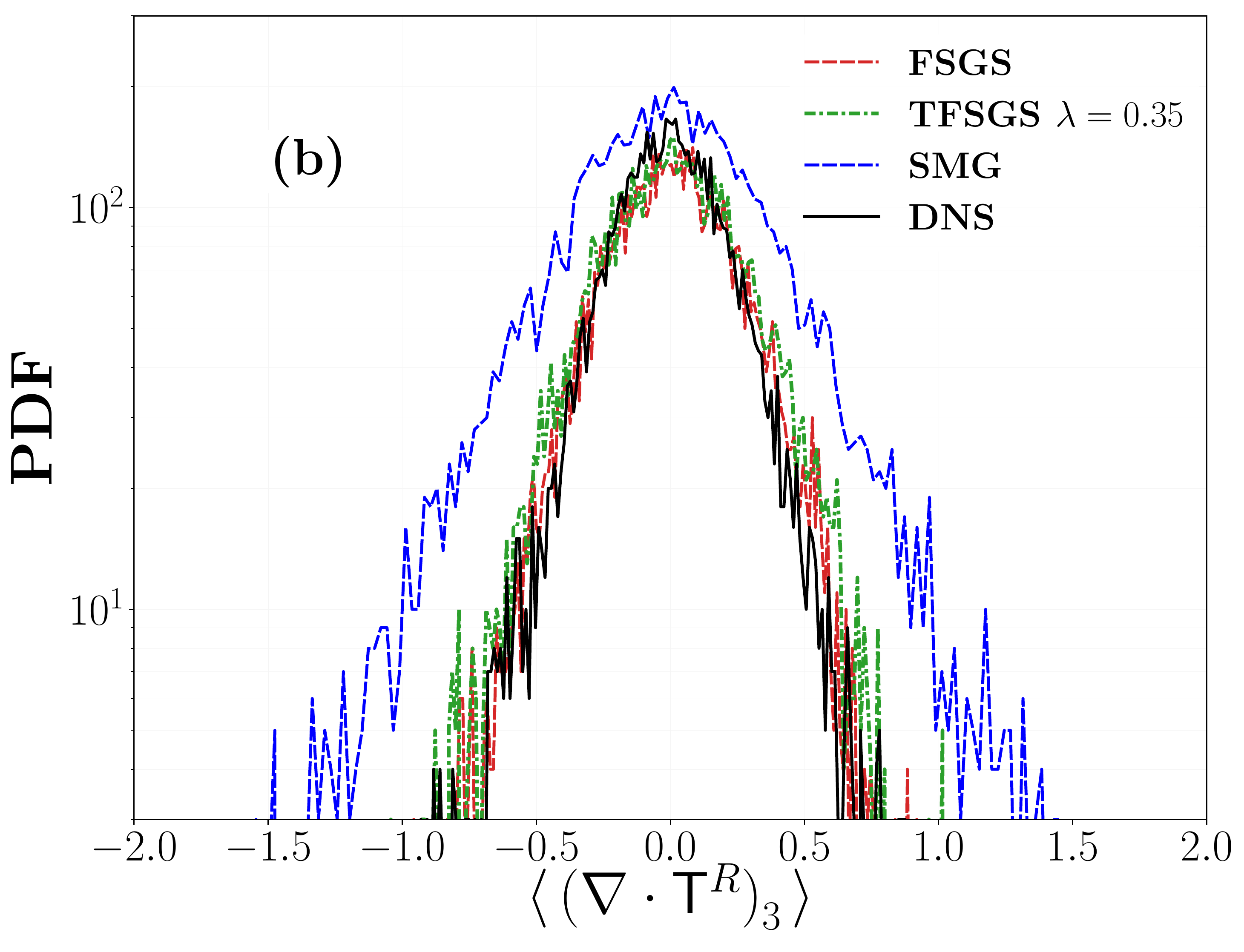}
		\subcaption*{} 
	\end{subfigure}
	\begin{subfigure}[b]{0.32\textwidth}
		\centering
		\includegraphics[width=1.0 \linewidth]{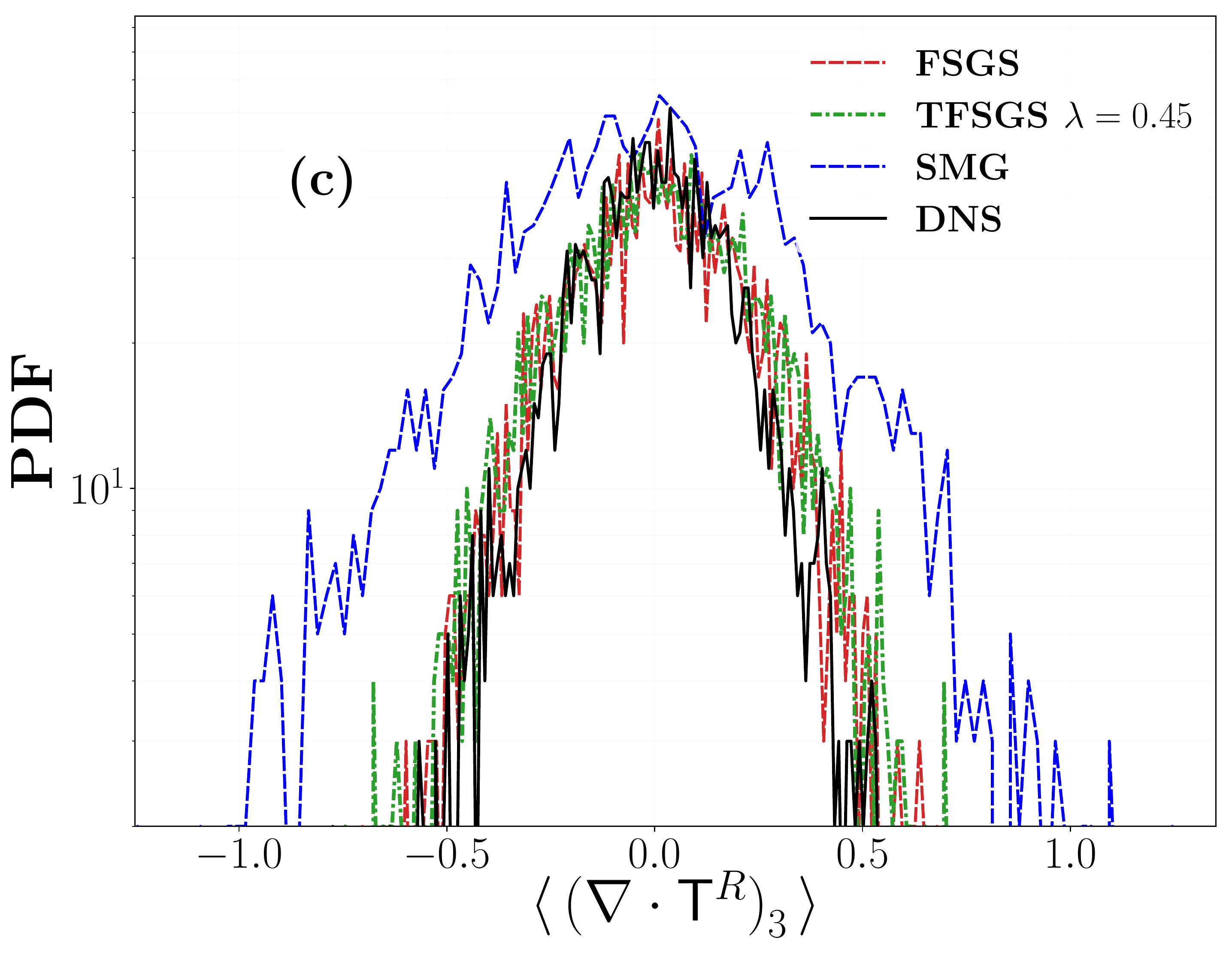}
		\subcaption*{} 
	\end{subfigure}
	\vspace{-0.25in}
	\caption{PDF of the ensemble-averaged $\big ( \nabla \cdot \mathsf{T} \big )_{i=3}$ for the optimized (tempered) fractional and the SMG models at \textbf{(a)} $\mathcal{L}_{\delta}=4$, \textbf{(b)} $\mathcal{L}_{\delta}=8$, and \textbf{(c)} $\mathcal{L}_{\delta}=12$.}
	\label{Figure 7}
\end{figure}

\noindent Within the proposed statistical framework, the last step in Algorithm \ref{alg:1} focuses on the PDFs of filtered DNS data. The key idea is to assess the performance of models and verify if the proposed model maintains the true statistics. In this context, we present the scatter plots of $\mathsfi{T}^{R,\,D}_{ii}$ against $\mathsfi{T}^{R,\,TF}_{ii}$ in Figure \ref{Figure 6} for three given filter widths and $i=3$. We should note that the present results are confined to $i=3$ due to the similarities in other directions. The slope in each plot is indicated by the corresponding correlation coefficients in Table \ref{Table2}. The most noticeable specific about these results is that the data points are bounded within a same order of magnitude on both axes. As a matter of fact, we achieve a roughly unit regression coefficient between $\mathsfi{T}^{R,\,D}_{ii}$ and $\mathsfi{T}^{R,\,TF}_{ii}$, where our optimization strategy targets for correct estimation of the SGS dissipation. This analysis can be extended to the PDF plots in Figure \ref{Figure 7}. With nearly the same correlation coefficients, the SMG model fails to reproduce the true statistics, while the optimum TFSGS model offers a great match with the true graphs. As pointed out previously, in $\mathcal{L}_{\delta}=4$ the FSGS model represents the equivalent form of the TFSGS with $\alpha^{opt}=0.76$ and $\lambda^{opt}\simeq 0$.


\begin{figure}
	\centering
	\begin{subfigure}[b]{0.32\textwidth}
		\centering
		\includegraphics[width=1.0 \linewidth]{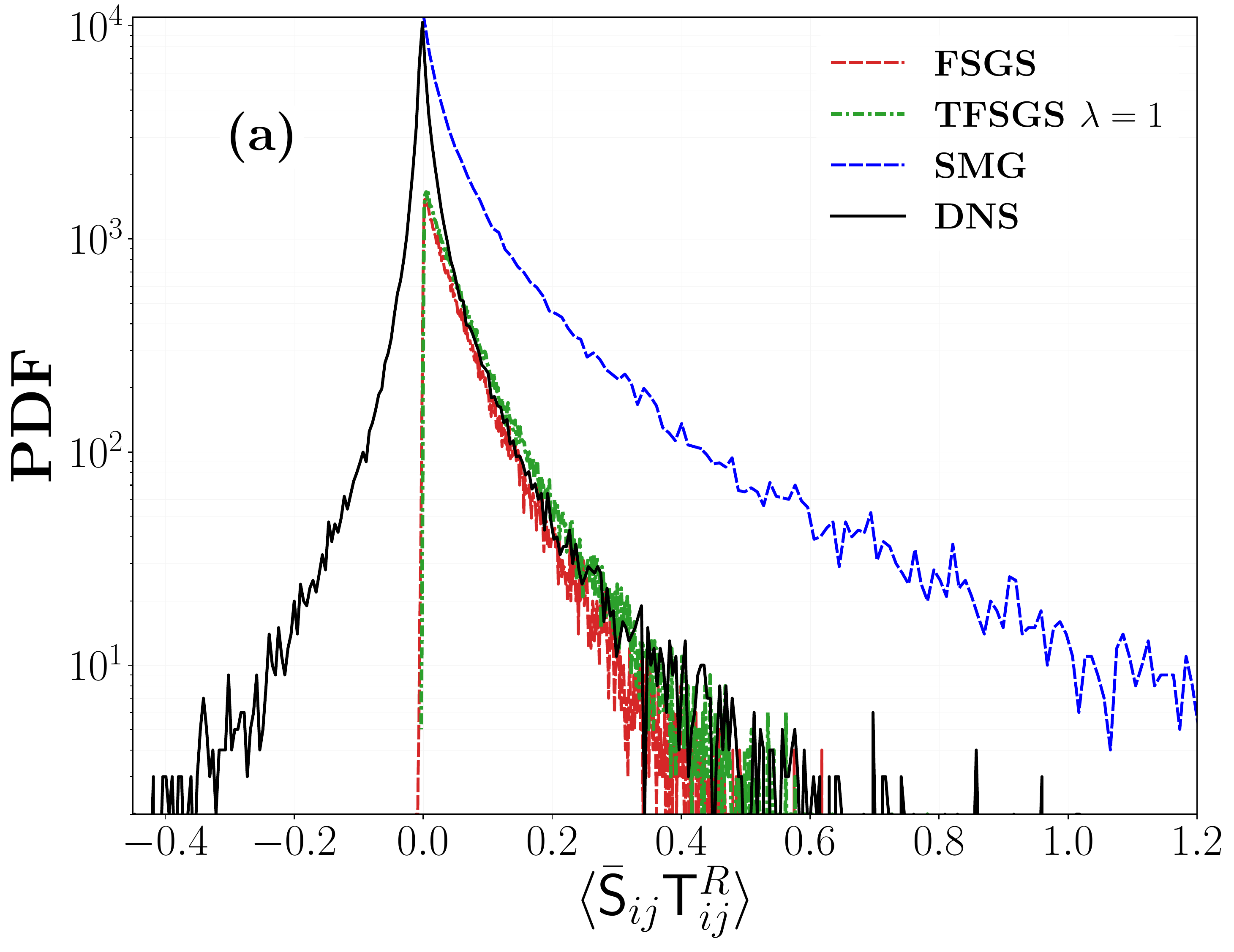}
		\subcaption*{}
	\end{subfigure}
	\begin{subfigure}[b]{0.32\textwidth}
		\centering
		\includegraphics[width=1.0 \linewidth]{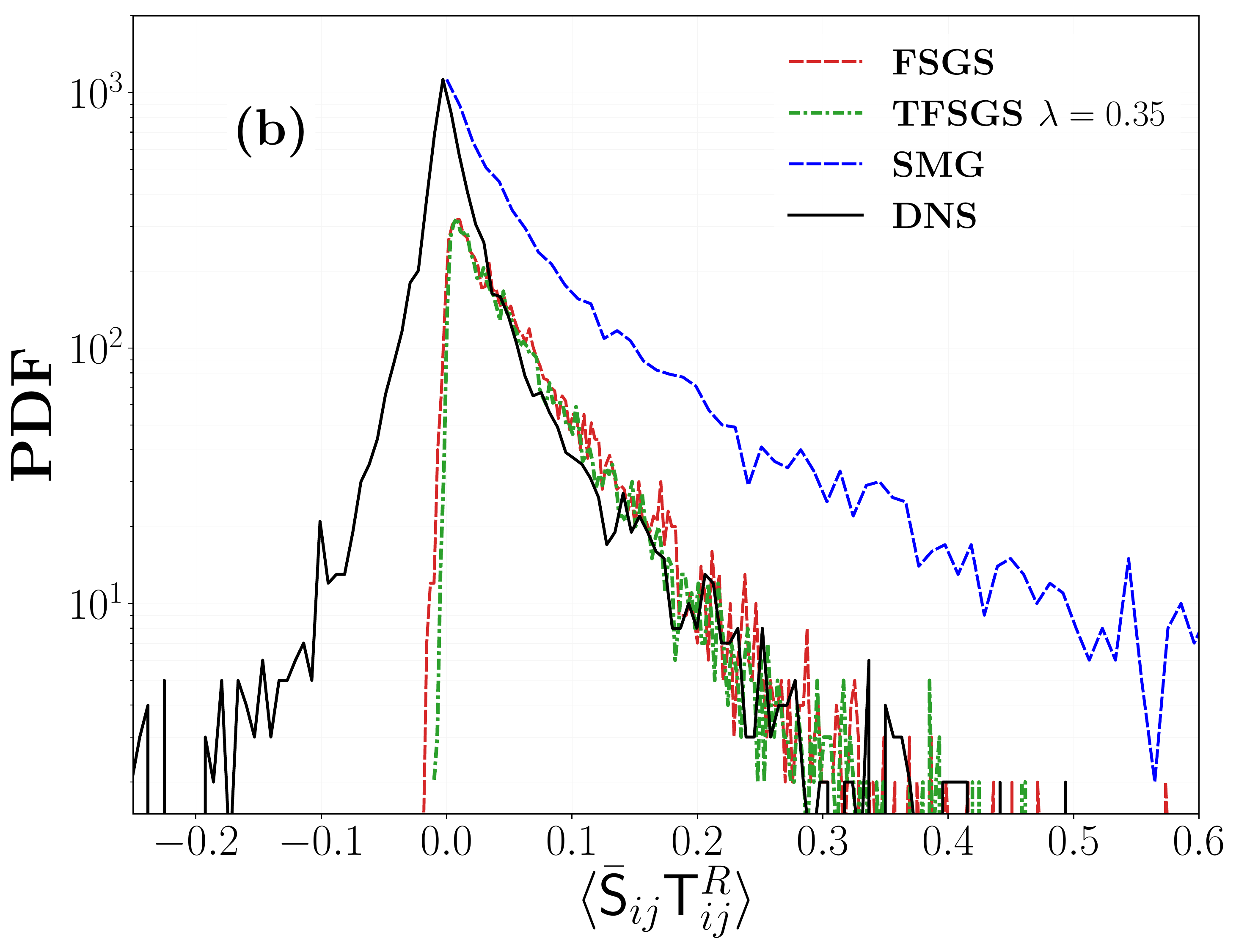}
		\subcaption*{} 
	\end{subfigure}
	\begin{subfigure}[b]{0.32\textwidth}
		\centering
		\includegraphics[width=1.0 \linewidth]{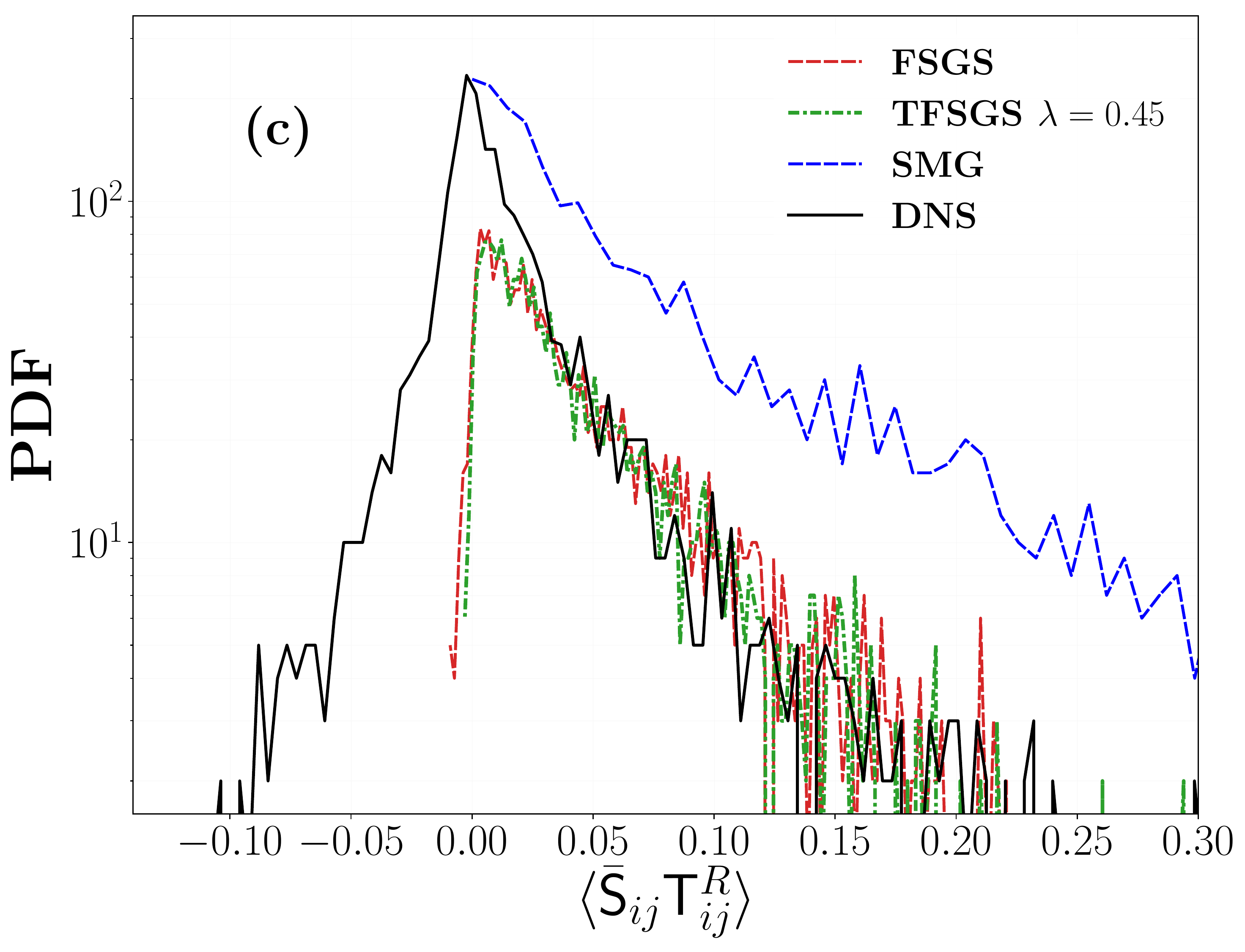}
		\subcaption*{} 
	\end{subfigure}
	\vspace{-0.25in}
	\caption{PDF of the ensemble-averaged SGS dissipation for the optimized (tempered) fractional and the SMG models at \textbf{(a)} $\mathcal{L}_{\delta}=4$, \textbf{(b)} $\mathcal{L}_{\delta}=8$, and \textbf{(c)} $\mathcal{L}_{\delta}=12$.}
	\label{Figure 8}
\end{figure}

From the understanding of energy cascading in turbulent flows, the SGS dissipation, $\epsilon$, is considered as an external parameter in two-point structure equations for describing small-scale motions. In the statistical sense, we compare the PDFs of $\epsilon$, implied by the models, with the true PDFs, obtained by the filtered DNS data for $\mathcal{L}_{\delta}=4,\, 8,\, 12$. As shown in Figure \ref{Figure 8}, the fractional models accurately predicts the forward scattering, associated with the positive dissipation, $\epsilon^{+}$, while the SMG model appears to be too dissipative due to its positive eddy viscosity. Furthermore, the TFSGS model presents an under-prediction of the backward scattering by producing a slim amount of negative dissipation, $\epsilon^{-}$. On the side of numerical analysis, this limitation results in preserving numerical stability by minimizing negative dissipation error. 
   
\subsection{\textit{A Posteriori} Analysis}
\label{Sec 5}

\noindent With a focus on numerical stability, we extend the statistical \textit{a priori} assessments to an \textit{a posteriori} analysis in order to evaluate the performance of proposed models in time. To outline the \textit{a posteriori} framework, we employ the flow solver, described in section \ref{Subsec 4-1}, on $40^3$ and $20^3$ grids for the corresponding $\mathcal{L}_{\delta}=4,\, 8$, respectively. The simulations are initiated with an instantaneous flow field, given from the filtered DNS of the stationary HIT flow. This analysis also allows for structural comparisons between the fractional models with the filtered DNS data through the resolved turbulent kinetic energy, $\overline{K}_{tot}(t) = \langle  \frac{\bar{V}_i\bar{V}_i}{2}\rangle_s$, and the resolved enstrophy, $\overline{\mathcal{E}}(t)= \langle \frac{\bar{\omega}_i\bar{\omega}_i}{2} \rangle_s$, where $\bar{\omega}_i(t,\,x_i)$ denotes the $i$th component of the instantaneous filtered vorticity field. It should be noted that in this section $\langle \cdot \rangle_{s}$ denotes the spatial averaging over the entire domain.

\begin{figure}
	\centering
	\begin{subfigure}[b]{0.45\textwidth}
		\centering
		\includegraphics[width=1.0 \linewidth]{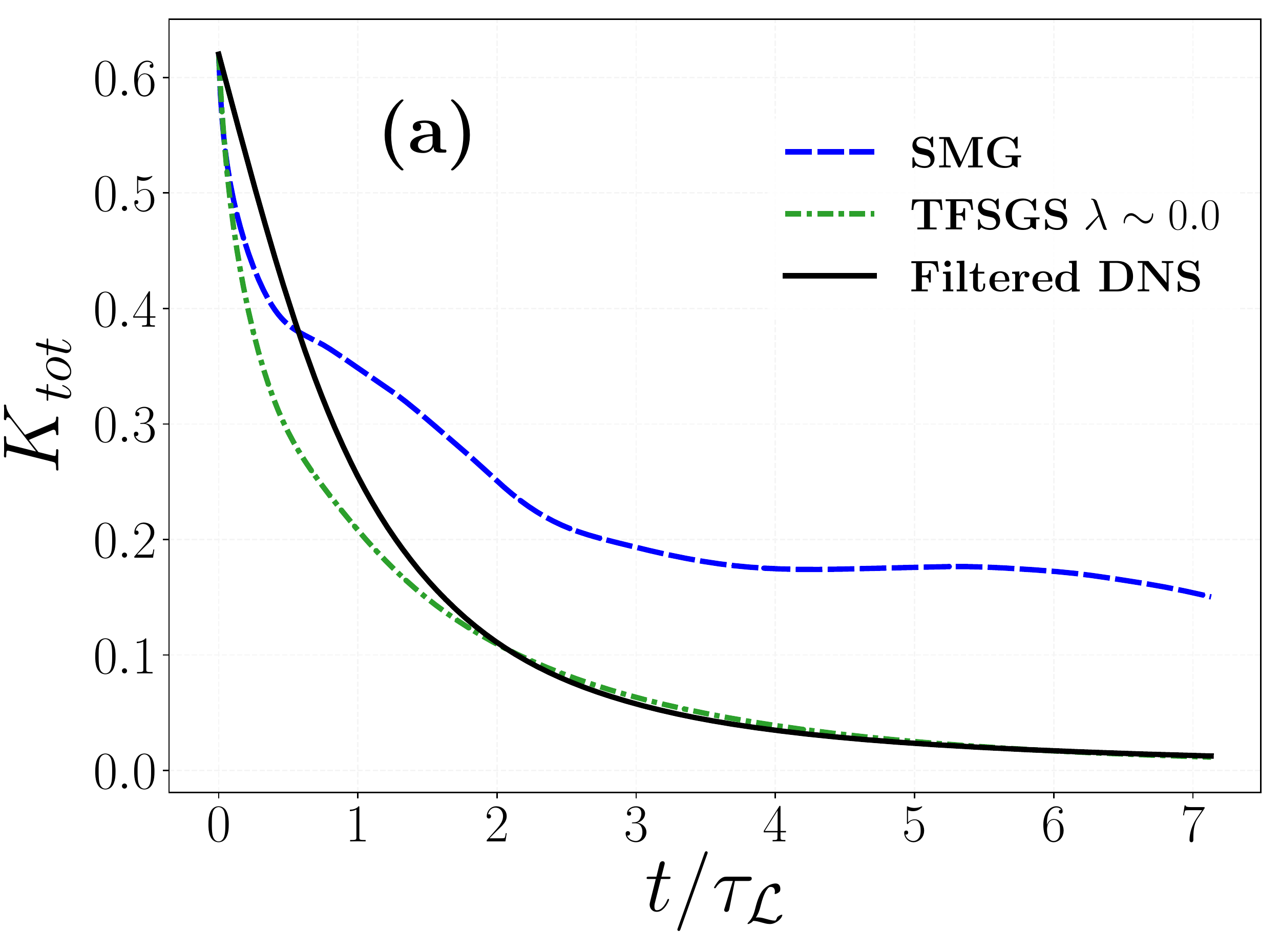}
		\subcaption*{}
	\end{subfigure}
	\begin{subfigure}[b]{0.45\textwidth}
		\centering
		\includegraphics[width=1.0 \linewidth]{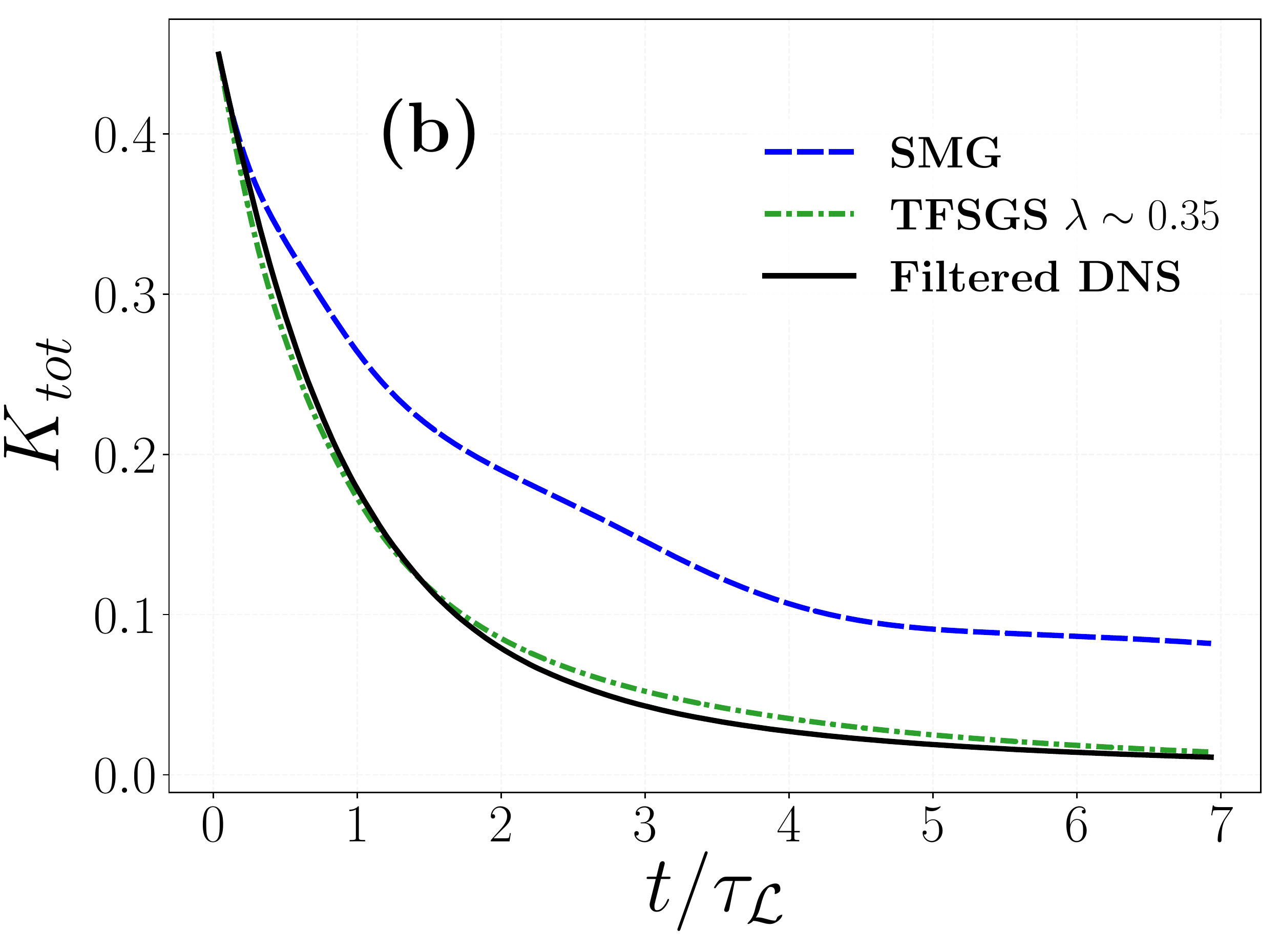}
		\subcaption*{} 
	\end{subfigure}
	\vspace{-0.25in}
	\caption{Decay of the resolved turbulent kinetic energy, $\overline{K}_{tot}$, for the optimized TFSGS and the SMG models with \textbf{(a)} $40^3$ and \textbf{(b)} $20^3$grid points.}
	\label{Figure 9}
\end{figure}

Figure \ref{Figure 9} displays the decay of kinetic energy in the subgrid-scale level, which verifies the computational stability for the fractional model. In most LES approaches, fidelity in representing spatial structures is essentially compromised to preserve numerical stability by inducing the excessive amount of energy dissipation. Nevertheless, the present results verify our findings in section \ref{Sec 3} that the TFSGS model provides stable LES solutions while preserving high-order structure functions in the \textit{a priori} tests. Comparing the results in Figure \ref{Figure 9}, the fractional model also exhibits an acceptable performance in predicting the time-evolution of kinetic energy after one or two eddy turn-over times, $\tau_{\mathscr{L}}$, for both resolutions.

\begin{figure}
	\centering
	\begin{subfigure}[b]{0.45\textwidth}
		\centering
		\includegraphics[width=1.0 \linewidth]{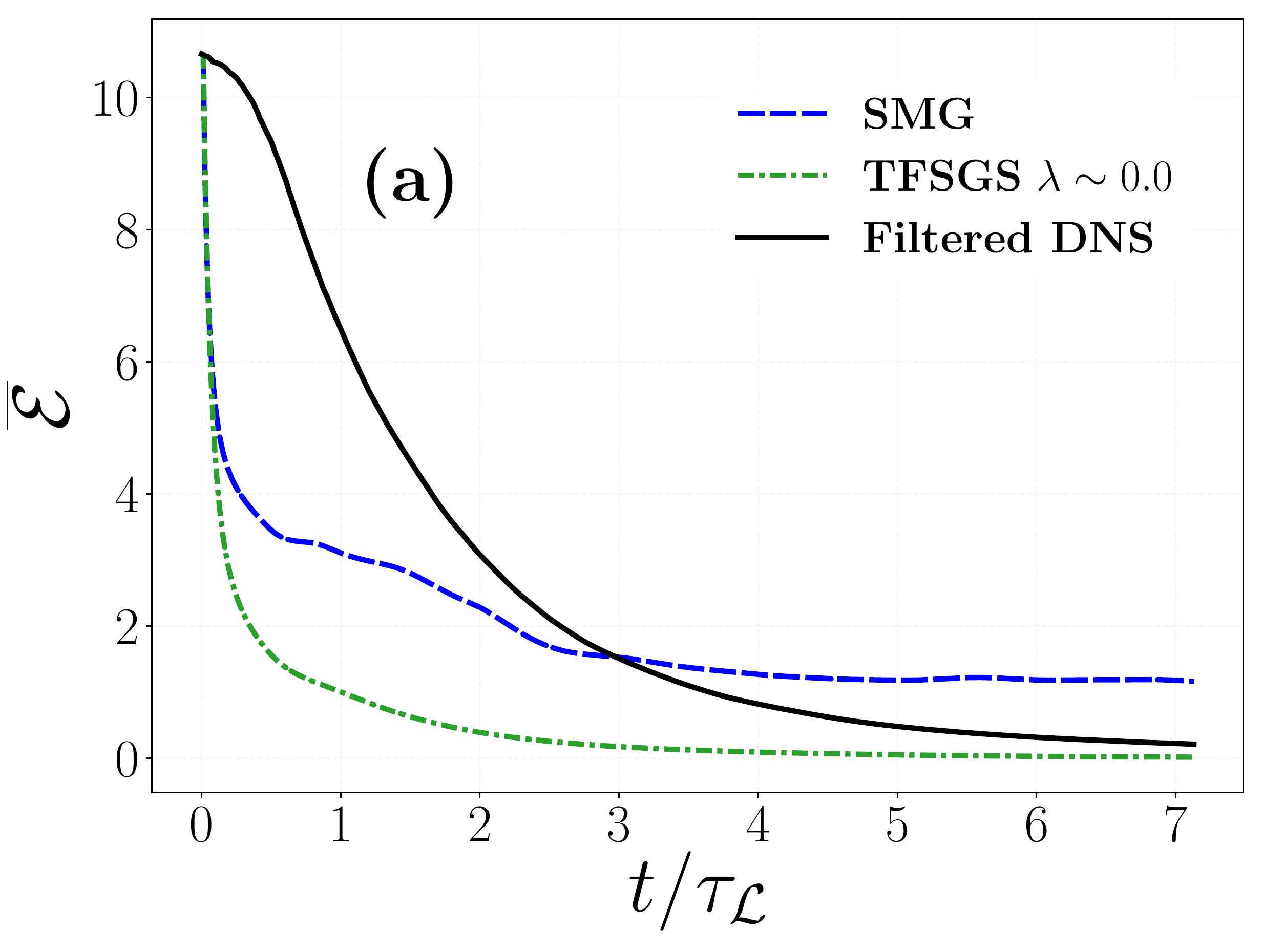}
		\subcaption*{}
	\end{subfigure}
	\begin{subfigure}[b]{0.45\textwidth}
		\centering
		\includegraphics[width=1.0 \linewidth]{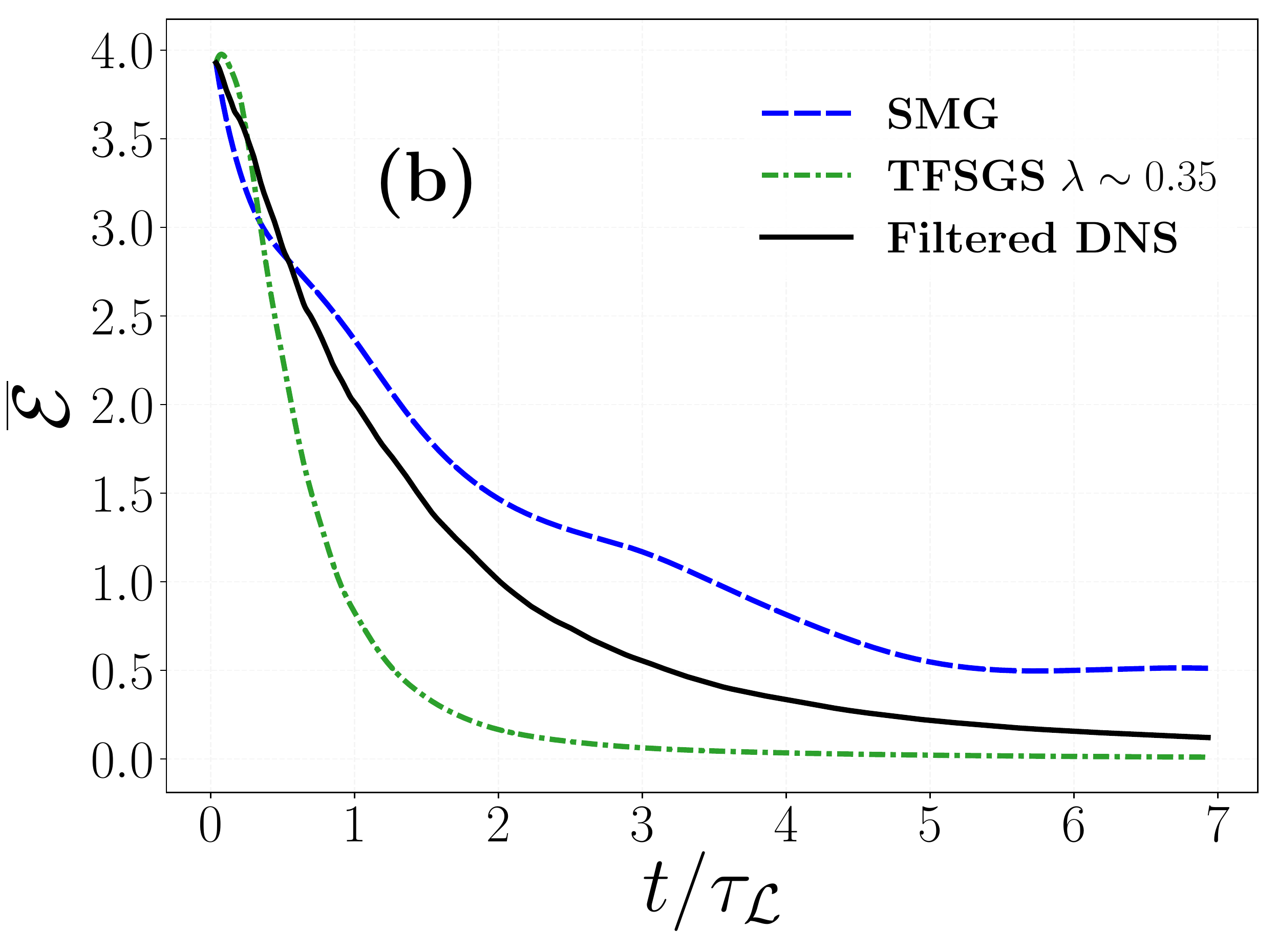}
		\subcaption*{} 
	\end{subfigure}
	\vspace{-0.25in}
	\caption{Time-evolution of the resolved enstrophy, $\overline{\mathcal{E}}$, for the optimized TFSGS and the SMG models with \textbf{(a)} $40^3$ and \textbf{(b)} $20^3$grid points.}
	\label{Figure 10}
\end{figure}

Given the analysis in section \ref{Sec 3}, third-order structure functions  are inherently connected with the third-order moments of filtered velocity derivatives and thereby the resolved enstrophy balance as described in \citep[][]{Meneveau1994}. Inspired by that, we also study the time evolution of $\overline{\mathcal{E}}$ in Figure \ref{Figure 10}, which evidently confirms numerical stability of the fractional LES solutions in both cases. 


\subsection{Merits, challenges, and future works}
\label{Subsec 5-1}

\noindent On the basis of \textit{a priori} and \textit{a posteriori} analyses, the present work provides a robust physics based framework for fractional LES modeling of SGS structures. The significance of this approach lies in:

\begin{enumerate}
	
	\vspace{0.05 in}
	\item \hspace{0.03 in} We treat the source of turbulent small-scale motions at the kinetic level, by employing a tempered heavy-tailed distribution in approximating $\overline{f^{eq}}$. This leads us to the tempered fractional operator in the filtered NS equations as a proper choice for modeling a power-law like behavior in the mid-range and a Gaussian tail in real-physics anomalous phenomena.

	\vspace{0.05 in}	
	\item \hspace{0.03 in} The proposed TFSGS model sets the ground for fulfilling essential statistical conditions as a relatively best approximation of an ideal LES model through fractional and tempering parameters. To achieve an optimized edition of the TFSGS model, we devise an optimization strategy, which involves conventional one-point correlation coefficients, two-point structures, and the SGS dissipation.
	
	\vspace{0.05 in}
	\item \hspace{0.03 in} The optimized TFSGS model presents reasonably accurate predictions of two-point structure functions for a range of filter widths while maintaining the expected correlations between the modeled and true SGS stresses.

	\vspace{0.05 in}	
	\item \hspace{0.03 in} The corresponding fractional LES solutions present a stable prediction of energy and enstrophy decays in the \textit{a posteriori} analysis. 
\end{enumerate}

\vspace{0.05 in}
Despite the theorertical and statistical achievements, we believe that as a viable and promising direction toward nonlocal modelings, this approach can be upgraded to attain better accuracy and involve higher-order statistical properties of turbulent flows. On the theoretical side, the current framework deserves a careful mathematical attention to be extended to anisotropic and inhomogeneous flows employing proper forms of distributions at the kinetic level. Moreover, further works should be undertaken to generalize the fractional model to a data-driven representation of spatial and temporal structures in more complex turbulent regimes.     

\section{Conclusions and Remarks}
\label{Sec 6}

\noindent Inspired by nonlocality, embedded in interactions between large and small scale motions, we developed a tempered fractional SGS model for LES of HIT flows. We began with modeling of turbulent effects at the kinetic level by closing the collision term in the filtered BT equation. To approximate multi-exponential behaviors of the filtered equilibrium distribution in the collision operator, we employed a tempered \textit{L\'evy}-stable distribution function, which presents a power-law at a moderate range and then converges to an exponential decay. By ensemble-averaging of the approximated Boltzmann transport, we derived the LES equations, in which the divergence of SGS stresses emerged as a summation of tempered fractional Laplacian, $(\Delta+\lambda)^{\alpha}(\cdot)$, where $\alpha \in (0,1)$, $\alpha\neq \frac{1}{2}$, and $\lambda > 0$. Interestingly, the FSGS is found to be a particular form of the TFSGS model when $\lambda$ approaches $0$. Moreover, we formulated the SGS stresses straightforwardly in terms of a combination of integer and fractional operators, which gives the advantage of being feasible and quite easy to implement in the Fourier space. The corollary on frame invariant property of the FSGS model were also extended to the current model, showing its physical and mathematical consistency.  

In a statistical framework, we constructed a structure based algorithm for optimizing the fractional models, which involved the closed essential conditions for a weaker sense of an ideal LES model. Following the optimization strategy, we inferred the optimum tempering parameter through a comparative study of two-point strain-stress correlation functions while the fractional exponent was fixed for maintaining reasonable values of correlation coefficients. Next, we quantified the fractional coefficient using SGS dissipation as a crucial factor in identifying high-order structures. The more profound analysis of dissipation structure functions emphasized on the central role of $\lambda$ in spanning the widening gap between the FSGS and the SMG models, especially at larger $\mathcal{L}_{\delta}$, when $\alpha^{opt}$ decreases. Regarding the KH equation, the optimum TFSGS model presented a great match with the true values of two-point velocity-stress correlation functions, which ensures the accurate prediction of third-order structure functions. 

The success of tempering mechanism in capturing structure correlation functions, particularly at larger $\mathcal{L}_{\delta}$, originated from the capabilities of our choice in fitting semi-heavy-tailed behavior of the filtered equilibrium distribution at the kinetic level. The inspection of statistical results also supported accuracy of the fractional model in keeping unit regression and capturing the corresponding PDF tails. As a notion of numerical stability, we demonstrated that the optimized TFSGS model well-predicted the true forward scattering in a statistical sense without generating any significant negative dissipation.

Lastly, the TFSGS model underwent the ultimate \textit{a posteriori} analysis, which verified numerically stable performance of the fractional model through tracking turbulent kinetic energy and enstrophy. With the emphasis on remarkable potentials and merits of the present work, we believe that this approach can be extended to more complex turbulent flows by employing a variety of rigorous fractional operators, derived from the statistical structures. 


\section*{Acknowledgment}

\noindent This work was financially supported by the MURI/ARO grant (W911NF-15-1-0562), the ARO Young Investigator Program (YIP) award (W911NF-19-1-0444), and partially by the National Science Foundation award (DMS-1923201). The high-performance computing resources and services were provided by the Institute for Cyber-Enabled Research (ICER) at Michigan State University.

\appendix

\section{}\label{appA}
As noted in \citep[][]{Deng2018, DiNezza2012}, the tempered fractional Laplacian operator can be represented in various equivalent forms, i.e.,
\begin{eqnarray}
\label{eqApp-1}
(\Delta+\lambda)^{\alpha} u(\boldsymbol{x}) &=& -C_{d,\alpha} \, \mathrm{P.V.} \int_{\mathbb{R}^d }\frac{u(\boldsymbol{x})-u(\boldsymbol{\mathfrak{s}})}{e^{\lambda\vert\boldsymbol{x}-\boldsymbol{\mathfrak{s}}\vert}\vert \boldsymbol{x}-\boldsymbol{\mathfrak{s}}\vert^{2\alpha+d}} d\boldsymbol{\mathfrak{s}} 
\nonumber
\\
\label{eqApp-1-1}
&=& \frac{C_{d,\alpha}}{2} \, \mathrm{P.V.} \int_{\mathbb{R}^d }\frac{u(\boldsymbol{x}+\boldsymbol{\mathfrak{s}})+u(\boldsymbol{x}-\boldsymbol{\mathfrak{s}})-2u(\boldsymbol{x})}{e^{\lambda \mathfrak{s}} \,  \mathfrak{s}^{2\alpha+d}} d\boldsymbol{\mathfrak{s}},
\end{eqnarray}
where $\mathfrak{s} = \vert \boldsymbol{\mathfrak{s}}\vert$, $\alpha \in (0,\frac{1}{2}) \cup (\frac{1}{2},1)$, and $\lambda>0$. By performing the Fourier transform of \eqref{eqApp-1-1}, we get
\begin{eqnarray}
\label{eqApp-2}
\mathcal{F} \big {[}(\Delta+\lambda)^{\alpha} u(\boldsymbol{x}) \big {]} (\boldsymbol{\xi})  &=& \frac{C_{d,\alpha}}{2} \int_{\mathbb{R}^d}^{} \frac{e^{\boldsymbol{\xi}\cdot \boldsymbol{\mathfrak{s}}}+e^{-\boldsymbol{\xi}\cdot \boldsymbol{\mathfrak{s}}}-2}{e^{\lambda \mathfrak{s}} \, s^{2\alpha+d}} d\boldsymbol{\mathfrak{s}}
\nonumber
\\
&=& - C_{d,\alpha} \int_{\mathbb{R}^d}^{} \frac{1-\cos(\boldsymbol{\xi}\cdot \boldsymbol{\mathfrak{s}})}{e^{\lambda \mathfrak{s}} \, \mathfrak{s}^{n+2\alpha}} d\boldsymbol{\mathfrak{s}} \, \mathcal{F} \big {[}u(\boldsymbol{x}) \big {]} (\boldsymbol{\xi}),
\end{eqnarray}
in which $\boldsymbol{\xi}$ denotes the Fourier numbers. 
For the sake of simplicity, we define 
$$I(\boldsymbol{\xi})=-\int_{\mathbb{R}^d}^{} \frac{1-\cos(\boldsymbol{\xi}\cdot \boldsymbol{\mathfrak{s}})}{e^{\lambda \mathfrak{s}} \, \mathfrak{s}^{n+2\alpha}} d\boldsymbol{\mathfrak{s}},$$
which appears to be rotationally invariant. Moreover, we introduce $\xi = \vert \boldsymbol{\xi}\vert$ and $\mathfrak{s}_{\theta} = \mathfrak{s} \cos(\theta)$. Without loss of generality, $\theta$ can be chosen such that  $\mathfrak{s} \, \cos(\theta)$ is aligned with the first primary direction. Therefore, $I(\boldsymbol{\xi})$ can be re-expressed by
\begin{eqnarray}
\label{eqApp-3}
I(\boldsymbol{\xi})=\int_{\mathbb{R}^d}^{} \frac{\cos(\boldsymbol{\xi}\cdot \boldsymbol{\mathfrak{s}})-1}{e^{\lambda \mathfrak{s}} \, \mathfrak{s}^{n+2\alpha}} d\boldsymbol{\mathfrak{s}} = \int_{\mathbb{R}^d}^{} \frac{\cos(\eta_{\theta})-1}{e^{\lambda \eta / \xi} (\eta / \xi)^{n+2\alpha}} \frac{d\boldsymbol{\eta}}{\xi^n} = \xi^{2\alpha} \int_{\mathbb{R}^d}^{} \frac{\cos(\eta_{\theta})-1}{e^{\lambda \eta / \xi} (\eta)^{n+2\alpha}} d\boldsymbol{\eta}, 
\end{eqnarray}
where $\boldsymbol{\eta}=\xi \boldsymbol{\mathfrak{s}}$, $\eta = \vert \boldsymbol{\eta} \vert$, and  $\eta_{\theta} = \xi \mathfrak{s}_{\theta}$. Due to the invariant properties of $I(\boldsymbol{\xi})$, we proceed the derivations with transforming \eqref{eqApp-3} into the corresponding spherical coordinate, $(\mathrm{r},\, \phi_1,\, \cdots,\, \phi_{d-1})$. 

In terms of the transformation, we let $\eta = \vert \boldsymbol{\eta} \vert = \mathrm{r}$ and $\eta_{\theta}=\eta \cos(\theta) = \mathrm{r} \cos(\phi_1)$. Then, in a general case for $d>1$, $d\boldsymbol{\eta}$ follows $$d\boldsymbol{\eta} = \mathcal{J}(\mathrm{r},\, \phi_1,\, \cdots,\, \phi_{d-1}) \, d\mathrm{r}\, d\phi_1 \cdots d\phi_{d-1},$$ where $$\mathcal{J}(\mathrm{r},\, \phi_1,\, \cdots,\, \phi_{d-1})=\left\vert det \frac{\partial x_i}{\partial(\mathrm{r} \phi_j)} \right\vert = \mathrm{r}^{d-1} \sin^{d-2}(\phi_1)\sin^{d-3}(\phi_2) \cdots \sin(\phi_{d-2})$$ for $i=1,\cdots,d$ and $j=1,\cdots,d-1$ \citep[see][]{Henderson2000}. Therefore, we find the general form of $I(\boldsymbol{\xi})$ as
\begin{eqnarray}
\label{eqApp-4}
I(\boldsymbol{\xi}) = \xi^{2\alpha}\, \bar{c} \,\int_{0}^{\infty} \int_{0}^{2\pi}  \frac{\cos \left(\mathrm{r} \cos(\phi_1) \right)-1}{e^{\lambda \mathrm{r} / \xi} (\mathrm{r})^{d+2\alpha}}\, \mathrm{r}^{d-1} \sin^{d-2}(\phi_1) \, d\phi_1 \,  d\mathrm{r},
\end{eqnarray}
where $\bar{c} =  \int_{0}^{\pi} \sin^{d-3}(\phi_2) \, d\phi_2 \cdots \int_{0}^{\pi} \sin(\phi_{d-1}) \, d\phi_{d-1} = \frac{2\pi^{(d-1)/2}}{\Gamma(\frac{d-1}{2})} $. It is shown by \cite{Deng2018} that
\begin{eqnarray}
\label{eqApp-5}
I(\boldsymbol{\xi}) &=& \bar{c} \, \xi^{2\alpha} \,\int_{0}^{\infty} \frac{e^{-\lambda \mathrm{r}/\xi}}{\mathrm{r}^{\beta+1}} \int_{0}^{2\pi} \big {[} \cos\big (\mathrm{r} \cos(\phi_1) \big )-1  \big {]}  \sin^{d-2}(\phi_1)\, d\phi_1 \, d\mathrm{r}
\nonumber
\\
\nonumber
&=& \frac{\bar{c} \, \Gamma(-2\alpha) \pi^{1/2} \Gamma(\frac{d-1}{2})}{\Gamma(\frac{d}{2})} \left[ \lambda^2 - (\lambda^2+\xi^2)^{\alpha} \prescript{}{2}{F}^{}_{1}(-\alpha, \frac{d+2\alpha-1}{2};\frac{d}{2};\frac{\xi^2}{\xi^2+\lambda^2}) \right],
\end{eqnarray}
where $\prescript{}{2}{F}^{}_{1}$ denotes a Gaussian hypergeometic function. Therefore, 
\begin{eqnarray}
\label{eqApp-6}
\mathcal{F} \left[ (\Delta+\lambda)^{\alpha} u(\boldsymbol{x}) \right] (\boldsymbol{\xi})  &=& \mathfrak{C}_{d,\alpha} \times \left[ \lambda^2 - (\lambda^2+\xi^2)^{\alpha} \prescript{}{2}{F}^{}_{1}(-\alpha, \frac{d+2\alpha-1}{2};\frac{d}{2};\frac{\xi^2}{\xi^2+\lambda^2}) \right], \qquad \qquad 
\end{eqnarray}
where $\mathfrak{C}_{d,\alpha} = C_{d,\alpha} \,\bar{c} \, \Gamma(-2\alpha) \frac{\pi^{1/2}\Gamma(\frac{d-1}{2})}{\Gamma(d/2)}=\frac{1}{ \prescript{}{2}{F}^{}_{1}(-\alpha, \frac{d+2\alpha-1}{2};\frac{d}{2};1)}$.

\section{}
\label{appB}
\noindent As we discussed in subsection \ref{Sec 2-3}, the SGS stresses are described by 
\begin{eqnarray}
\label{app2-1}
\mathsfi{T}_{ij}^{R} &=& \frac{\rho \, \mathrm{c}_{\beta,\lambda}}{U^3} \int_{0}^{\infty}  \int_{\mathbb{R}^3} (u_i-\bar{V}_i)(u_j-\bar{V}_j) \left(F^{\beta,\lambda}(\bar{\Delta}_{s})-F^{\beta,\lambda}(\bar{\Delta})\right) e^{-s} d\boldsymbol{u}\,  ds
\end{eqnarray}
where $F^{\beta,\lambda}(\bar{\Delta})$ represents a tempered \textit{L\'evy} $\beta$-stable distribution. Let consider $\beta = -\alpha -\frac{3}{2}$ for $\alpha \in (0,\frac{1}{2}) \cup (\frac{1}{2},1)$. Regarding the equivalent \textit{Pareto}-like behavior of \textit{L\'evy} distributions \citep{Weron2001} at $\bar{\Delta} > 1$, we decompose the domain of kinetic momentum such that $\mathbb{R}^3 = \mathcal{I}_{\epsilon} \cup \left( \mathbb{R}^3 \setminus \mathcal{I}_{\epsilon} \right)$, where $\mathcal{I}_{\epsilon}=\{ u\in \mathbb{R}^d\, s.t. \, \vert \bar{\Delta} \vert < \epsilon \}$ and  $\epsilon \ll 1$. This allows for the following approximation as
\begin{equation}
\label{app2-1-1}
F^{\alpha,\lambda}(\bar{\Delta}) \simeq  C_{\alpha}
\begin{cases}
0, & \boldsymbol{u} \in \mathcal{I}_{\epsilon}  \\
e^{-\lambda \, \bar{\Delta}^{\frac{1}{2}}} \bar{\Delta}^{-(\alpha+\frac{3}{2})}, &  \boldsymbol{u} \in \mathbb{R}^3 \setminus \mathcal{I}_{\epsilon}
\end{cases}  
\end{equation}
where $C_{\alpha} = \frac{-\Gamma(\frac{3}{2})}{2 \pi^{\frac{3}{2}}\Gamma(-2\alpha)} \frac{1}{\prescript{}{2}{F}^{}_{1}(-\alpha,1+\alpha;\frac{3}{2};1)}$. It is worth mentioning that $F^{\alpha,\lambda}(\bar{\Delta})$ reduces exponentially in a close proximity of $\bar{\Delta} = 0$. With all this in mind, the approximated function of $F^{\alpha,\lambda}(\bar{\Delta})$ in \eqref{app2-1-1} can properly capture the heavy-tailed behavior of the filtered collision term. Evidently, by replacing $e^{-\lambda \bar{\Delta}_s^{\frac{1}{2}}}$ with $e^{-\lambda \bar{\Delta}^{\frac{1}{2}}}$ for  $\bar{\Delta} >1$, we arrive at the following expression
$$F^{\alpha,\lambda}(\bar{\Delta}_{s})-F^{\alpha,\lambda}(\bar{\Delta}) = C_{\alpha} \left( \frac{e^{-\lambda \bar{\Delta}_s^{\frac{1}{2}}}} {\bar{\Delta}_s^{(\alpha+\frac{3}{2})}} - \frac{e^{-\lambda \bar{\Delta}^{\frac{1}{2}}}}{\bar{\Delta}^{(\alpha+\frac{3}{2})}} \right) \simeq C_{\alpha} \, e^{-\lambda \bar{\Delta}^{\frac{1}{2}}} \left( \frac{1}{\bar{\Delta}^{(\alpha+\frac{3}{2})}_s }-  \frac{1}{\bar{\Delta}^{(\alpha+\frac{3}{2})}} \right),$$ 
and accordingly, 
\begin{eqnarray}
\label{app2-2}
\mathsfi{T}_{ij}^{R} &=& \frac{\rho \, \mathrm{c}_{\alpha,\lambda}}{U^3} \, C_{\alpha} \int_{0}^{\infty}  \int_{\mathbb{R}^3 \setminus \mathcal{I}_{\Delta}} (u_i-\bar{V}_i)(u_j-\bar{V}_j) e^{-\lambda \bar{\Delta}^{\frac{1}{2}}} \left( \frac{1}{\bar{\Delta}^{(\alpha+\frac{3}{2})}_s }-  \frac{1}{\bar{\Delta}^{(\alpha+\frac{3}{2})}} \right) e^{-s} d\boldsymbol{u}\,  ds,
\end{eqnarray}
where $ \mathrm{c}_{\alpha,\lambda}$ is a real-valued constant. As a continuous differentiable function for $\bar{\Delta} >1$, we proceed with the Taylor expansion of $\bar{\Delta}_s^{-(\alpha+\frac{3}{2})}$ according to $$\bar{\Delta}_s^{-(\alpha+\frac{3}{2})} -  \bar{\Delta}^{-(\alpha+\frac{3}{2})} \approx \frac{\partial\bar{\Delta}^{-(\alpha+\frac{3}{2})} }{\partial \bar{\Delta}} (\bar{\Delta}_s-\bar{\Delta}) = -(\alpha+\frac{3}{2}) \,  C_{\alpha}\, \frac{(\bar{\Delta}_s-\bar{\Delta})}{\bar{\Delta}^{\alpha+5/2}}.$$ In terms of the assumptions in remark \ref{remark 1}, we use the same argument, presented by \citep[][Appendix]{SamieePoF2020}, on approximating $\bar{\Delta}_s-\bar{\Delta}$ for $\bar\Delta \gg 1$, which allows for $u_i -\bar{V}_i \approx u_i$ and thus 
\begin{equation}
\label{AP_10_1}
\bar{\Delta}_s-\bar{\Delta} \approx 2 \sum_{k=1}^{3}\frac{ u_k(\bar{V}_k(\boldsymbol{x}^{\prime})-\bar{V}_k(\boldsymbol{x}))}{U^2}.
\end{equation}
Reminding the definition of $\boldsymbol{u} = \frac{\boldsymbol{x} - \boldsymbol{x}^{\prime}}{s\, \tau}$ from section \ref{Sec 2-2}, we plug \eqref{AP_10_1} into \eqref{app2-2} and obtain
\begin{eqnarray}
\label{app2-3}
\mathsfi{T}_{ij}^{R} &=& (2\alpha+3)(\rho\, \mathrm{c}_{\alpha,\lambda}\, C_{\alpha} \tau^{2\alpha-1} U^{2\alpha})  \times 
\\
\nonumber
&& \qquad \qquad \qquad \int_{0}^{\infty} \frac{e^{-s}}{s^{1-2\alpha}} \int_{\mathbb{R}^3 \setminus \mathcal{I}_{\Delta}} (x_i-x_i^{\prime}) (x_j-x_j^{\prime}) \frac{(\boldsymbol{x} -\boldsymbol{x}^{\prime})\cdot (\bar{\boldsymbol{V}}(\boldsymbol{x})- \bar{\boldsymbol{V}} (\boldsymbol{x}^{\prime}) ) }{\vert \boldsymbol{x} -\boldsymbol{x}^{\prime}\vert^{2\alpha+5} e^{\lambda \frac{\vert \boldsymbol{x} -\boldsymbol{x}^{\prime}\vert}{s\tau U}}} d\boldsymbol{x}^{\prime} ds. 
\end{eqnarray}

In order to evaluate the outer integral in \ref{app2-3} and find the corresponding coefficient, our approach is to dissociate the temporal element by employing the bionomial series of $e^{\lambda \frac{\vert \boldsymbol{x} -\boldsymbol{x}^{\prime}\vert}{s\tau U}}$ as follows:
\begin{eqnarray}
e^{ -\frac{\lambda \vert \boldsymbol{x} -\boldsymbol{x}^{\prime}\vert}{s\tau U}} &=& (1-1+e^{- \frac{\lambda \vert \boldsymbol{x} -\boldsymbol{x}^{\prime}\vert}{s\tau U}})^{\frac{1}{s}} = \sum_{k=0}^{\infty} \binom{\frac{1}{s}}{k} \, (e^{\bar{\lambda}\vert \boldsymbol{x} -\boldsymbol{x}^{\prime}\vert}-1)
\nonumber
\\
&=& 1 + \frac{1}{s} (e^{\bar{\lambda}\vert \boldsymbol{x} -\boldsymbol{x}^{\prime}\vert}-1) + \frac{\frac{1}{s}(\frac{1}{s}-1)}{2!}  (e^{\bar{\lambda}\vert \boldsymbol{x} -\boldsymbol{x}^{\prime}\vert}-1) + \cdots 
\nonumber
\\
&=& \sum_{k=0}^{\infty} W_{k,\infty}(s) \, e^{-\bar{\lambda}_k \vert \boldsymbol{x} -\boldsymbol{x}^{\prime}\vert }
\nonumber
\\
&\simeq& \sum_{k=0}^{\mathcal{K}} W_{k,\mathcal{K}}(s) \, e^{-\bar{\lambda}_k \vert \boldsymbol{x} -\boldsymbol{x}^{\prime}\vert },
\end{eqnarray}
where $\bar{\lambda}=\frac{\lambda}{\tau U}$ and $\bar{\lambda}_k=\bar{\lambda}$. Under the assumption of $\lambda > 0.01$, we can approximate the binomial series with the first two leading terms, which yields $W_{0,1}=1-\frac{1}{s}$ and $W_{1,1}=\frac{1}{s}$ for $\mathcal{K}=1$. Accordingly, by defining $\bar{\nu}_{\alpha} = (2\alpha+3)(\rho \, C_{\alpha} \, \tau^{2\alpha-1} U^{2\alpha})$ and 
\begin{equation}
\label{app2-3-1}
\bar{\phi}_{k}^{\mathcal{K}}(\alpha) = \int_{0}^{\infty} \frac{e^{-s}}{s^{1-2\alpha}} \, W_{k,\mathcal{K}}(s) \, ds,
\end{equation} 
we obtain the closed form of $\mathsfi{T}_{ij}^{R}$ as
\begin{equation}
\label{app2-4}
\mathsfi{T}_{ij}^{R} =  \mathrm{c}_{\alpha,\lambda} \, \bar{\nu}_{\alpha} \sum_{k=0}^{\mathcal{K}} \bar{\phi}_{k}^{\mathcal{K}}(\alpha) \int_{\mathbb{R}^d-B_{\epsilon}} \underbrace{(x_i -x_i^{\prime}) \, (x_j -x_j^{\prime}) \frac{(\boldsymbol{x} -\boldsymbol{x}^{\prime})\cdot (\bar{\boldsymbol{V}}- \bar{\boldsymbol{V}}^{\prime} ) }{\vert \boldsymbol{x} -\boldsymbol{x}^{\prime}\vert^{2\alpha+5} e^{\bar{\lambda}_k \vert \boldsymbol{x} -\boldsymbol{x}^{\prime} \vert}}}_{\mathcal{I}_{ij}}     d\boldsymbol{x}^{\prime}.
\end{equation} 

To ensue the proper form of the SGS stresses in the filtered NS equations, we take the derivative of $\mathcal{I}_{ij}$ term by term, which yields
\begin{eqnarray}
\label{app2-5}
\nonumber
\frac{\partial \mathcal{I}_{ij}}{\partial x_i} = \int_{\mathbb{R}^d-B_{\epsilon}} \sum_{i=1}^{3} \Big {\{} && -(x_j -x_j^{\prime}) \frac{(\boldsymbol{x} -\boldsymbol{x}^{\prime})\cdot (\bar{\boldsymbol{V}}- \bar{\boldsymbol{V}}^{\prime} ) }{\vert \boldsymbol{x} -\boldsymbol{x}^{\prime}\vert^{2\alpha+5} \, e^{\bar{\lambda}_k \vert \boldsymbol{x} -\boldsymbol{x}^{\prime} \vert}}
\\
\nonumber
&& 
- (x_i -x_i^{\prime}) \, \delta_{ij} \, \frac{(\boldsymbol{x} -\boldsymbol{x}^{\prime})\cdot (\bar{\boldsymbol{V}}- \bar{\boldsymbol{V}}^{\prime} ) }{\vert \boldsymbol{x} -\boldsymbol{x}^{\prime}\vert^{2\alpha+5} \, e^{\bar{\lambda}_k \vert \boldsymbol{x} -\boldsymbol{x}^{\prime} \vert}} 
\nonumber
\\
\nonumber
&&  - \frac{(x_j -x_j^{\prime}) (x_j -x_j^{\prime}) (\bar{V}_i - \bar{V}^{\prime}_i ) }{\vert \boldsymbol{x} -\boldsymbol{x}^{\prime}\vert^{2\alpha+5} \, e^{\bar{\lambda}_k \vert \boldsymbol{x} -\boldsymbol{x}^{\prime} \vert}}
\\
\nonumber
&& 
-(x_i -x_i^{\prime}) (x_j -x_j^{\prime})(x_k -x_k^{\prime}) \frac{\partial \bar{V}_k}{\partial x_i} \frac{e^{-\bar{\lambda}_k \vert \boldsymbol{x} -\boldsymbol{x}^{\prime} \vert}}{\vert \boldsymbol{x} -\boldsymbol{x}^{\prime}\vert^{2\alpha+5}}
\\
\nonumber
&&(2\alpha+5) (x_i -x_i^{\prime})^2 (x_j -x_j^{\prime}) \frac{(\boldsymbol{x} -\boldsymbol{x}^{\prime})\cdot (\bar{\boldsymbol{V}}- \bar{\boldsymbol{V}}^{\prime} ) }{\vert \boldsymbol{x} -\boldsymbol{x}^{\prime}\vert^{2\alpha+5} \, e^{\bar{\lambda}_k \vert \boldsymbol{x} -\boldsymbol{x}^{\prime} \vert}} 
\\
\nonumber
&&
+ \bar{\lambda}_k (x_i -x_i^{\prime}) (x_j -x_j^{\prime}) \frac{(\boldsymbol{x} -\boldsymbol{x}^{\prime})\cdot (\bar{\boldsymbol{V}}- \bar{\boldsymbol{V}}^{\prime} ) }{\vert \boldsymbol{x} -\boldsymbol{x}^{\prime}\vert^{2\alpha+5} \, e^{\bar{\lambda}_k \vert \boldsymbol{x} -\boldsymbol{x}^{\prime} \vert}}
\Big {\}} \, d\boldsymbol{x}^{\prime},
\end{eqnarray}
which is clearly simplified to
\begin{eqnarray}
\label{app2-6}
\frac{\partial \mathcal{I}_{ij}}{\partial x_i} &=& (\bar{\lambda}_k + 2\alpha +5-3-1-1) \int_{\mathbb{R}^d-B_{\epsilon}}  (x_j -x_j^{\prime})  \frac{(\boldsymbol{x} -\boldsymbol{x}^{\prime})\cdot (\bar{\boldsymbol{V}}- \bar{\boldsymbol{V}}^{\prime} ) }{\vert \boldsymbol{x} -\boldsymbol{x}^{\prime}\vert^{2\alpha+5} \, e^{\bar{\lambda}_k \vert \boldsymbol{x} -\boldsymbol{x}^{\prime} \vert}} d\boldsymbol{x}^{\prime}.
\end{eqnarray}

Following the derivations in \citep[][]{SamieePoF2020, Epps2018}, \eqref{app2-6} can be formulated in the form of a tempered fractional Laplacian by performing the technique of integration-by-parts for \eqref{app2-6} as $\int A dB = AB - \int B dA$. We consider 
\begin{equation}
A=(x_j -x_j^{\prime})(\bar{V}_k - \bar{V}^{\prime}_k ) \, e^{-\bar{\lambda}_k \vert \boldsymbol{x} -\boldsymbol{x}^{\prime} \vert}, \qquad \qquad dB = \frac{(x_k -x_k^{\prime})}{\vert \boldsymbol{x} -\boldsymbol{x}^{\prime}\vert^{2\alpha+5}} \, d\boldsymbol{x}^{\prime},
\end{equation}
which directly leads to $AB|_{u\in \mathbb{R}^3} \simeq 0$. Therefore, we get $\int A \, dB = - \int B \, dA$, in which
\begin{eqnarray}
\label{app2-6-2}
B&=&\frac{-1}{(2\alpha+3) \vert \boldsymbol{x} -\boldsymbol{x}^{\prime}\vert^{2\alpha+3}}, \qquad \qquad 
\nonumber
\\
\nonumber
dA &=& \delta_{jk} (\bar{V}_k - \bar{V}^{\prime}_k )e^{-\bar{\lambda}_k \vert \boldsymbol{x} -\boldsymbol{x}^{\prime} \vert} + (x_j -x_j^{\prime})\left(\frac{\partial \bar{V}_k}{\partial x_k} \right)e^{-\bar{\lambda}_k \vert \boldsymbol{x} -\boldsymbol{x}^{\prime} \vert}- \bar{\lambda}_k (x_j -x_j^{\prime})(\bar{V}_k - \bar{V}^{\prime}_k ) e^{-\bar{\lambda}_k \vert \boldsymbol{x} -\boldsymbol{x}^{\prime} \vert} d\boldsymbol{x}^{\prime}.
\end{eqnarray}
We can make even more simplifications by eliminating the second term of $dA$ due to the incompressibility assumption, i.e., $\partial \bar{V}_k/\partial x_k=0$. Moreover, by evaluating $\int B \, dA$ the last term vanishes since it represents an odd function of $\boldsymbol{x}^{\prime}$. Therefore, the ultimate form of the TFSGS model is found to be  
\begin{eqnarray}
\label{app2-7}
(\nabla \cdot \mathsfbi{T}^{\mathcal{R}})_j &=&  \mathrm{c}_{\alpha,\lambda} \, \bar{\nu}_{\alpha}  \sum_{k=0}^{\mathcal{K}} \frac{(2\alpha+\bar{\lambda}_k)}{(2\alpha+3)} \, \bar{\phi}_{k}^{\mathcal{K}}(\alpha) \int_{\mathbb{R}^d-B_{\epsilon}} \frac{(\bar{V}_j -\bar{V}_j^{\prime}) }{\vert \boldsymbol{x} -\boldsymbol{x}^{\prime}\vert^{2\alpha+3} \, e^{\bar{\lambda}_k \vert \boldsymbol{x} -\boldsymbol{x}^{\prime} \vert}}  d\boldsymbol{x}^{\prime},
\nonumber
\\
&=& \nu_{\alpha} \sum_{k=0}^{\mathcal{K}} \phi_{k}^{\mathcal{K}}(\alpha,\bar{\lambda}_k) (\Delta+\bar{\lambda}_k)^{\alpha} \, \bar{V}_j,
\end{eqnarray}
where $ \nu_{\alpha} = \mathrm{c}_{\alpha,\lambda} \, \bar{\nu}_{\alpha} $.


\bibliographystyle{jfm}
\bibliography{Ref}

\end{document}